\documentclass[letterpaper]{aa}
\input psfig
\begin{document}

	
%
   \title{The planet search program at the ESO Coud\'e Echelle 
	spectrometer.
   \thanks{Based on observations collected at the 
           European Southern Observatory, La~Silla}}
   \subtitle{III. The complete Long Camera survey results.}
   \author{M. Endl \inst{1,} \inst{2,} \inst{3} \and
	   M. K\"urster \inst{3,} \inst{4} \and
           S. Els \inst{5,} \inst{4,} \inst{3,} \inst{7} \and
	   A. P. Hatzes \inst{4} \and 
	   W.D. Cochran \inst{2} \and
	   K. Dennerl \inst{6} \and
	   S. D\"obereiner \inst{6} 
         }
 
   \offprints{M.~Endl: mike@astro.as.utexas.edu}

   \institute{Universit\"at Wien, Institut f\"ur Astronomie,
		T\"urkenschanzstr.~17, A-1180 Wien, Austria
	 \and McDonald Observatory, The University of Texas at Austin,
	      Austin, TX 78712-1083, USA, \\ 
	 e-mail M.Endl:{\it mike@astro.as.utexas.edu}, W.D.Cochran: {\it wdc@shiraz.as.utexas.edu}		
	 \and European Southern Observatory, Casilla 19001, Vitacura,
              Santiago 19, Chile
	 \and Th\"uringer Landessternwarte Tautenburg, Sternwarte 5,
	      07778 Tautenburg, Germany, \\ 
	      e-mail M.K\"urster: {\it martin@tls-tautenburg.de}, A.P.Hatzes: {\it artie@jupiter.tls-tautenburg.de} 
	 \and Isaac Newton Group of Telescopes, Apartado de Correos 321,
		E-38700 Santa Cruz de La Palma, Spain, \\
		e-mail S.Els: {\it sels@ing.iac.es}
 	 \and Max--Planck--Institut f\"ur extraterrestrische Physik,   
   	      Giessenbachstr., D-85748 Garching, Germany \\
		e-mail K.Dennerl: {\it kod@mpe.mpg.de}  
 	 \and Universit\"at Heidelberg, Institut f\"ur Theoretische
              Astrophysik, Tiergartenstr.~15, D--69121 Heidelberg, Germany, \\ 
	}

   \date{Received date / accepted date}

   \titlerunning{The planet search program at the ESO CES III.}
   \authorrunning{M. Endl et al.}



\abstract{
We present the complete results of the planet search program carried out at the ESO
Coud\'e Echelle Spectrometer (CES) on La Silla, using the Long Camera from Nov.\,1992 to April 1998.
The CES survey has monitored $37$ late-type (F8V -- M5V) stars in the southern
hemisphere for variations in their differential radial velocities ($RV$) in order to detect
Doppler reflex motions caused by planetary companions. 
This led to the discovery of the first extrasolar planet in an Earth-like orbit around the
young (ZAMS) and active G0V star $\iota$ Horologii (K\"urster et al. \cite{martin00}).  
Here we present the $RV$ results for all survey stars and perform a statistical examination of
the whole data-set. Each star is tested for $RV$ variability, $RV$ trends (linear and non-linear) and
significant periodic signals. $\beta$~Hyi and $\epsilon$~Ind are identified as long-term, low-amplitude
$RV$ variables. Furthermore, for $30$ CES survey stars we determine quantitative upper mass-limits 
for giant planets based on our long-term $RV$ results. We find that 
the CES Long Camera survey would have detected short-period (``51 Peg-type'') planets 
around $all$ $30$ stars but no planets with $m\sin i < 1~{\rm M}_{\rm Jup}$ at orbital
separations larger than 2 AU. Finally, we demonstrate that the CES planet search can be
continued without applying velocity corrections to the $RV$ results coming from the currently installed
Very Long Camera at the CES.  
}

\maketitle

%
\section{Introduction}

The exciting discoveries of giant planets orbiting solar-type stars by precise Doppler
searches have caused a shift 
in our paradigm of the structure and formation of planetary systems. Although we now know that planets 
have also formed around stars other than the Sun, their orbital characteristics turned out
to be quite exotic (for an overview see e.g. Marcy, Cochran \& Mayor~\cite{proto}).   

Extrasolar giant planets were detected in very close-by orbits around
their host stars with periods on the order of a few days, while orbital eccentricities 
at longer periods appear to be distributed quite uniformly. To date no Solar System
analogue has been detected which is primarily due to the insufficient time baseline and long-term 
$RV$ precision of present Doppler surveys. However, the detection of Jovian-mass companions
with $P>10$ yrs will become possible in the near future.     

One of these long-term $RV$ surveys is the planet search program at the 
Coud\'e Echelle Spectrometer (CES)
at ESO La Silla, which was begun in Nov. 1992 using the 1.4 m CAT telescope. 
The highlight of this program so far was the
discovery of an extrasolar giant planet in an Earth-like orbit around the young (ZAMS) and modestly
active G0V star $\iota$ Horologii (K\"urster et al.~\cite{martin00}).   

It is important to set such discoveries into the context of the complete
results obtained by planet search programs. 
The pioneering study by Walker et al. (\cite{walker}) first
presented long-term (12 years) $RV$ results for a sample of 21 stars and discussed the
implications of their non-detections on the occurrence of Jovian-type planets around solar-type stars. 
In an even earlier work, Murdoch et al. (\cite{murdoch}) presented an analysis of their $RV$
measurements for 29 stars over 2.5 years, finding no brown dwarf companions within 10 AU in their
sample. Since then $RV$ measurement precision (e.g. Butler et al.~\cite{butler96}) and the 
size of target samples has increased dramatically. Extrasolar giant planets, which
can be detected by present Doppler searches, exist around $\approx 3 - 5\%$ of the observed 
solar-type stars. Another study of the long-term $RV$ behaviour of a sample of stars was presented 
by Cumming et al. (\cite{cumming}). These authors examined 11 years of $RV$ data collected 
by the Lick survey for 76 F-, G-, and K-type stars and derived companion limits for these stars.

With this work we present all $RV$ measurements of the CES survey and a complete analysis thereof over
the time period of November 1992 to April 1998. During that time observations were performed with
the same telescope and spectrograph configuration and thus form a homogenous data set. After April 1998
the CES instrument underwent major modifications and the results based on data collected after 
that point of time will be presented in an upcoming paper.    

The structure of this paper is the following: Sect. 2 gives an overview of the CES planet
search program, Sect. 3 presents the complete $RV$ results of the CES targets (appendix 
A displays the $RV$ measurements graphically for each star),
Sect. 4 is a statistical examination of the CES $RV$ data where we
perform tests to identify variable stars, the presence of linear and non-linear trends and periodic signals,
in section 5 we set quantitative upper mass-limits for orbiting planets based on the $RV$ results 
(appendix B shows the derived limits for each survey star) and finally Sects. 6 and 7 contain the
discussion and summary.               

\section{The Coud\'e Echelle Spectrometer planet search program on La Silla}

The search for extrasolar planets in the southern hemisphere using the
Coud\'e Echelle Spectrometer at ESO La Silla was started in November 1992
(K\"urster et al.~\cite{martin94}; Hatzes et al.~\cite{artie96}).
At the beginning, the CES survey was a ``classical'' $RV$ planet search program in the sense that at a
time when no extrasolar planet had been found, the common expectation was to discover
planets similar to Jupiter. Thus the observing strategy was tailored for long-period
and low amplitude signals. Observations were performed on an irregular temporal basis, starting with
2-night runs performed every other month.
The sampling density was later increased, after the discovery of the short-period
planet around $51$~Peg by Mayor \& Queloz (\cite{mayor}), 
to assure also detection capability for planets of this type.

\subsection{The target sample}

At the beginning of the survey targets were selected according to the following criteria:
Stars with $V < 6$ (with few exceptions) to attain a sufficient S/N-ratio, spectral type F8V -
M5V, Stars with declination $< 10^{\circ}$ to avoid overlap with surveys in the northern
hemisphere (again with some exceptions), known (at that time) close binaries were rejected (with
the $\alpha$ Centauri system being another exception) and known active stars were neglected.

\begin{figure}
\centering{
       \vbox{\psfig{figure=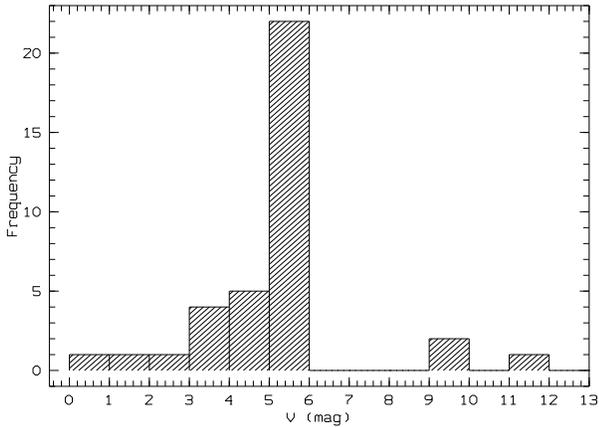,width=9.0cm,height=6.0cm,angle=270}}
       \par
       }
       \caption[]{Histogram of $V$-magnitudes of the CES targets. The distribution peaks
        in the magnitude range of 5 - 6 mag while the 3 M-dwarfs form the faint ``tail''
        on the right side.
       }
       \label{maghist}
\end{figure}

\begin{figure}
 \centering{
   \vbox{\psfig{figure=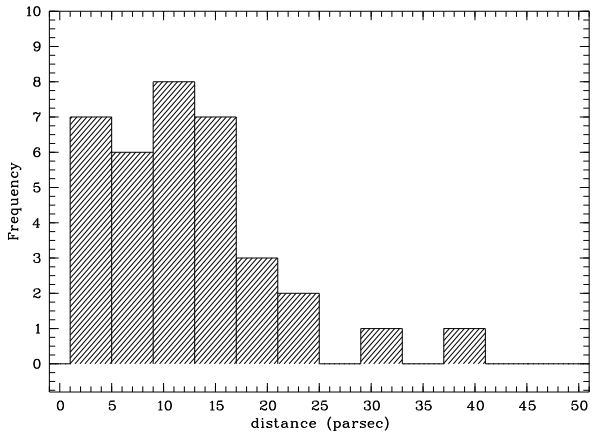,width=9.0cm,height=6.0cm,angle=270}}
   \par
        }
  \caption[]{Histogram of the distances in the CES sample. Not shown are the two
        stars with distances of more than 100 pc: HR~3677 (192.3 pc) \& HR~8883
        (100.8 pc). The rest of the targets are located within 50 pc, with the
        bulk lying closer than 20 pc. Distances are based on $Hipparcos$ parallaxes
	(ESA (\cite{hipp})).
        }
  \label{disthist}
\end{figure}

The final target list consists of 37 bright late-type stars
(mostly) in the southern hemisphere: 6 F-, 21 G-, 7 K-, and 3 M-type stars.
Table 1 summarizes all targets with their HR, HD and GJ catalogue number,
spectral classification, $V$-magnitudes, distance in parsecs, chromospheric emission indices
(log$R^{/}_{\rm HK}$) and their X-ray luminosity ($L_{X}$).
Distances are based on the $Hipparcos$ parallaxes (ESA (\cite{hipp})), the log$R^{/}_{\rm HK}$-values 
are taken from the survey of Ca II H \& K emission in southern solar-type
stars by Henry et al. (\cite{henry96}) and the $L_{X}$ values are coming from the
RASS (ROSAT All Sky Survey) results (H\"unsch et al.~\cite{huensch3},\cite{huensch1}).
Chromospheric emission in the cores of the Ca II H\&K lines and the
X-ray luminosities serve as stellar activity indicators since increased chromospheric Ca II H\&K 
and coronal X-ray emission is a common sign of active stars.
Stellar activity can produce intrinsic $RV$ variability and adds an additional noise source into the RV
measurement (e.g. Saar \& Donahue~\cite{saar1}). It can even mimic the $RV$ signature 
of a short-period planet like in the case of HD 166435 (Queloz et al.~\cite{queloz}). 
In 1992 the Ca II and X-ray results were not known, which would have probably led to the
exclusion of stars like $\kappa$~For, $\iota$~Hor, $\alpha$~For and HR 8883. 
Fig.~\ref{maghist} and Fig.~\ref{disthist} show the $V$-magnitude and distance
histograms of the CES target sample.

Not all targets were monitored since November 1992: three targets, HR 753, HR 5568 and GJ 433, were
added to the sample in May 1997 as $Hipparcos$ candidates for having 
short-period substellar companions (H.-H. Bernstein \& U. Bastian priv. comm.), 
while HR 7373 was only observed for a short time in
1996 and 1997 searching for a short-period planetary companion. 

Although listed as a target, $\tau$ Cet was mainly included as a reference object since it is a
known long-term RV-constant star (Campbell et al.~\cite{campbell}; Walker et al.~\cite{walker})
and two other stars with known extrasolar planets, 51 Peg (Mayor \& Queloz~\cite{mayor}) and 70 Vir
(Marcy \& Butler~\cite{marcy96}), were also observed after 1995 to serve as $RV$ precision check
stars.

\begin{table*}
\begin{center}
\begin{tabular*}{0.99\textwidth}{@{\extracolsep{\fill}}cccccrrlr}
\hline
\hline
 & & & & & & & \\
HR & HD & GJ & Name & Sp.type & $V$ & d & log$R^{/}_{\rm HK}$ & $L_{X}$ \\
 & & & & & [mag] & [pc] & & $10^{27}~{\rm [erg/s]}$ \\ 
\hline
77 & 1581 & 17 & $\zeta$ Tuc & F9V & 4.2 & 8.59 & -4.85 & \\
98 & 2151 & 19 & $\beta$~Hyi & G2IV & 2.8 & 7.47 & -4.99 & 6.4 \\
209 & 4391 & 1021 & & G1V & 5.8 & 14.94 & -4.55 & 55.8 \\
370 & 7570 & 55 & $\nu$ Phe & F8V & 4.96 & 15.05 & -4.95 & 14.3 \\
448 & 9562 & 59.2 & & G2IV & 5.76 & 29.66 & -5.10 & \\
506 & 10647 & 3109 & & F9V & 5.52 & 17.35 & & \\
509 & 10700 & 71 & $\tau$ Cet & G8V & 3.5 & 3.65 & -4.96 & 1.1 \\
695 & 14802 & 97 & $\kappa$ For & G0V & 5.19 & 21.93 & & 397.1 \\
753 & 16160 & 105A & & K3V & 5.82 & 7.21 & -4.85 & 1.8 \\
810 & 17051 & 108 & $\iota$ Hor & G0V & 5.41 & 17.24 & -4.65 & 68.3 \\
963 & 20010 & & $\alpha$ For & F8V & 3.87 & 14.11 & & 524.6 \\
1006 & 20766 & 136 & $\zeta^{1}$Ret & G2.5V & 5.54 & 12.12 & -4.65 & 5.8 \\
1010 & 20807 & 138 & $\zeta^{2}$ Ret & G1V & 5.24 & 12.08 & -4.79 & \\
1084 & 22049 & 144 & $\epsilon$ Eri & K2V & 3.73 & 3.22 & -4.47 & 20.9 \\
1136 & 23249 & 150 & $\delta$ Eri & K0IV & 3.51 & 9.04 & -5.22 & 0.9 \\
2261 & 43834 & 231 & $\alpha$ Men & G6V & 5.1 & 10.15 & -4.94 & 2.9 \\
2400 & 46569 & 1089 & & F8V & 5.58 & 37.22 & & \\
2667 & 53705 & 9223A & & G3V & 5.54 & 16.25 & -4.93 & \\
3259 & 69830 & 302 & & G7.5V & 5.95 & 12.58 & & 3.0 \\
3677 & 79807 & & & G0III & 5.86 & 192.31 & & \\
4523 & 102365 & 442A & & G3V & 4.91 & 9.24 & -4.95 & \\  
4979 & 114613 & 9432 & & G3V & 4.85 & 20.48 & -5.05 & 23.0 \\
5459 & 128620 & 559A & $\alpha$ Cen A & G2V & -0.01 & 1.347 & -5.00 & 2.2 \\
5460 & 128621 & 559B & $\alpha$ Cen B & K1V & 1.33 & 1.347 & -4.92 & 2.2 \\
5568 & 131977 & 570A & & K4V & 5.74 & 5.91 & -4.48 & 3.5 \\
6416 & 156274 & 666A & & G8V & 5.47 & 8.79 & -4.94 & 1.9 \\
6998 & 172051 & 722 & & G4V & 5.86 & 12.98 & -4.89 & 4.3 \\
7373 & 182572 & 759 & & G8IV & 5.16 & 15.15 & & 3.9 \\
7703 & 191408 & 783A & & K3V & 5.31 & 6.05 & -4.99 & \\
7875 & 196378 & 794.2 & $\phi^{2}$ Pav & F8V & 5.12 & 24.19 & & \\ 
8323 & 207129 & 838 & & G0V & 5.58 & 15.64 & -4.80 & 11.2 \\
8387 & 209100 & 845 & $\epsilon$~Ind & K4.5V & 4.69 & 3.63 & -4.56 & 1.6 \\
8501 & 211415 & 853A & & G3V & 5.33 & 13.61 & -4.86 & 12.2 \\
8883 & 220096 & & & G4III & 5.66 & 100.81 & & 34175 \\
 & & 699 & Barnard & M4V & 9.54 & 1.82 & & 0.1 \\
 & & 433 & & M2V & 9.79 & 9.04 & & \\
 & & 551 & Prox Cen & M5Ve & 11.05 & 1.29 & & 1.7 \\  
\hline
\hline
\end{tabular*}
\caption[]{
        Target list of the CES planet search program. HR, HD and GJ catalogue numbers,
spectral type, visual magnitude $V$, distance in parsec (based on $Hipparcos$ parallaxes), 
chromospheric emission index from Henry et al. (\cite{henry96})  
and X-ray luminosity $L_{X}$ from H\"unsch et al. (\cite{huensch3},\cite{huensch1})
are given.
        }
\end{center}
\hspace{5cm}
\label{targets}
\end{table*}
%
%

\subsection{Observations}

All stars were observed with the 1.4 m Coud\'e Auxiliary Telescope (CAT) 
on La Silla which fed the CES via a direct beam from the telescope. 
All spectra were taken in a single echelle order centered at a
wavelength of $5389~{\rm \AA}$. The Long Camera yields
a resolving power of $R=100,000$ and a small spectral range of $48.5$~{\rm \AA}.    
After the installation of the Very Long Camera in April 1998 the  
resolving power of the CES was raised to $R=230,000$ but the spectral
coverage was even reduced (depending on which CCD was used).
The results presented in this work all refer to the Long Camera configuration
($R=100,000$) prior to this modification, which thus form a homogeneous
data set. 

In order to assure the necessary high long-term precision for $RV$ measurements 
all CES spectra are self-calibrated by a superimposed absorption spectrum
of molecular iodine (I$_2$) vapor. This is achieved by passing the starlight through 
a temperature controlled cell filled with I$_2$ (see also K\"urster et al. 1994). 

Typical exposure times of the CES survey were $10$ to $15$ minutes, 
and the S/N-ratios of the
obtained spectra were in the range of $100$ to $250$. For the brightest targets
exposure times were much shorter (in the order of $10$ to $30$ seconds), while
for the faint M-dwarfs we set a maximum exposure time of $30$ minutes, in order
to minimize timing uncertainties and subsequent systematic errors in 
the barycentric velocity correction.   

\subsection{Data analysis}

To extract the $RV$ information from I$_2$ self-calibrated spectra it is necessary 
to perform a full spectral modeling. For the analysis of the CES planet search data we 
employ the $Austral$ code which establishes a model of the observation based on 
high resolution templates of the stellar and the I$_2$ spectrum. For a detailed
description of this analysis technique we refer the reader to paper I of this series (Endl et
al.~\cite{michl00}). The modeling 
process includes the reconstruction of the shape and asymmetry of the spectrograph 
instrumental profile (IP) as well as Maximum Entropy Method deconvolution to obtain 
a higher resolved stellar template spectrum. 
All computations are carried out on an oversampled sub-pixel grid and a 
multi-parameter $\chi^{2}$-optimization is performed to achieve a best-fit model.   
The algorithm follows in general the modeling idea
first outlined by Butler et al. (\cite{butler96}) and IP reconstruction techniques by
Valenti et al. (\cite{valenti}). 

The main limiting factor for the achievable $RV$ precision with the CES is the 
small spectral bandwidth of $48.5~{\rm \AA}$. Using different test scenarios we  
demonstrated in Endl et al. (\cite{michl00}) that a long-term $RV$ precision of 
$8 - 15~{\rm m\,s}^{-1}$ was attained.  

\section{Radial velocity results}

\begin{figure}
\centering{
  \vbox{\psfig{figure=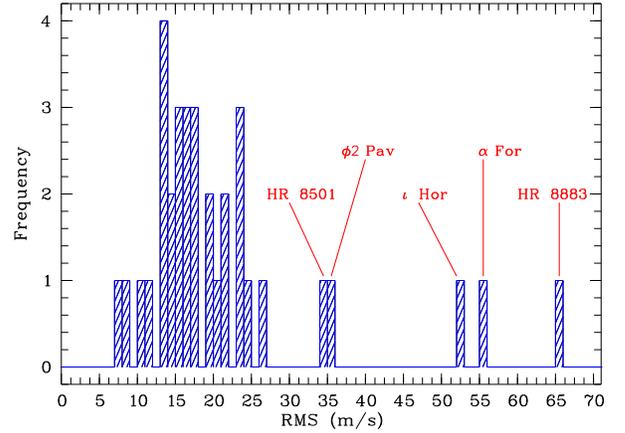,width=9.0cm,height=6.0cm,angle=270}}
   \par
        }
 \caption{Histogram of the $RV$ scatter of all stars with rms $< 100~{\rm m\,s}^{-1}$ and without the
        3 faint M-dwarfs. The distribution peaks at $14~{\rm m\,s}^{-1}$, the stars with higher scatter
	are: HR 8883 ($65.2~{\rm m\,s}^{-1}$), $\alpha$ For ($55.2~{\rm m\,s}^{-1}$),
	$\iota$ Hor ($52.5~{\rm m\,s}^{-1}$), $\phi^{2}$ Pav ($35.3~{\rm m\,s}^{-1}$) and HR 8501
	($34.0~{\rm m\,s}^{-1}$).
}
 \label{histo1}
\end{figure}

By analysing the complete data set of the CES planet search program until 
April 1998 with the $Austral$ code we obtained precise differential radial velocities for all 37
survey stars. 
Table~\ref{all_stars} summarizes the $RV$ results by giving the total $RV$ rms-scatter, the average
internal error for each star, the mean S/N-ratio of the CES spectra and the duration of monitoring
by the CES survey. The internal $RV$ measurement error is the uncertainty of the mean value of the $RV$ distribution
along one CES spectrum of the typically 90-pixel long spectral segments, for which the modeling is performed
independently (see Endl et al.~\cite{michl00} for a detailed description).
A histogram of the $RV$ scatter is shown
in Fig.~\ref{histo1}, with the exclusion of binaries and the 3 fainter M-dwarfs. 

\begin{figure}
\centering{
  \vbox{\psfig{figure=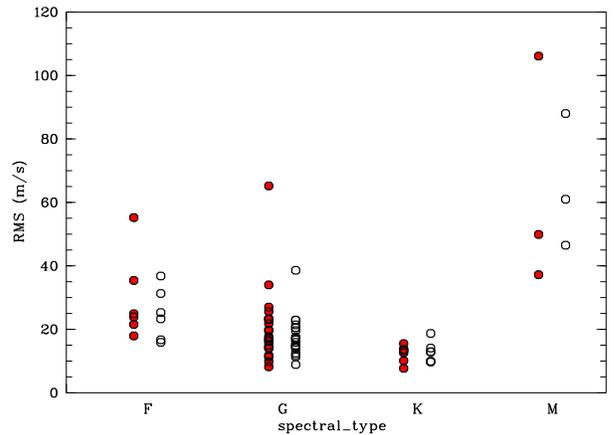,width=9.0cm,height=6.0cm,angle=270}}
   \par
        }  
 \caption[]{
        $RV$ scatter (full circles) of the 37 target stars as a function of spectral type. 
        Open circles represent the mean internal measurement error for the stars in each bin. Minimum
        for both distributions are K-type stars, which can be explained by their higher intrinsic
        line density, while the increase at faint M-type stars is due to the weak signal.
        }
 \label{specrms}
\end{figure}

The average $RV$ rms-scatter of the complete target sample (37 stars) is $24.1~{\rm m\,s}^{-1}$ (in the
cases of $\iota$~Hor (see next section), $\kappa$~For, HR~2400, HR~3677 (three new binaries, see
section 3.2) and $\alpha$~Cen A \& B (see Endl et al.~\cite{michl01}) we take the $RV$ residuals after 
subtraction of either the planetary or stellar secondary signal).  
The dependence of the $RV$ scatter on spectral type is demonstrated in Fig.~\ref{specrms}. 
The average $RV$ scatter for F-type stars is $29.8~{\rm m\,s}^{-1}$ (7 stars),
for G-type stars $20.7~{\rm m\,s}^{-1}$ (21 stars), for K-type stars $12.3~{\rm m\,s}^{-1}$ (7 stars) and
for the M-dwarfs $64.4~{\rm m\,s}^{-1}$ (3 stars). The scatter declines from spectral type F to K, which
can be explained as the functional dependence of the measurement precision on the spectral line
density (velocity information content) in the CES bandpass.
In the case of the short CES spectra
the $RV$ precision is clearly depending on the total number of spectral lines within this bandpass.
Since the
line density is higher for stars with later spectral type, one can expect the highest achievable $RV$
precision for K or M stars. This is the case for K-type stars as demonstrated in Fig.~\ref{specrms}.
The strong increase of scatter and
internal error for the 3 M-type stars is caused by the low S/N-ratio of the obtained spectra (they are
all fainter than $V>9.5$), which degrades the measurement precision despite their higher line density.

\begin{table}
\begin{center}
\begin{tabular}{cr|rr|rr}
\hline
\hline
Star & N & rms & m.int.err. & S/N & T \\
 & & [${\rm m\,s}^{-1}$] & [${\rm m\,s}^{-1}$] & & [days] \\
\hline
$\zeta$ Tuc & 51 & 21.6 & 16.7 & 257 & 1889\\
$\beta$~Hyi & 157 & 23.3 & 20.5 & 161 & 1888\\
HR 209 & 35 & 23.1 & 19.6 & 151 & 1573\\
$\nu$ Phe & 58 & 17.9 & 15.9 & 212 & 1927\\
HR 448 & 24 & 17.1 & 20.5 & 129 & 439\\
HR 506 & 23 & 23.9 & 23.3 & 173 & 1574\\
$\tau$ Cet & 116 & 11.3 & 14.1 & 196 & 1889\\
$\kappa$ For & 40 & 780.9 & 14.8 & 199 & 1890\\
HR 753 & 6 & 10.1 & 18.7 & 118 & 64\\
$\iota$ Hor & 95 & 52.5 & 17.4 & 163 & 1976\\
$\alpha$ For & 65 & 55.2 & 36.8 & 197 & 1890\\
$\zeta^{1}$Ret & 14 & 17.7 & 15.9 & 109 & 185\\
$\zeta^{2}$ Ret & 58 & 21.8 & 16.9 & 180 & 1977\\
$\epsilon$ Eri & 66 & 13.7 & 9.7 & 174 & 1890\\
$\delta$ Eri & 48 & 15.5 & 12.9 & 189 & 1889\\
$\alpha$ Men & 41 & 9.8 & 11.3 & 170 & 1853\\
HR 2400 & 53 & 254.9 & 25.3 & 150 & 1925\\
HR 2667 & 66 & 16.5 & 21.4 & 144 & 1935\\
HR 3259 & 35 & 16.2 & 14.2 & 124 & 1852\\  
HR 3677 & 34 & 486.1 & 16.7 & 145 & 1925\\   
HR 4523 & 27 & 15.0 & 14.5 & 210 & 1925\\
HR 4979 & 52 & 14.0 & 12.5 & 185 & 1934\\  
$\alpha$ Cen A & 205 & 165.3 & 11.9 & 225 & 1853\\
$\alpha$ Cen B & 291 & 205.1 & 9.9 & 206 & 1853\\
HR 5568 & 40 & 7.7 & 12.9 & 114 & 384\\
HR 6416 & 57 & 25.6 & 15.0 & 154 & 1845\\
HR 6998 & 51 & 19.6 & 22.9 & 137 & 1789\\
HR 7373 & 8 & 8.2 & 8.9 & 209 & 266\\
HR 7703 & 30 & 13.3 & 14.1 & 162 & 1042\\
$\phi^{2}$ Pav & 90 & 35.4 & 31.3 & 184 & 1969\\
HR 8323 & 20 & 19.8 & 17.3 & 147 & 1068\\
$\epsilon$~Ind & 73 & 13.5 & 9.9 & 203 & 1889\\
HR 8501 & 66 & 34.0 & 17.5 & 184 & 1890\\
HR 8883 & 31 & 65.2 & 38.6 & 137 & 1259\\  
Barnard & 24 & 37.2 & 46.5 & 31 & 1414\\ 
GJ 433 & 15 & 49.9 & 61.0 & 26 & 337\\ 
Prox Cen & 65 & 106.1 & 88.0 & 18 & 1728\\ 
\hline
\hline
\end{tabular}
\end{center}
\caption[]{Radial velocity results of all survey stars. N is the total number
of analysed spectra, rms is the total scatter of the $RV$s, while m.int.err. gives the
mean internal measurement error, the mean S/N-ratio of the spectra and 
T denotes the duration of monitoring (i.e. the timespan from first to last observation of this star).}  
\label{all_stars}
\end{table}
%
%
%

Appendix A (Fig.~\ref{rvsfig1} - Fig.~\ref{rvsfig10}) presents the $RV$ results for all stars, 
plotted for
comparison in the same time frame (JD 2,448,800 to JD 2,451,000). 
The near sinusoidal $RV$ variation caused by the orbiting planet around $\iota$~Hor clearly
stands out of the rest of the sample (see Fig.~\ref{rvsfig3}).

For the faint M dwarf Prox Cen ($V=11.05$) the larger rms-scatter is caused by the insufficient
S/N-ratio of the CES spectra obtained with the 1.4 m CAT telescope (the average S/N-ratio
of the Prox Cen spectra is only $18$).    
The results for the inner binary (components A \& B) of the $\alpha$~Centauri system were
already presented in Endl et al. (\cite{michl01}).
The large scatter seen in the $RV$ results for $\kappa$~For, HR~2400 and HR~3677 is caused by apparent
binary orbital motion and will be discussed in detail.

\subsection{The planet orbiting $\iota$ Horologii}

The G0V star $\iota$ Hor (HR 810, $V = 5.4$) has been earlier identified as an $RV$ variable star
and thus as a ``hot candidate'' in the CES survey for having a planetary companion
(K\"urster et al.~\cite{martingemini}, K\"urster et al.~\cite{martin99}).
A possible eccentric Keplerian signal with a period of $600$ days was found, but with a low confidence
level. 

\begin{figure}
\centering{
  \vbox{\psfig{figure=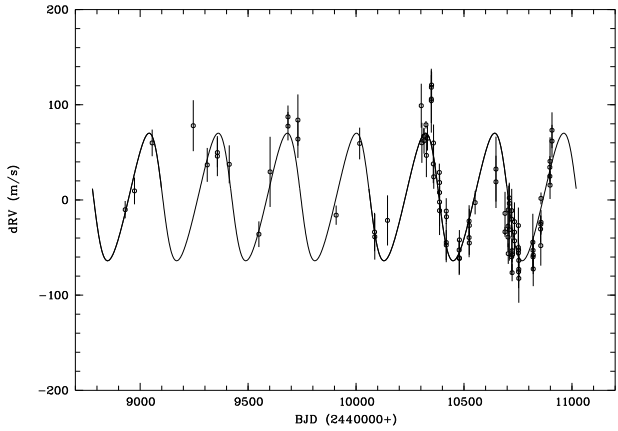,width=9.0cm,height=6.0cm,angle=270}}
   \par
        }
  \caption[]{
        Keplerian orbital solution for $\iota$ Hor (solid line) plotted along the
        $RV$ data. The $RV$ semi-amplitude $K$ is $67.0~{\rm m\,s}^{-1}$, the orbital period
        $320.1$ days and the eccentricity $e = 0.16$. The planet has an $m\sin i$ value of
        $2.26~{\rm M}_{\rm Jup}$.
        The residual rms scatter around this
        best-fit orbit is $27.0~{\rm m\,s}^{-1}$.
        }
 \label{orbit} 
\end{figure}

After the analysis of all $95$ spectra of $\iota$~Hor using the $Austral$ code the resulting $RV$s have
a total rms scatter of $52.5~{\rm m\,s}^{-1}$, an average internal error of $17.4~{\rm m\,s}^{-1}$ and reveal a
near sinusoidal variation which is apparent during the last 2 years of monitoring (see
Fig.~\ref{rvsfig3}). The $95$ spectra were taken between November
1992 and April 1998 and have an average S/N-ratio of $163$.
A period search within this time series using the Lomb-Scargle periodogram (Lomb~\cite{lomb},
Scargle~\cite{scargle}) detected a highly significant signal with a period of $320$ days and a very low
False Alarm Probability (FAP) of $<~10^{-11}$. It was possible to find a Keplerian orbital solution
for these $RV$ data and thus successfully detect an orbiting extrasolar planet. We presented
this discovery already in K\"urster et al. (\cite{martin00}) and we refer the reader to
this earlier paper for a more detailed description. Here we want to summarize the orbital, planetary and
stellar properties. Fig.~\ref{orbit} displays the found Keplerian orbital solution and 
Table~\ref{orb_par} lists the parameters of the planet and its orbit (note that in 
K\"urster et al.~\cite{martin00} the time of maximum $RV$ was given wrong by one day due to a
typo).  
\begin{table}
\begin{center}
\begin{tabular}{ll}
\hline
\hline
 & \\
Minimum planet mass     & $m\sin i=2.26\pm 0.18~{\rm M}_{\rm Jup}$ \\
Orbital period          & $P=320.1\pm 2.1~{\rm d}$ \\
Orbital semi-major axis & $a=0.925\pm 0.104~{\rm AU}$\\
Orbital eccentricity    & $e=0.161\pm 0.069$ \\
RV semi-amplitude       & $K=67.0\pm 5.1~{\rm m\,s}^{-1}$ \\
Time of maximum $RV$    & $T_{\circ }={\rm BJD} 2,450,306.0\pm 3.0$ \\
Periastron angle        & $\omega = 83^{\circ }\pm 11^{\circ }$ \\
 & \\
\hline
\hline
\end{tabular}
\caption[]{Parameters of the planet and its orbit around $\iota$ Hor.}
\label{orb_par}
\end{center}
\end{table}

$\iota$ Hor b was the first planet to be detected residing entirely within the so-called
``habitable zone'' (as defined in Kasting et al.~\cite{kasting}) of its parent star.
The residual rms scatter around the orbit is $27.0~{\rm m\,s}^{-1}$, larger than the error expected
from the $RV$ precision tests in Endl et al. (\cite{michl00}).
A lot of this excess scatter is probably caused by stellar activity as it turned out that $\iota$ Hor
is a quite young (ZAMS) and active star. Both the $RV$ variation caused by the planet as well as the
excess scatter have been confirmed in the meantime by Butler et al.~(\cite{butler01}) and 
Naef et al.~(\cite{dominic}). 

There are indications that the $\iota$ Hor system might host 
additional planetary companions: the periodogram of the $RV$ residuals (after subtraction of
the orbit) reveals a peak at $P\approx620$ days. This 
looks intriguing especially after the detections of extrasolar planets moving in near-resonance 
orbits, e.g. the two companions of HD 83443 in a $10:1$ resonance (Mayor et al.~\cite{mayor2}),
the planetary pair around GJ 876 in a $2:1$ resonance (Marcy et al.~\cite{marcygj876}), 
and the two planets orbiting 47 UMa in a $5:2$ resonance (Fischer et al.~\cite{fischer}). 
We demonstrate in K\"urster et al.~(\cite{martin00}) that the $P\approx620$ days peak is not 
due to spectral leakage from the $P=320$ days signal (see panel (d) of Fig.1 in K\"urster 
et al.~\cite{martin00}). 
This could indicate the presence of 
a second planet located close to the $2:1$ resonance. However, the FAP of this peak is still
above $0.1\%$ and we cannot confirm yet the presence of a second companion. After the 
replacement of the Long Camera at the CES with the Very Long Camera we continued to monitor 
$\iota$ Hor using the same I$_2$-cell for self-calibration. The analysis of the
new data and merging it with the Long Camera data set might allow us in the near future to
verify the existence of the second planet.    

The CES Long Camera results also contributed to another extrasolar planet detection:
our $RV$ data for the nearby ($3.22$ pc) K2V star $\epsilon$ Eri add 
to the evidence for a long-period ($P\approx 6.9$ yrs) planet, as presented in
Hatzes et al. (\cite{artie00}). 

\subsection{Three new spectroscopic binaries: $\kappa$ For, HR 2400, and HR 3677}

$\kappa$~For, HR~2400 and HR~3677 were found to be single-lined spectroscopic
binaries, their large $RV$ scatter (see Table~\ref{all_stars})
is the direct result of huge $RV$ trends induced by high mass (stellar) companions. 
These trends were already discovered by an earlier analysis of a fraction of the data of these
3 stars (Hatzes et al.~\cite{artie96}). Now the analysis of the entire Long Camera
data of $\kappa$~For and HR~3677 exhibits a curved shape of the $RV$ trends and - in the case of
$\kappa$~For - allows us to find a preliminary
Keplerian orbital solution, while the very long period for HR~3677 and the linearity of the $RV$ trend 
for HR~2400 prohibits this.

\begin{figure}
\centering{
  \vbox{\psfig{figure=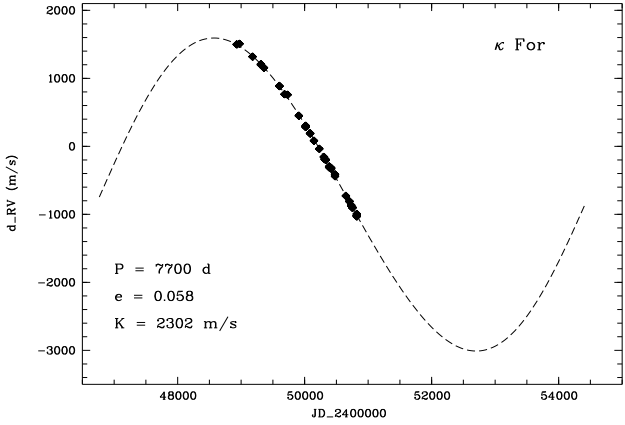,width=9.0cm,height=6.0cm,angle=270}}
   \par 
        }
  \caption[]{
        Preliminary Keplerian orbital solution for $\kappa$~For, the best-fit orbit
        (dashed line, $\chi^{2}_{\rm best}=38.3$, $\chi^{2}_{\rm red}=1.13$)
	is plotted along with the $40$ $RV$ measurements (diamonds).   
        }
 \label{kapforfit}
\end{figure}
\begin{figure}
\centering{
  \vbox{\psfig{figure=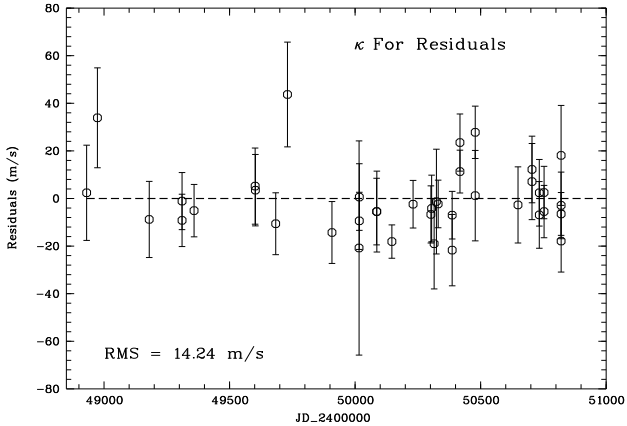,width=9.0cm,height=6.0cm,angle=270}}
   \par
        }
  \caption[]{
        Residual $RV$ scatter of $\kappa$~For after subtraction of the best-fit
        Keplerian orbit (Fig.~\ref{kapforfit}). The rms scatter of $14.24~{\rm m\,s}^{-1}$
        agrees well with the mean internal error of $14.8~{\rm m\,s}^{-1}$.
        }
 \label{kapforresid}
\end{figure}
The G0V star $\kappa$~For has the largest $RV$ scatter (rms=$780.9~{\rm m\,s}^{-1}$) of all stars in the
CES sample. 
We find a preliminary Keplerian orbital solution (see Fig.~\ref{kapforfit}) with the
following parameters: orbital period $P=7700$ days, time of periastron passage
$T=2454466$ JD, a low eccentricity $e=0.0576$, an $RV$ 
semi-amplitude $K=2302~{\rm m\,s}^{-1}$ and periastron angle $\omega=269.07^{\circ}$.  
This fit to the 40 $RV$ measurements gives a $\chi^{2}_{\rm best}=38.3$, and a reduced 
$\chi^{2}_{\rm red}=1.13$ (with 34 degrees of freedom) and $P_{\chi}(\chi^{2})=0.28$.
In other words the found preliminary Keplerian orbit represents a good fit to the $RV$ data.
By changing the value of $P$ (and letting the remaining orbital parameters vary until
$\chi^{2}=\chi^{2}_{\rm best}+1$) we determined the uncertainty of the period to be $\pm295$ days.
Since our $RV$ data cover only a fraction of one orbital cycle and do not constrain the orbit well enough, 
it was not possible to find a simultaneous solution for all orbital parameters and derive the
error-range for the remaining 5 parameters. 
The mass function is $f(m)=(9.67\pm3.17)\times10^{-3}~{\rm M}_\odot$ and the orbital period
transforms to $a\approx 8.5$ AU. 
The scatter around this orbit is $14.24~{\rm m\,s}^{-1}$ (Fig.~\ref{kapforresid}) and consistent
with the mean internal error of $14.8~{\rm m\,s}^{-1}$. 
In the $Hipparcos$ catalogue $\kappa$~For was given a double/multiple systems annex flag G, 
meaning that higher-order terms were necessary to find an adequate astrometric solution. 
This is an indication that $\kappa$~For is a long-term ($P>10$ yrs) astrometric binary, consistent
with our results. The $RV$ variabilty of $\kappa$~For was also noted by Nidever et al.~(\cite{nidever})
who find a linear $RV$ slope of $-1.73~{\rm m\,s}^{-1}$ per day for their 7 measurements of this star. 

\begin{figure}
\centering{  
  \vbox{\psfig{figure=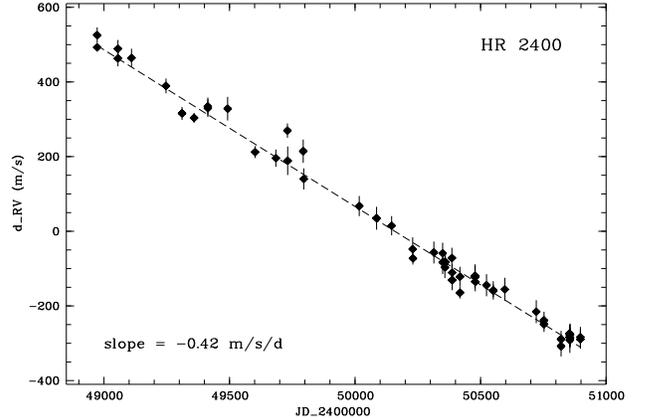,width=9.0cm,height=6.0cm,angle=270}}
   \par
        }
  \caption{
        Linear fit (dashed line) to the $RV$ data of HR~2400 (diamonds with errorbars),
	the residuals are shown in Fig.~\ref{hr2400res}.  
	This best fit linear trend has a slope of $-0.42 \pm 0.005~{\rm m\,s}^{-1}{\rm d}^{-1}$.
        }
 \label{hr2400fit}
\end{figure}

\begin{figure}
\centering{  
  \vbox{\psfig{figure=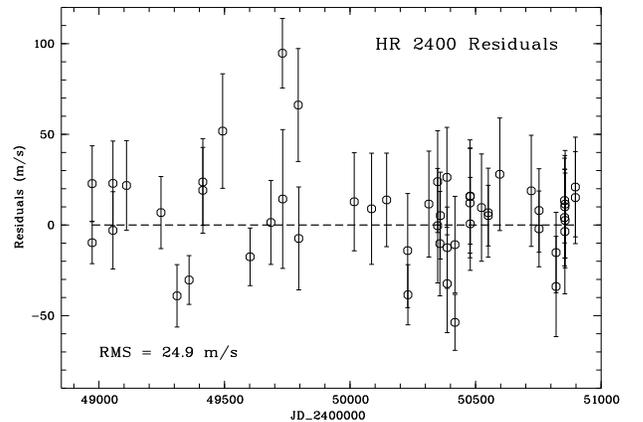,width=9.0cm,height=6.0cm,angle=270}}
   \par
        }
  \caption{
	$RV$ residuals of HR 2400 after subtraction of the linear slope (Fig.~\ref{hr2400fit}). 
        The residual rms scatter of $24.9~{\rm m\,s}^{-1}$ is consistent with the 
	mean internal error of $25.3~{\rm m\,s}^{-1}$.
        }
 \label{hr2400res}
\end{figure}

HR~2400 (F8V) reveals a linear trend in its $RV$ data indicating a high mass companion in a long-period
orbit which does not allow us to find a Keplerian orbital solution.
Fig.~\ref{hr2400fit} shows the best-fit linear function with a slope of $-0.42 \pm 0.005~{\rm m\,s}^{-1}{\rm
d}^{-1}$ (the error range of the slope is determined by varying the value of the slope, whereby for 
each slope the zero-point is always fitted, until $\chi^{2}=\chi^{2}_{\rm best}+1$). 
The residual rms-scatter around this slope is $24.9~{\rm m\,s}^{-1}$ which is of the same order
as the average internal error of $25.3~{\rm m\,s}^{-1}$ (see Fig.~\ref{hr2400res}). 
The linearity of the trend does not allow an
estimate of the mass or period of the secondary. Moreover, HR~2400 does not possess a double/multiple 
systems annex
flag in the $Hipparcos$ catalogue indicating that the period is indeed very long compared to the 
monitoring time spans of both programs ($Hipparcos: 3.2$ yrs, CES Long Camera: $5.2$ yrs).  

\begin{figure}
\centering{
  \vbox{\psfig{figure=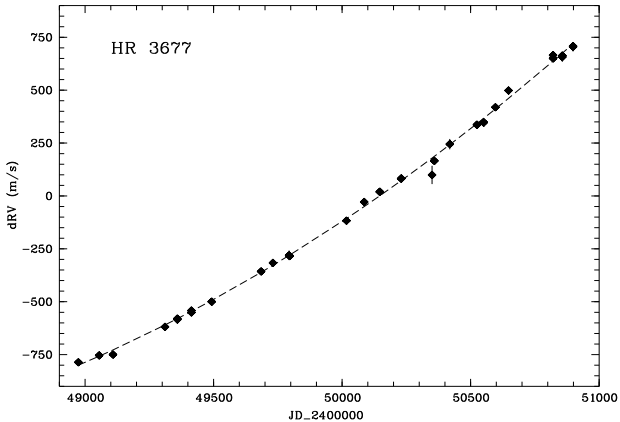,width=9.0cm,height=6.0cm,angle=270}}
   \par
        }
  \caption[]{
	Best parabolic fit for HR 3677 (G0III) indicating an orbital period much longer
	than the monitoring time ($Hipparcos$ astrometry gives a period of $\approx75$ years). 
	The best-fit curved trend is plotted as
	dashed line along with our $RV$ data (diamonds). The $\chi^{2}_{\rm red}$ of this fit is $0.99$,
	indicating a good fit. See Fig.~\ref{hr3677resid} for the residuals.  
        }
 \label{hr3677fit}
\end{figure}

\begin{figure}
\centering{
  \vbox{\psfig{figure=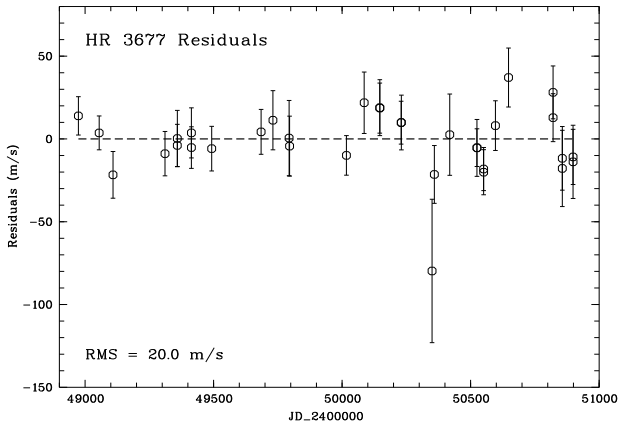,width=9.0cm,height=6.0cm,angle=270}}
   \par
        }
  \caption[]{
        Residual $RV$s of HR~3677 after subtraction of the best-fit curved trend (shown in
        Fig.~\ref{hr3677fit}). The rms-scatter is $20.0~{\rm m\,s}^{-1}$ slightly larger than the 
	mean internal error of $16.7~{\rm m\,s}^{-1}$. This larger residual scatter is primarily
	caused by the one outlier at JD 2,450,350; without this data-point the scatter is reduced to 
	$14.7~{\rm m\,s}^{-1}$ and is consistent with our measurement errors. 
        }
 \label{hr3677resid}
\end{figure}

The giant HR~3677 (G0III) is - with a distance of $192.31$ pc - by far the most distant star
in the CES sample. 
A parabolic fit to the $RV$ results is shown in Fig.~\ref{hr3677fit}. This fit gives an
acceptable description of the data with
a $\chi^{2}_{\rm red}$ of $0.99$ and $P_{\chi}(\chi^{2})=0.48$.
Fig.~\ref{hr3677resid} shows the residuals after subtraction of this best-fit curved trend.
The residual rms-scatter around this orbit is $20.0~{\rm m\,s}^{-1}$, slightly larger than the
average internal error of $16.7~{\rm m\,s}^{-1}$.
From the $Hipparcos$ measurements of HR~3677 a two-component astrometric solution was derived.
The angular separation of the components is given as $0.131\pm0.010$ arcseconds which corresponds at the
distance of $192.31$ pc to a minimum orbital separation of $\approx 25\pm2$ AU.
The orbital period would be around $75$ years, too long to determine a Keplerian solution, but
it seems to be consistent with the $RV$-variation we find for HR~3677.  

%

\section{Statistical analysis}

In the following section we perform a thorough statistical examination of the 
complete $RV$ data set of the CES Long Camera survey. We test each star for
variability, linear and curved $RV$ slopes and significant periodic signals.
 
\subsection{Probing for variability}

To test if a star in the sample is variable we first apply the $F$-test
($F = \sigma_{1}^{2} / \sigma_{2}^{2}$).
We start by comparing the total variance of the $RV$ results of one star
($\sigma_{1}^{2}$) with the rest of the sample ($P(F)_{1}$ in Table~\ref{ftest}).
Since, for the small CES bandwidth, the $RV$ precision and hence the
scatter is a function of spectral type, we compare the stars with the mean scatter
($\sigma_{2}^{2}$) of the remaining stars within the same spectral type bin
(as described in Sect. 3).  

For a second $F$-test, we take as the
$\sigma_{2}^{2}$ the individual mean internal $RV$ error and determine the $P(F)$ 
($P(F)_{2}$ in Table~\ref{ftest}). With this we test if the star is more variable (or
more constant) than its internal $RV$ error would suggest.

Finally, we fit a constant to the $RV$ results (with the zero-point as free
parameter) and use the $\chi^{2}$-statistic to check whether the constant model
delivers a good description of the data.
Again a small value of $P_{\chi}(\chi^{2})$
means that the $RV$ results for a star have a larger scatter than expected
from their internal errors, which then can be interpreted as a sign for variability.

For the binaries ($\kappa$~For, HR~2400, HR~3677, $\alpha$~Cen A \& B) the $RV$ residuals
after subtraction of either the known binary orbit ($\alpha$~Cen A \& B) or the preliminary orbit
we have found are taken for the variability tests. For $\alpha$~Cen A \& B with the known
binary orbit, HR~2400 with its linear trend and HR~3677 with its curved trend, 
$\nu$ the degrees of freedom equals $N-1$ (the velocity zero-point is adjusted), so that 
the F-test results for these stars are
strictly valid only under the assumption that the binary orbit and the velocity trends are
precisely known. In the cases of $\kappa$~For and $\iota$~Hor we take $\nu=N-6$.  
 
As criterion for a
significant result we adopt the $99\%$-level (i.e. $P < 0.01$).
Table~\ref{ftest} summarizes the results of these variability tests for all survey stars.

\begin{table}
\begin{center}
\begin{tabular}{c|ll|cl}
\hline
\hline
Star & $P(F)_{1}$ & $P(F)_{2}$ & $\chi^{2}_{\rm red}$ & $P_{\chi}(\chi^{2})$ \\
\hline
$\zeta$ Tuc & 0.023 & 0.073 & 1.92 & 0.0001 \\
$\beta$~Hyi & 0.13 & 0.12 & 1.51 & 3.5E-05 \\
HR 209 & 0.5 & 0.35 & 0.98 & 0.50 \\ 
$\nu$ Phe & 3.6E-05 & 0.367 & 1.17 & 0.18 \\   
HR 448 & 0.34 & 0.40 & 0.70 & 0.85 \\ 
HR 506 & 0.23 & 0.90 & 2.02 & 0.003 \\
$\tau$ Cet & 6.6E-11 & 0.016 & 0.73 & 0.98 \\  
$\kappa$ For$_{\rm Res}$ & 0.03 & 0.83 & 1.09 & 0.33 \\  
HR 753 & 0.63 & 0.203 & 0.32 & 0.90 \\ 
$\iota$ Hor$_{\rm Res}$ & 0.009 & 4.9E-19 & 4.8 & 5.6E-45 \\
$\alpha$ For & 1.E-09 & 0.0014 & 1.82 & 6.9E-05 \\
$\zeta^{1}$Ret & 0.57 & 0.71 & 1.55 & 0.09 \\
$\zeta^{2}$ Ret & 0.67 & 0.05 & 1.69 & 0.0009 \\
$\epsilon$ Eri & 0.32 & 0.0059 & 1.86 & 3.5E-05 \\
$\delta$ Eri & 0.06 & 0.22 & 1.16 & 0.21  \\
$\alpha$ Men & 3.5E-06 & 0.37 & 0.90 & 0.64 \\
HR 2400$_{\rm Res}$ & 0.13 & 0.91 & 1.51 & 0.012 \\
HR 2667 & 0.06 & 0.04 & 0.68 & 0.67 \\
HR 3259 & 0.14 & 0.45 & 2.2 & 7.2E-05\\
HR 3677$_{\rm Res}$ & 0.87 & 0.32 & 0.94 & 0.57 \\
HR 4523 & 0.09 & 0.85 & 1.57 & 0.03 \\
HR 4979 & 0.004 & 0.41 & 1.18 & 0.18 \\
$\alpha$ Cen A$_{\rm Res}$ & 0.0001 & 0.78 & 0.70 & 0.94 \\
$\alpha$ Cen B$_{\rm Res}$ & 0.88 & 0.13 & 1.26 & 0.12 \\
HR 5568 & 0.001 & 0.002 & 0.38 & 0.99 \\
HR 6416 & 0.09 & 0.0001 & 3.63 & 1.4E-18 \\
HR 6998 & 0.70 & 0.28 & 0.55 & 0.99 \\
HR 7373 & 0.02 & 0.83 & 0.77 & 0.62 \\
HR 7703 & 0.65 & 0.75 & 0.81 & 0.76 \\
$\phi^{2}$ Pav & 0.05 & 0.25 & 1.59 & 0.0003 \\
HR 8323 & 0.85 & 0.56 & 1.02 & 0.43 \\
$\epsilon$~Ind & 0.39 & 0.0094 & 1.73 & 0.0001 \\
HR 8501 & 3.1E-05 & 2.7E-07 & 2.26 & 3.7E-08 \\
HR 8883 & 6.8E-10 & 0.005 & 4.21 & 9.3E-14 \\
Barnard & 0.0007 & 0.29 & 0.57 & 0.95 \\
GJ 433 & 0.19 & 0.46 & 0.80 & 0.68 \\
Prox Cen & 2.2E-11 & 0.14 & 1.61 & 0.001 \\
\hline
\hline
\end{tabular}
\end{center}
\caption[]{F-test and $\chi^{2}$-test results for the complete sample. $P(F)_{1}$ is the result of the
        F-test comparing the $RV$ scatter with the average scatter within a spectral type bin, while
        $P(F)_{2}$ gives the results of the comparison with the mean internal $RV$ error. $\chi^{2}_{\rm
	red}$ is the best (reduced) $\chi^{2}$-value after fitting a constant to the
        $RV$ data and $P_{\chi}(\chi^{2})$ the corresponding probability.
        }
\label{ftest}
\end{table}

For $5$ stars in the CES sample all $3$ tests yielded $P<0.01$ and are thus clearly identified as RV
variables: $\iota$~Hor (residuals), $\alpha$~For, HR~6416, HR~8501 and HR~8883.
For $\iota$~Hor the cause of the residual variability was already discussed (stellar activity and
possible second planet), while in the cases of
$\alpha$~For ($L_{X}=525 \times 10^{27}~{\rm erg/s}$) and HR~8883 ($L_{X}=34175 \times 10^{27}~{\rm erg/s}$) 
also a high level of stellar activity appears to be
responsible for the detected variability. We found strong Ca II H\&K emission
for HR~8883 by taking a spectrum with the FEROS instrument and the 1.5 m telescope on La Silla.
We will show later that we can also determine
the cause of variability for HR~6416 and HR~8501 (we will examine the presence
of linear and curved trends in the $RV$ data). The case of $\alpha$~For will be further discussed in more
detail.

For $17$ stars (HR~209, HR~448, $\kappa$~For, HR~753, $\zeta^{1}$~Ret, $\delta$~Eri, HR~2400 (residuals),
HR~2667, HR~3677 (residuals), HR~4523, HR~4979, $\alpha$~Cen B (residuals), HR~6998, HR~7373, HR~7703, HR~8323
and GJ~433) all 3 tests resulted in $P>0.01$ and thus no sign of variability whatsoever can be found
for these stars. $5$ stars are less variable than the rest of the sample (which of
course also results in a low $P(F)_{1}$): $\nu$~Phe, $\tau$~Cet, $\alpha$~Cen A (residuals), HR~5568 
and Barnard's star.

$\epsilon$ Eri and $\epsilon$~Ind both have a $P(F)_{1}>0.01$ (i.e. their overall $RV$ scatter is quite
typical for the CES sample) but a low $P(F)_{2}$ and $P_{\chi}(\chi^{2})$, making them
candidates for low-amplitude variations.

$\zeta$ Tuc, $\beta$~Hyi, HR 506, $\zeta^{2}$ Ret and $\phi^{2}$ Pav all appear unsuspicious 
in both F-tests but their $RV$ data is not well fit by a constant function. 
Maybe an inclined linear or curved trend will describe these $RV$ data better.

Finally, Prox Cen is more variable than the rest of the CES M-dwarf sample (of same spectral
type) but not if compared to its own intrinsic mean $RV$ error, however, for this star the
$P_{\chi}(\chi^{2})$ for a constant fit is still less than $0.01$.

\subsection{$RV$ trends: linear and curved slopes}

As the next step in the statistical analysis we examine whether the $RV$ data of
those stars which showed up as variable by the previous tests can be better described by a linear slope
or a curved trend. For this purpose we determine the best-fit linear and parabolic function by
$\chi^{2}$-minimization and again compute $P_{\chi}(\chi^{2})$ for the following $12$ stars:
$\zeta$~Tuc, $\beta$~Hyi, HR~506, $\alpha$~For, $\zeta^{2}$~Ret, $\epsilon$~Eri,
HR~6416, $\phi^{2}$~Pav, HR~8501, HR~8883, $\epsilon$~Ind and Prox Cen.

Table~\ref{slopcurv} shows the results of the slope and curvature tests. For $7$ stars of the sample
($\zeta$~Tuc, HR~506, $\zeta^{2}$~Ret, $\epsilon$~Eri, $\phi^{2}$~Pav, HR 8883, Prox Cen) 
the resulting probabilities $P_{\chi}(\chi^{2})$ remain below $0.01$. For these stars no linear or curved
trend delivers a satisfactory explanation for their $RV$ variations.

\begin{figure}
\centering{
        \vbox{\psfig{figure=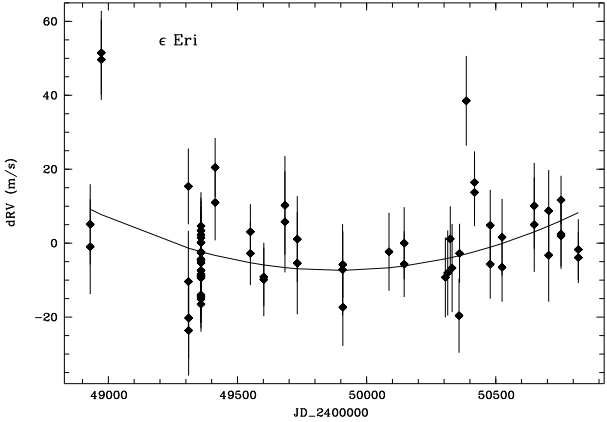,width=9.0cm,height=6.0cm,angle=270}}
        \par
        }
        \caption[]{Best-fit curved $RV$ trend for $\epsilon$~Eri.
        With $\chi^{2}_{\rm red}=1.7$ and
        $P_{\chi}(\chi^{2})_{\rm curv}=0.0005$ this fit is still a poor description of the
        $RV$ results. However, $P_{\chi}(\chi^{2})_{\rm curv}$ is a magnitude higher than for
	a linear fit. The presence of the highly eccentric Keplerian signal of the planet described in
        Hatzes et al. (\cite{artie00}) is already indicated here. 
        }
\label{epsericurv}
\end{figure}

For $\epsilon$~Eri the $P_{\chi}(\chi^{2})$-value for a curved trend is
a magnitude higher than for a linear trend. 
As described in detail in Hatzes et al. (\cite{artie00}) there is evidence
for a long-period ($P\approx 2500$ days), low amplitude ($K\approx19~{\rm m\,s}^{-1}$) planet orbiting
$\epsilon$~Eri.
The difference we find between
the linear and curved trend fits (see Fig.~\ref{epsericurv}) might indicate the presence of the
$RV$ signature of this planet. 
The rms-scatter around the curved
trend is with $12.9~{\rm m\,s}^{-1}$ still larger than the mean internal error of $9.7~{\rm m\,s}^{-1}$,
which is not surprising for the modestly active star $\epsilon$~Eri. 

\begin{figure}
\centering{
  \vbox{\psfig{figure=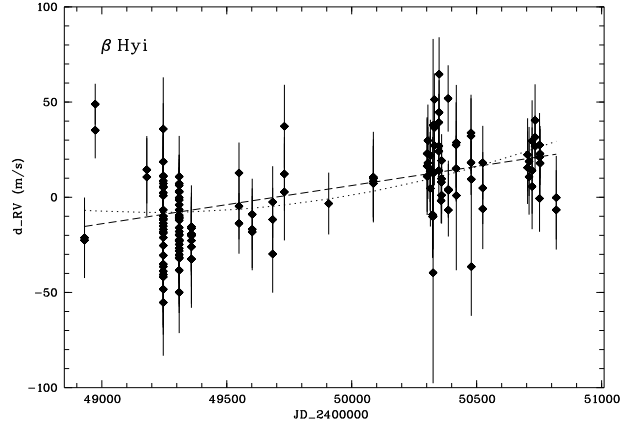,width=9.0cm,height=6.0cm,angle=270}} 
   \par
        }
  \caption[]{
        $\beta$~Hyi $RV$ data and the best-fit linear slope (dashed line) and
        curved trend (dotted line). The rms scatter around these trends is
        $19.3~{\rm m\,s}^{-1}$ and $19.5~{\rm m\,s}^{-1}$ respectively. The found linear
        trend is $+0.02\pm0.0024~{\rm m\,s}^{-1}{\rm d}^{-1}$.
        }
 \label{bethyislop}
\end{figure}

\begin{figure}  
\centering{
  \vbox{\psfig{figure=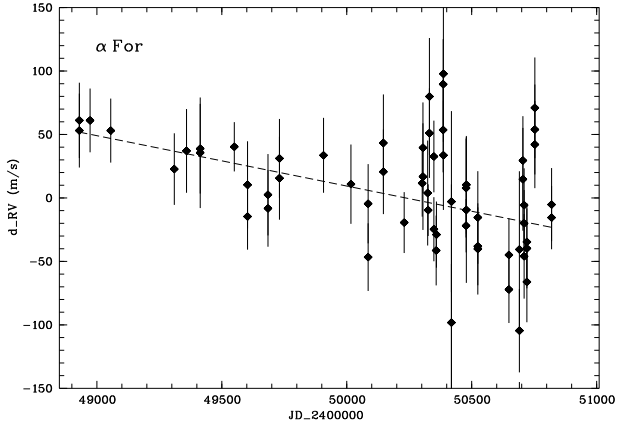,width=9.0cm,height=6.0cm,angle=270}}
   \par
        }
  \caption[]{ 
        Linear trend in the $RV$s of $\alpha$ For. The slope of
        $-0.04\pm0.007~{\rm m\,s}^{-1}{\rm d}^{-1}$ has a residual scatter of $50.9~~{\rm m\,s}^{-1}$.
        }
 \label{alpforslop}
\end{figure}

\begin{figure}
\centering{  
  \vbox{\psfig{figure=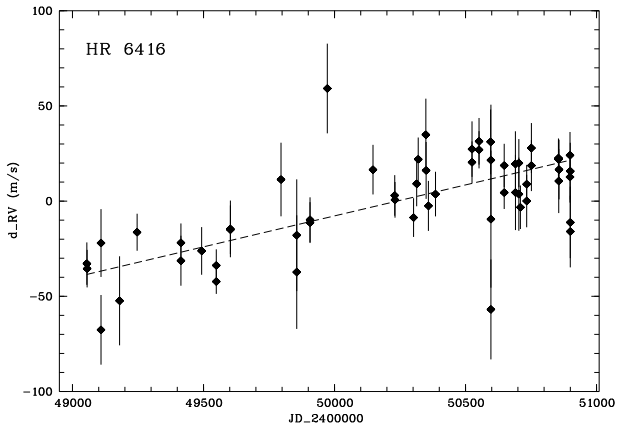,width=9.0cm,height=6.0cm,angle=270}}
   \par
        }
  \caption[]{
        Best-fit linear trend for HR 6416, the $RV$ shift per day is $+0.032\pm0.003~{\rm m\,s}^{-1}$
        and the rms scatter around this slope is $19.4~{\rm m\,s}^{-1}$.
        }  
 \label{hr6416slop}
\end{figure}

\begin{figure}
\centering{
  \vbox{\psfig{figure=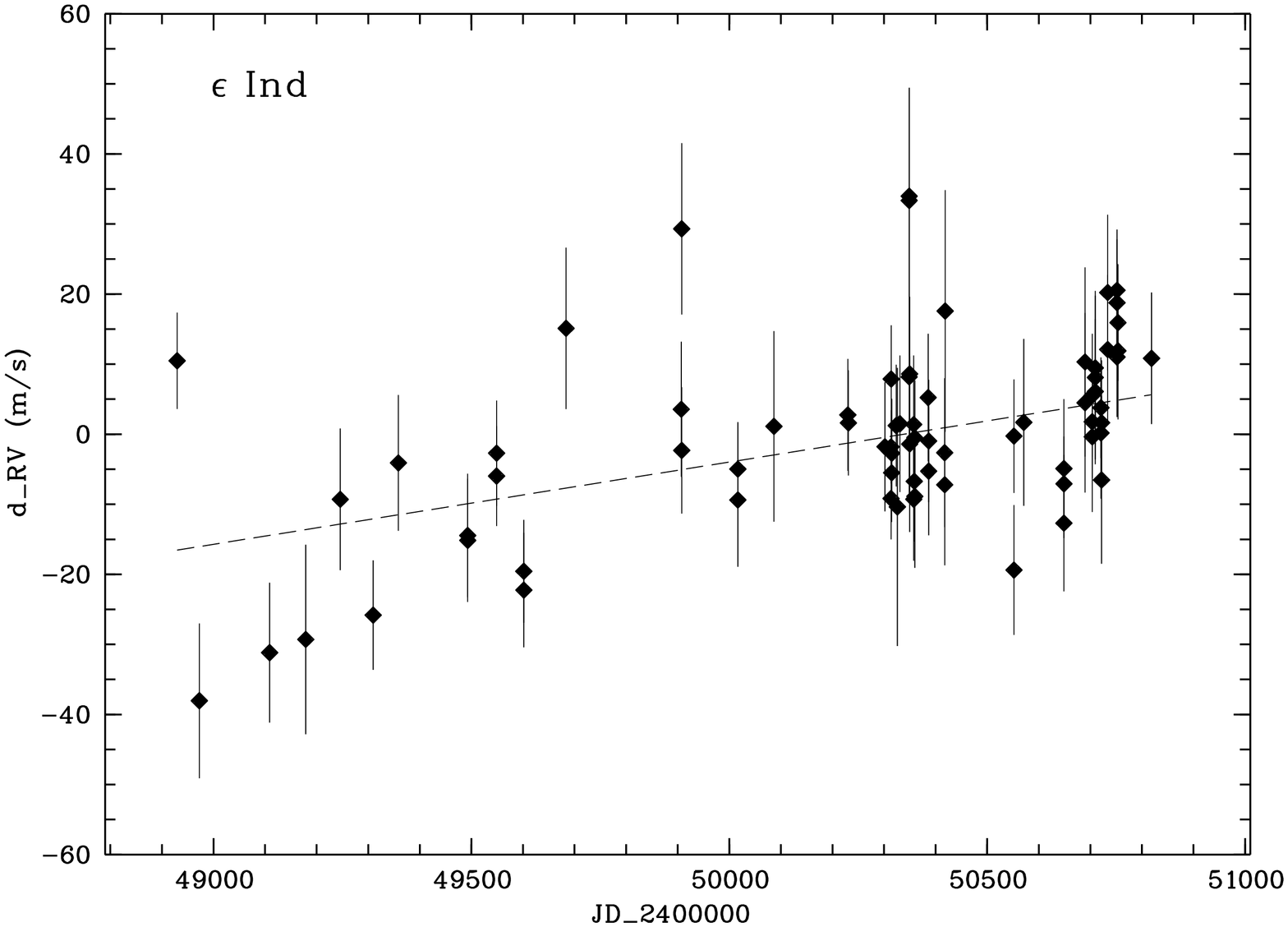,width=9.0cm,height=6.0cm,angle=270}}
   \par
        }
  \caption[]{
        Best-fit linear trend for $\epsilon$~Ind, the $RV$ shift per day is $+0.012\pm0.002~{\rm
	ms}^{-1}$ and the rms scatter around this slope is $11.6~{\rm m\,s}^{-1}$.
        }
 \label{epsindslop} 
\end{figure}

\begin{figure}
 \centering{
  \vbox{\psfig{figure=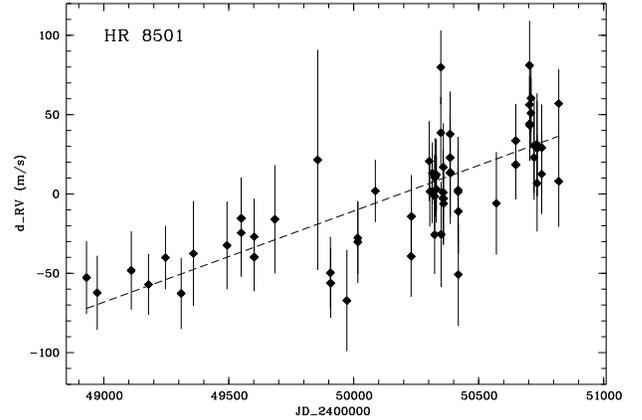,width=9.0cm,height=6.0cm,angle=270}}
   \par
        }
  \caption[]{
        Best-fit linear trend for HR 8501, the $RV$ shift per day is $+0.057\pm0.006~{\rm m\,s}^{-1}$
        and the rms scatter around this slope is $23.4~{\rm m\,s}^{-1}$.
        }
 \label{hr8501slop}
\end{figure}

For 5 stars ($\beta$~Hyi, $\alpha$~For, HR~6416, $\epsilon$~Ind and HR~8501) the linear slope
tests resulted in a significant increase in $P_{\chi}(\chi^{2})$. This is a true indication
for the presence of a linear trend in their $RV$ behavior. However, for none of them, the probability of
a curved trend is found to be significantly higher than for a linear one, except for $\beta$~Hyi where
the difference is 5\%. Table~\ref{linearslops} gives the values of $RV$ shift per day of the found linear
trends and the remaining $RV$ scatter around those trends. With the exception of $\alpha$ For the $RV$
trends are all positive, with HR~8501 having the strongest slope of $+0.057~{\rm m\,s}^{-1}{\rm d}^{-1}$
and $\epsilon$~Ind with the smallest variation of $+0.011~{\rm m\,s}^{-1}{\rm d}^{-1}$.
Fig.~\ref{bethyislop} to \ref{hr8501slop} display the best-fit trends in the $RV$ results for these 5
stars.

\begin{table*}
\begin{center}
\begin{tabular*}{0.99\textwidth}{@{\extracolsep{\fill}}c|l|ll|ll}
\hline
\hline
Star & $P_{\chi}(\chi^{2})_{\rm constant}$ & $P_{\chi}(\chi^{2})_{\rm linear}$ &
$\chi^{2}_{\rm red}$ & $P_{\chi}(\chi^{2})_{\rm curvature}$ & $\chi^{2}_{\rm red}$ \\
\hline
$\zeta$ Tuc     & 0.0001        & 0.0001  & 1.93 & 7.6E-05   & 1.96     \\
$\beta$~Hyi     & 3.5E-05       & 0.19    & 1.10 & 0.24      & 1.07     \\
HR 506          & 0.003         & 0.005   & 1.98 & 0.003     & 2.07     \\
$\alpha$ For    & 6.9E-05       & 0.03    & 1.36 & 0.025     & 1.38     \\
$\zeta^{2}$ Ret & 0.0009        & 0.001   & 1.67 & 0.001     & 1.69     \\
$\epsilon$ Eri  & 3.5E-05       & 3.3E-05 & 1.86 & 0.0005    & 1.70     \\	 
HR 6416         & 1.4E-18       & 0.056   & 1.32 & 0.059     & 1.32     \\
$\phi^{2}$ Pav  & 0.0003        & 0.0004  & 1.59 & 0.0003    & 1.60     \\  
$\epsilon$~Ind  & 0.0001        & 0.035    & 1.33 & 0.038     & 1.32     \\
HR 8501         & 3.7E-08       & 0.86    & 0.81 & 0.93      & 0.75     \\
HR 8883         & 9.3E-14       & 1.5E-13 & 4.32 & 9.3E-14   & 4.37     \\
Prox Cen        & 0.001         & 0.002   & 1.59 & 0.0025    & 1.59     \\
\hline
\hline
\end{tabular*}
\end{center}
\caption[]{Results of linear slope and curvature tests. Probabilites $P_{\chi}(\chi^{2})$ and
        $\chi^{2}_{\rm red}$-values are given for the best-fit linear and parabolic functions.
        The first column ($P_{\chi}(\chi^{2})_{\rm constant}$) is repeated from
	Table~\ref{ftest} for comparison purposes.}
\label{slopcurv}
\end{table*}

Since HR~6416 and HR~8501 are known binary stars, the detected linear $RV$ trend can be attributed
to the binary orbital motion. The $Hipparcos$ catalogue gives an angular separation of $8.658$
arcseconds for the HR~6416 binary. 
At a distance of $8.79$ pc this implies a minimum separation of $76.1$ AU.
In the Gliese catalogue of nearby stars HR 6416 is listed as GJ 666A and the secondary GJ 666B as an
M0V dwarf. We adopt $0.89$~M$_{\odot}$ as mass value for the G8V primary and for the secondary
$0.52$~M$_{\odot}$ (after Gray~\cite{gray}). Assuming a circular orbit and the minimum separation 
as the true separation and using Kepler's third law we 
can find an estimate for the $RV$ acceleration for HR 6416. The orbital period is $\approx 565$ yrs and
the $RV$ semi-amplitude $K \approx 1442~{\rm m\,s}^{-1}$. Since the observing time span is not even a
$1/100$ of one orbital cycle we lineary interpolate to find an {\it average} acceleration of $\approx
0.028~{\rm m\,s}^{-1}{\rm d}^{-1}$ which is in
good agreement with the detected trend of $0.032\pm0.003~{\rm m\,s}^{-1}{\rm d}^{-1}$.

The angular separation for the HR 8501 binary (GJ 853A \& B) is given by the Gliese catalogue as $3.4$
arcseconds. This transforms into a minimum orbital separation of $46.3$ AU at the distance of $13.61$
pc. The spectral type of the secondary (GJ 853B) is unknown and therefore no mass estimate is possible.
Assuming an M0V companion the average linear $RV$ trend for the G3V primary ($1.04$~M$_{\odot}$) would be
$\approx 0.075~{\rm m\,s}^{-1}{\rm d}^{-1}$, slightly larger than the found trend of 
$+0.057\pm0.006~{\rm m\,s}^{-1}{\rm d}^{-1}$.
The difference can easily be explained either by a larger orbital separation than the projected minimum
value, the $\sin i$-effect, a lower mass of the secondary, the orbital phase corresponds to a 
steeper part of the $RV$-curve, or a combination of all these effects.
\begin{table}
\begin{center}
\begin{tabular}{llll}
\hline
\hline
Star & \hspace{1cm} & $RV$ trend & rms \\
 & & [${\rm m\,s}^{-1}{\rm d}^{-1}$] & [${\rm m\,s}^{-1}$] \\
\hline
$\beta$~Hyi & & $+0.02 \,\,\, \pm 0.0024$ & $19.3$ \\
$\alpha$ For & & $-0.04 \,\,\, \pm 0.007$ & $50.9$ \\
HR 6416 & & $+0.032 \pm 0.003$ & $19.4$ \\
$\epsilon$~Ind & & $+0.012 \pm 0.002$ & $11.6$ \\
HR 8501 & & $+0.057 \pm 0.005$ & $23.4$ \\
\hline
\hline
\end{tabular}
         
\caption[]{Linear $RV$ trends and the rms scatter around these slopes.
Uncertainties of the slopes correspond to the $\chi^{2}=\chi^{2}_{\rm best}+1$ range.}
\label{linearslops}
\end{center}
\end{table}

$\alpha$~For is also a known binary with an angular separation of $4.461$ arcseconds (from $Hipparcos$
catalogue). At a distance of $14.11$ pc this means $a\approx63$ AU. Adopting a mass for the F8V primary
of $1.2$~M$_{\odot}$ (after Gray~\cite{gray}) and assuming again an M0V secondary (the spectral type of
the companion is unknown) we find an average acceleration of $0.04~{\rm m\,s}^{-1}{\rm d}^{-1}$ which is
exactly the value of the $RV$ trend we observe for $\alpha$ For. However, the rms scatter around this
trend is large with $\approx 51~{\rm m\,s}^{-1}$. Most of this residual scatter is probably caused by the
high stellar activity of $\alpha$ For. Another possible explanation for an increase of the $RV$ scatter
can be contamination of the spectra of the primary by the secondary. 
Due to field rotation at the Nasmyth focus of the CAT telescope, 
light from the close secondary also entered the CES during some observations when the slit 
was aligned with the binary axis. 
This could at least to some degree have contributed to the larger
scatter (the same is true for HR~8501 with an even smaller angular separation).

        Neither $\beta$~Hyi nor $\epsilon$~Ind are known binary stars. The detected linear $RV$ trends are
thus caused by previously unknown companions. The linearity of both trends points towards distant
stellar companions. However, both stars also represent candidates for having very 
long-period ($P>20$ yrs) planetary companions. Follow up observations using the upgraded CES, now
equipped with the Very Long Camera (VLC), and the 3.6 m telescope were already performed and are still
in progress. Analysis of the new data and the combination with the $RV$ results presented here will
show whether the linearity of both trends continues.

\subsection{Period search}

To search for periodic signals in the complete $RV$ results of the CES sample we again use the
Lomb-Scargle periodogram (Lomb~\cite{lomb}, Scargle~\cite{scargle}). 
We estimate the False Alarm Probability (FAP) of a peak in the power spectrum by employing a 
bootstrap randomization method.
In this bootstrap approach the actual $RV$ measurements are randomly redistributed while keeping the
times of observations fixed (K\"urster et al. \cite{martinabdor}, Murdoch et al. \cite{murdoch}). 
The major advantage of this method is the fact that the FAP-levels can be derived without any
assumptions on the underlying noise distribution (like e.g. a Gaussian). 
We will continue to follow this bootstrap philosophy also in determining upper mass-limits for planets
in Sect. 5.  

The search interval is $2$ to $5000$ days.
For each star we perform 10,000 bootstrap randomization runs to estimate the FAP of the maximum peak
in the power spectrum. Since the validity of the FAP resulting from random redistribution is lower in
the case of high temporal concentration at one or several points (i.e. ``data clumping'' 
when during one night a large number of measurements were taken while in other nights this number is
significantly lower) we also perform the analysis on the $RV$ set binned in nightly averages.

        Table~\ref{periods} summarizes the periodogram results for all survey stars (in the cases of
$\kappa$~For, HR~2400, HR~3677 and $\alpha$~Cen A \& B the period search was performed on the 
$RV$ residuals after subtraction of the binary orbit).

\begin{table*}
\begin{center}
\begin{tabular*}{0.99\textwidth}{@{\extracolsep{\fill}}l|rrrl|rrrl}
\hline
\hline
Star & N$_{1}$ & Period$_{1}$  & Power$_{1}$ & FAP$_{1}$
 & N$_{2}$ & Period$_{2}$ & Power$_{2}$ & FAP$_{2}$  \\
 & & [days] & & & &[days] & & \\
\hline
$\zeta$ Tuc                & 51  & 9.7  & 8.5  & 0.009  & 36 & 9.7  & 6.9 & 0.10 \\
$\beta$~Hyi                & 157 & 2000 & 28.9 &$<$0.0001& 39 & 3.2  & 6.4 & 0.684 \\
HR 209                     & 35  & 30.0 & 5.7  & 0.458  & 24 & 4.7  & 5.0 & 0.932 \\
$\nu$ Phe                  & 58  & 6.5  & 7.9  & 0.085  & 41 & 6.5  & 5.7 & 0.672 \\
HR 448                     & 24  & 2.0  & 3.0  & 0.79   & 13 & 6.8  & 4.8 & 0.134 \\
HR 506                     & 23  & 7.6  & 5.8  & 0.36   & 18 & 7.6  & 5.7 & 0.447 \\
$\tau$ Cet                 & 116 & 5.2  & 14.3 & 0.0002 & 32 & 5.6  & 7.2 & 0.158 \\
$\kappa$ For$_{\rm Res}$   & 40  & 75.8 & 6.3  & 0.381  & 30 &14.0  & 5.4 & 0.72 \\
HR 753                     & 6   & 3.3  & 2.2  & 0.20   & 3  &10.7  & 1.0 & 0.827 \\   
$\iota$ Hor                & 95  & 322.6& 26.7 & $<$0.0001& 57 &322.6 &16.2 &$<$0.0001 \\
$\alpha$ For               & 65  & 4.3  & 11.0 & 0.0013 & 36 &4.3   & 7.1 & 0.265 \\
$\zeta^{1}$Ret             & 14  & 6.2  & 3.8  & 0.275  & 8  &33.1  & 3.0 & 0.368 \\
$\zeta^{2}$ Ret            & 58  & 6.4  & 6.7  & 0.5    & 44 &2.0   & 6.5 & 0.638 \\
$\epsilon$ Eri             & 66  & 4.4  & 10.2 & 0.0025 & 28 &2.7   & 5.1 & 0.739 \\
$\delta$ Eri               & 48  & 2.3  & 5.9  & 0.011  & 27 &9.5   & 6.8 & 0.089 \\
$\alpha$ Men               & 41  & 4.5  & 8.0  & 0.067  & 29 &4.5   & 7.6 & 0.051 \\
HR 2400$_{\rm Res}$        & 53  & 3.9  & 7.0  & 0.284  & 38 &5.3   & 6.5 & 0.405 \\
HR 2667                    & 66  & 13.5 & 8.9  & 0.066  & 43 &5.8   & 6.6 & 0.719 \\
HR 3259                    & 35  & 3.8  & 3.3  & 0.307  & 26 &3.8   & 2.6 & 0.931 \\
HR 3677$_{\rm Res}$        & 34  & 5.0  & 7.3  & 0.05   & 26 &178.6 & 5.4 & 0.266 \\
HR 4523                    & 27  & 2.4  & 6.4  & 0.309  & 21 &2.4   & 5.5 & 0.768 \\
HR 4979                    & 52  & 16.8 & 8.6  & 0.055  & 33 &16.8  & 6.7 & 0.252 \\
$\alpha$ Cen A$_{\rm Res}$ & 205 & 16.4 & 19.4 & $<$0.0001& 48 &16.4  & 7.7 & 0.114 \\
$\alpha$ Cen B$_{\rm Res}$ & 291 & 11.6 & 31.8 & $<$0.0001& 43 &17.3  & 7.1 & 0.042 \\
HR 5568                    & 40  & 5.1  & 4.7  & 0.604  & 19 &2.5   & 4.8 & 0.675 \\
HR 6416                    & 57  & 5000 & 11.4 & 0.001  & 38 &5000  & 8.7 & 0.034 \\
HR 6998                    & 51  & 6.7  & 8.2  & 0.039  & 35 &6.7   & 7.7 & 0.116 \\
HR 7373                    & 8   & 4.5  & 3.1  & 0.4651 & 7  &4.5   & 2.7 & 0.955 \\
HR 7703                    & 30  & 7.2  & 6.2  & 0.14   & 21 &7.2   & 5.8 & 0.295 \\
$\phi^{2}$ Pav             & 90  & 7.0  & 14.4 & 0.0001 & 41 &7.0   &10.1 & 0.003 \\
HR 8323                    & 20  & 2.1  & 4.5  & 0.464  & 13 &2.1   & 4.7 & 0.21 \\
$\epsilon$~Ind             & 73  & 2.6  & 10.0 & 0.017  & 44 &2.6   & 7.0 & 0.429 \\
HR 8501                    & 66  & 5000 & 19.0 & $<$0.0001& 44 &5000  &13.1 & 0.0001 \\  
HR 8883                    & 31  & 36.2 & 8.7  & 0.013  & 21 &36.2  & 5.8 & 0.391 \\
Barnard                    & 24  & 2.3  & 5.0  & 0.29   & 15 &2.3   & 5.1 & 0.015 \\
GJ 433                     & 15  & 5.4  & 3.5  & 0.417  & 8  &5.4   & 3.2 & 0.591 \\
Prox Cen                   & 65  & 3.2  & 10.5 & 0.0012 & 35 &3.2   & 6.8 & 0.05 \\
\hline
\hline
\end{tabular*}
\caption[]{Periodogram results. Power$_{1}$, Period$_{1}$ and FAP$_{1}$ are the results of the
        original $RV$ data set (N$_{1}$), while Power$_{2}$, Period$_{2}$ and FAP$_{2}$ gives the
        values for the nightly averages (N$_{2}$).
 }
\label{periods}
\end{center}
\end{table*}

The periodogram analysis of the original $RV$ data found periods with FAP$_{1}$ (for the original
un-binned $RV$ data set) below $0.001$ at: $\beta$~Hyi, $\tau$~Cet, $\iota$~Hor, $\alpha$~Cen A \& B 
residuals, $\phi^{2}$~Pav and HR~8501. However, none of them, with
the exception of $\iota$~Hor and HR~8501, reveal a significant signal in their nightly averaged $RV$ data set.
In the cases of $\beta$~Hyi and $\alpha$~Cen A \& B the
significance of the low FAP$_{1}$ is decreased by high temporal concentration of their $RV$ data
(the periodogram results for $\alpha$~Cen A \& B are discussed in detail in 
Endl et al.~\cite{michl01}). The significant $P=5000$ day signal for HR~8501 is caused
by the found linear $RV$ trend (see Fig.~\ref{hr8501slop}), since 5000 days is the maximum
period searched for. The same happens in the case of HR~6416, although the $P=5000$ day signal is
no longer significantly recovered in the nightly averaged data set.     

Thus $\iota$~Hor remains the single case in the CES Long Camera survey which shows a convincing periodic 
$RV$ variation of statistical significance due to an orbiting planet. 

The results for
$\phi^{2}$~Pav are interesting, since the FAP$_{2}$ of the $P=7$ day signal 
is still low with $0.003$ and just marginally above the 1 permill threshold. We will
discuss the case of $\phi^{2}$~Pav later.     

\subsection{Activity induced $RV$ excess scatter}

Several studies (Saar \& Donahue (\cite{saar1}), Saar et al. (\cite{saar3}), Santos et al. (\cite{santos}),
Paulson et al. (\cite{diane})) have investigated and discussed the relationship
between stellar activity and $RV$ scatter induced by activity related phenomena like 
surface inhomogenities (spots) and variable granulation pattern. Saar et al. (\cite{saar3})
found a correlation between an excess $RV$ scatter (exceeding the expected scatter due
to internal measurement uncertainties) and the rotational speed of the G and K
dwarfs in the sample of the Lick planet search program. Interestingly, 
Paulson et al. (\cite{diane}) measured simultaneously the Ca II H\&K emission and $RV$s  
of members of the Hyades cluster and showed that only for a few stars (5 out of 82) the
chromospheric activity is correlated with an $RV$ excess.       

\begin{figure}
\centering{  
  \vbox{\psfig{figure=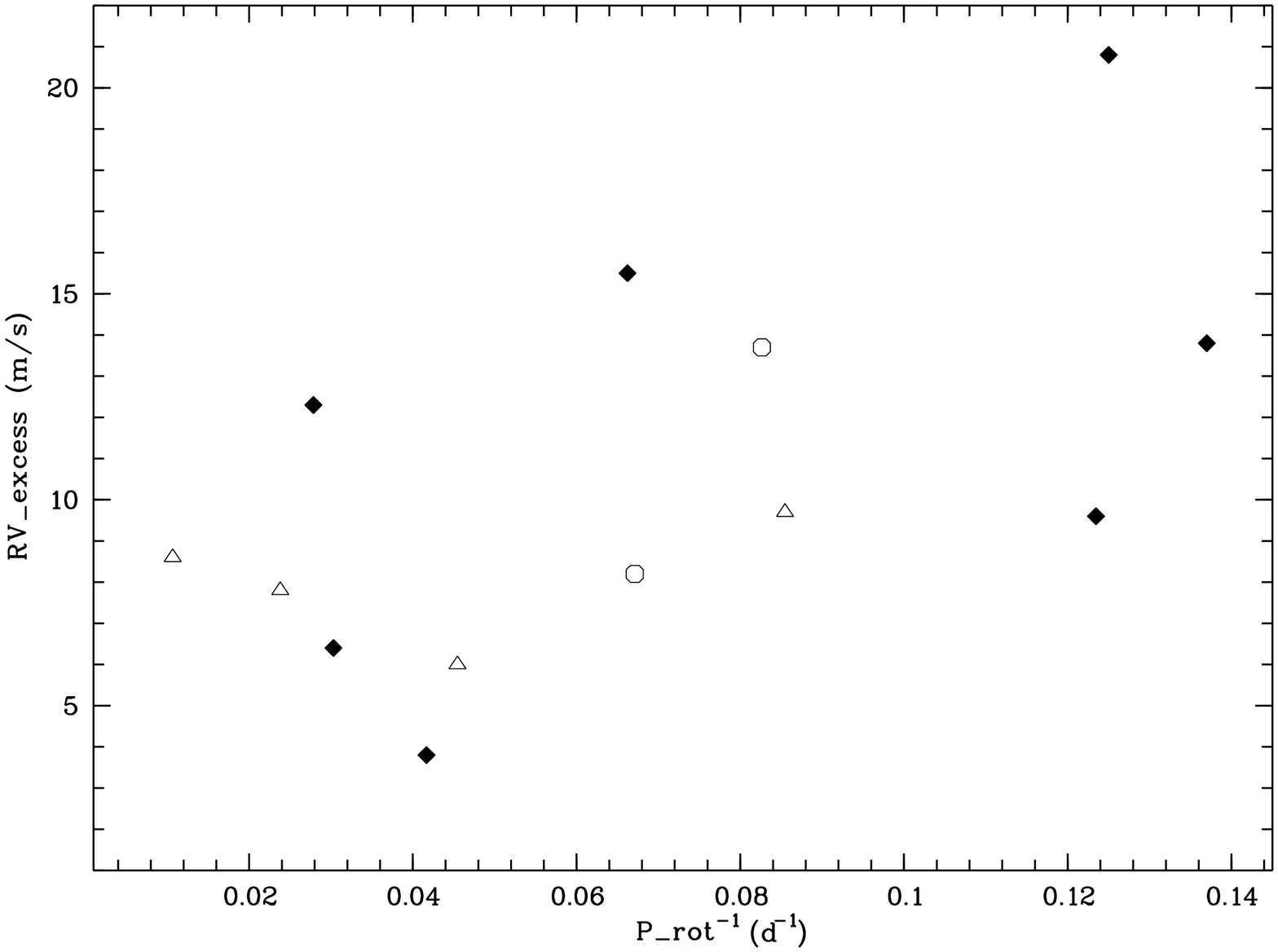,width=9.0cm,height=6.0cm,angle=270}}
   \par
        }
  \caption[]{
	Correlation between $RV$ excess scatter and stellar rotational period.
	F-type stars are shown as circles, G-type stars as full diamonds and 
	K-type as triangles. The excess scatter increases with decreasing 
	$P_{\rm Rot}$ (with $\iota$~Hor having the highest value of excess scatter). 
	}  
 \label{excess}
\end{figure}

In the case of the CES Long Camera survey we can estimate $P_{\rm Rot}$ for $13$ stars, 
which reveal an excess $RV$ scatter compared to their internal errors, and 
examine whether this excess scatter is correlated with intrinsic stellar activity. 
Stellar rotational periods are either taken from Saar \& Osten (\cite{saar2}) or
estimated using Eqs.(3) \& (4) in Noyes et al. (\cite{noyes}) and the Ca II H\&K
results from Henry et al. (\cite{henry96}).     
Fig.~\ref{excess} displays the $RV$ excess scatter plotted vs. $P_{\rm Rot}$, which
shows a general increase in the excess scatter with shorter rotational periods.
The sample of stars included in the diagram consist of $2$ F-type stars ($\zeta$~Tuc, $\nu$~Phe), 
$7$ G-type stars ($\iota$~Hor, $\zeta^{2}$~Ret, HR~4523, HR~4979, HR~6416, HR~8323, 
HR~8501) and $4$ K-type stars ($\epsilon$~Eri, $\delta$~Eri, $\alpha$~Cen B, $\epsilon$~Ind).
Although we do not have $P_{\rm Rot}$-values for $\alpha$~For and HR~8883, the observed
large excess scatter for these two stars is probably also due to fast rotation and 
enhanced stellar activity. As already mentioned, both stars are bright X-ray sources and
HR 8883 displays prominent Ca II H\&K emission. This result confirms the general
picture of increased $RV$-jitter due to enhanced stellar activity.

\section{Limits for planetary companions}

We now continue with the quantitative determination
of the sensitivity of the CES Long Camera survey for the discovery of planets orbiting the 
target stars.
With this we ask the question: which planets would we have detected if they existed? Since they are not
detected we can exclude their presence and hence set constraints.
Starting from the $null-hypothesis$ that the observed $RV$ scatter is mainly caused by measurement
uncertainties and/or additional intrinsic stellar effects, we establish the planet detection threshold
for each star. This detection limit for each individual target is determined by
numerical simulations of planetary orbits of varying amplitudes, periods and phase angles.
These simulated planetary signals can either be recovered by a periodogram significantly or not and
thus deliver the quantitative upper limits. 

\subsection{The method}

Companion limits for different planet search samples were presented by Murdoch et al.
(\cite{murdoch}), Walker et al. (\cite{walker}) and Cumming et al. (\cite{cumming}) for the Mount John
Observatory, CFHT and Lick Observatory surveys, respectively. Nelson \& Angel (\cite{nelson}) derived
an analytical expression for detection limits and re-examined the CFHT data set. The sensitivity of $RV$ 
surveys for outer planets with orbital periods exceeding the survey duration was studied in
Eisner \& Kulkarni (\cite{eisner}).  
They all used different approaches, most of them are based on the periodogram,  
to determine the detection capabilities of individual surveys.
In K\"urster et al. (\cite{martinprox}) we have already determined these limits for one target of the
CES program, namely Prox Cen, using also a different method based on a Gaussian noise term.

The method we apply to the CES data was already described in Endl et al. (\cite{michl_man}) and
in greater detail in Endl et al. (\cite{michl01}). The latter paper also includes a comparison with
a method based on Gaussian noise terms and discusses the differences and the advantage of the 
bootstrap approach. Our bootstrap-based method can be summarized as follows: for each star we perform 
numerical simulations of planetary orbits with $K$ (the $RV$ semi-amplitude), $\phi$ the orbital phase
and $P$ (the orbital period) as model parameters (orbital eccentricity $e$ is set to $0$).  
The maximum period represents the time span of CES observations of the target star (with a typical
duration of $\approx 2000$ days). For each set of $K$ and $P$ 8 different orbits are created with their
phase angles shifted by $\pi/4$. These signals are added to the $actual$ $RV$ measurements (i.e. we use
our own data set as noise term) and perform a period search in the range of $2 - 5000$ days (using the
Lomb-Scargle periodogram). We define a planet as $detectable$ if the period is found by the periodogram
and the statistical significance of its peak in the power spectrum is higher than $99\%$ (i.e. its
false alarm probability (FAP) is lower than $0.01$). If the FAP equals or exceeds $0.01$ at only one of
the trial phases, a planet is not classified as detectable at these $P$ and $K$ values.
Again the FAP is estimated by using the bootstrap randomization method and for each simulated signal we
perform 1000 bootstrap runs. By increasing the $K$ parameter until the FAP is less than $1\%$ for all
orbital phases at a given $P$ value we obtain quantitative upper limits for planetary companions.
By taking as the noise-term the obtained $RV$ results for single stars (and the $RV$ residuals for binary
stars) we assure that the noise distribution is equal to the measurement errors of the CES survey for
this star (obviously, the temporal sampling of the simulated signals are identical with the
real monitoring of each individual star by the CES survey).
The value of the parameter $K$ can be transformed immediately into an $m\sin i$ value
and the parameter $P$
into an orbital separation $a$ for the companion, thus we derive an $m\sin i - a$ companion limit
function.
The simulations are performed in the period range of $P = 3$ days and the total duration of
observation for each individual target. This period range is sampled in such a manner that
companion limits are computed every $0.25$ AU (i.e. a $P^{2/3}$ spacing).
One exception is the $1$-year window ($P = 365$ days)
where the companion limit is determined additionally in every case.
The derived upper mass-limits are strictly correct only at the $P \& K$-values where the simulations
are performed (circles in Figs.~\ref{limfig1} - \ref{limfig7}), the interpolated limits (dashed lines in the
figures) are not mathematically stringent,
since the frequency-space is too large to be sampled completely (i.e. two times for each independent
frequency = $Nyquist$ criterion).

The following stars are excluded from the limit determination due to insufficient observations: HR 753
(with only $6$ $RV$ measurements), HR 7373 ($8$ $RV$ measurements), and $\zeta^{1}$Ret ($14$ RV
measurements).
Another exclusion is $\iota$ Hor, where the planet was discovered, and the $\alpha$ 
Centauri system, in which case companion limits based on the CES
data have already been presented in K\"urster et al. (\cite{martinprox}) for Proxima and in
Endl et al. (\cite{michl01}) for $\alpha$ Cen A \& B.

\subsection{Stellar masses}
In order to transform the obtained $K$ value of detectable $RV$ signals into an $m\sin i$ value for
planetary companions using Kepler's third law, we have to
assume a certain mass for the host star. The stellar mass values we take for the $m\sin i$ calculations
are summarized in Table.~\ref{masstab}. As an example for the dependence of the $m\sin i$ values on the
stellar mass value, we take the case of $\beta$~Hyi ($1.1~{\rm M}_{\odot}$).
For an orbital period of $P = 44$ days an error of
$0.1~{\rm M}_{\odot}$ leads to a small uncertainty in the $m\sin i$ of $\pm 0.03~{\rm M}_{\rm Jup}$ and
at a period of $P = 1890$ days the corresponding value is larger with $\pm 0.13~{\rm M}_{\rm Jup}$.
 
\begin{table}
\begin{center}
\begin{tabular}{ll|ll}
 \hline
 \hline
 Star & Mass [${\rm M}_{\odot}$] & Star & Mass [${\rm M}_{\odot}$] \\
 \hline
 $\zeta$ Tuc & $1.06^{1}$ &  HR 3259 & $0.9$ \\
 $\beta$~Hyi & $1.1^{2}$ & HR 3677 & $2.1$ \\
 HR 209 & $1.1$ & HR 4523 & $1.04$ \\
 $\nu$ Phe & $1.2$ & HR 4979 & $1.04$  \\
 HR 448 & $1.23^{1}$ & HR 5568 & $0.71$  \\
 HR 506 & $1.17$ & HR 6416 & $0.89$  \\
 $\tau$ Cet & $0.89$ & HR 6998 & $1.0$  \\
 $\kappa$ For & $1.12$ & HR 7703 & $0.74$  \\
 $\alpha$ For & $1.2$ & $\phi^{2}$ Pav & $1.1^{1}$  \\
 $\zeta^{2}$ Ret & $1.1$ & $\epsilon$~Ind & $0.7$ \\
 $\epsilon$ Eri & $0.85^{3}$ & HR 8501 & $1.04$ \\
 $\delta$ Eri & $1.23^{1}$ & HR 8883 & $2.1$ \\
 $\alpha$ Men & $0.95$ & Barnard & $0.16^{4}$ \\
 HR 2400 & $1.2$ & GJ 433 & $0.42$ \\
 HR 2667 & $1.04$ & & \\
 \hline
 \hline
\end{tabular}
\end{center}
\caption[]{Assumed stellar masses for the companion limit determination.
        If not otherwise stated the mass estimates are taken from Gray (\cite{gray}) for the
        according spectral type.
        Other references: $1$. Porto de Mello, priv. comm.,
        $2$. Dravins et al. (\cite{dravins98}),
        $3$. Drake \& Smith (\cite{drake})
        $4$. based on the mass-luminosity relation from Henry et al. (\cite{henry}).
        }
\label{masstab}
\end{table}

\subsection{Windows of non-detectability}
 
There are two special cases which can render a simulated signal totally undetectable
(using the criteria described above). These windows of non-detection are displayed in the figures as
vertical lines which bracket a gap in the companion limits.

\subsubsection{Insignificant signal}

In the first case the data structure (total number of observation and sampling density) leads to the effect
that for a certain phase angle the FAP of the input signal always exceeds the $1\%$ level regardless of
the increase of the $K$ parameter. This is the case for the gaps seen in the limits for instance for HR
209 at $2.25$ AU (Fig.~\ref{limfig1}) and for HR 506 at the same separation (Fig.~\ref{limfig2}).
        
\subsubsection{Aliasing}
        
However, for most cases of non-detecability we find that due to aliasing the maximum
power in the periodogram for certain phase angles 
is not located at the input period, again, regardless of the increase in $K$.
The seasonal 1 year period
is a typical window interval for every astronomical observing program and
the problem of aliasing will occur if the signal is not properly sampled within this
interval (in general sampling on irregular time basis can reduce the occurance of aliasing).
 
This occurs for instance in the cases of HR~3259, $\beta$~Hyi, $\tau$~Cet, $\alpha$~Men
and HR~4523, where the maximum power is found at the $1/2~P$-value for input signals with
$P$ close to the seasonal 1 year period. The peaks in the power spectrum are significant (i.e. FAP $>
0.01$) but at the wrong $P$-values. In other cases the maximum power is found at a completely different
$P$ value for a certain orbital phase (e.g. at $2.9$ days instead of the input value of $1425$ days for
HR 4523).  

In a conservative approach we declare these cases as non-detections since the correct parameters
of the planetary signal were not recovered (remember that one undetected signal out of the eight trial phases 
is sufficient to define a planet as $undetectable$ although seven out of eight signals were 
successfully recovered). 
One would conclude though that a planet with the wrong
orbital period is present and the correct period of its orbit remains unknown (though continued
monitoring of this star would eventually lead to the correct orbital parameters). 

\subsection{Resulting companion limits}

The derived upper mass-limits for planets orbiting the CES survey stars are displayed in
appendix B for each individual star. The horizontal dotted line in each plot shows (for better comparison) 
the $m\sin i=1.0~{\rm M}_{\rm Jup}$ border. For most of the stars the CES limit line crosses this
border at orbital separations less than 1 AU. This clearly demonstrates the need for a longer time
baseline as well as a better $RV$ measurement precision to detect 'real' Jupiters at 5.2 AU. 
         
The $\alpha$ Centauri system represents a special case in this limit-analysis: it
allows the combination of observational constraints for planetary companions with dynamical limitations for
stable orbits within the binary. In Endl et al.~(\cite{michl01}), paper II of this series,
we have combined the CES limits with the results from the dynamical stability study of  
Wiegert \& Holman (\cite{wiegert}) which led to strong constraints for the presence of giant planets 
orbiting $\alpha$ Cen A or B. Upper limits for giant planets around the third member of
the $\alpha$ Centauri system, Prox Cen, based on the results of a different $RV$ analysis of the CES
data were presented in K\"urster et al.~(\cite{martinprox}). We therefore didn't include the plots
for these 3 stars in this paper and refer the reader to the former articles.

In the cases of $\beta$~Hyi and $\tau$ Cet we have nights where a great number of spectra were taken in a
short consecutive time. In order to distribute the sampling more evenly and to reduce the total number of
RV measurements to save CPU time, we averaged the numerous $RV$ measurements in those nights, to get a
maximum number of $3$ observations per night. This reduced the total number of measurements for $\beta$~Hyi
to $94$ (instead of $157$) and for $\tau$ Cet to $62$ (instead of $116$).

The results for HR 448 are limited by the short monitoring time span of $438$ d and the small
number of observations ($24$ $RV$ measurements). For this star only companion limits for orbital separations
of $a < 0.15$ AU were found.

The candidate for a planetary companion to $\epsilon$ Eri (from Hatzes et al.~\cite{artie00}) is also
indicated in Fig.~\ref{limfig3} by an asterisk. 
It lies well inside the non-detectable region of the CES survey for this star.
As described in Hatzes et al.~(\cite{artie00}) the combination of several different $RV$ data sets was necessary
to find the signal of this companion. Moreover, the orbital period of 7 years is longer than the
duration of the CES Long Camera survey. 

For the binaries $\kappa$~For, HR~2400 and HR~3677 the simulated planetary signal is added to the orginial $RV$ set,
which still possess the huge variation due to the binary orbits. Thus, one additional step in the 
limit determination has to be performed.
Before the periodogram analysis is started we subtract the binary motion (the preliminary
Keplerian solution and trends from Sect. 3) by minimization of the $\chi^{2}$-function. 
The same is done for the low-amplitude trends of $\alpha$~For, HR~6416 and HR~8501 caused by their 
stellar secondaries. Again the found trends are subtracted (by $\chi^{2}$-minimization) from the synthetic 
$RV$ sets and the periodogram analysis is performed on the $RV$ residuals.  

HR~5568 was only observed for $384$ days and companion signals with periods $> 250$ days ($a \approx
0.7$ AU) were not detected at all orbital phases. The CES monitoring of HR~7703 spans over $1042$ days and
no gaps were found in the detectability of planetary companions within this time-frame (maximum separation
$a \approx 1.75$ AU).

For HR~8323 we detected only $3$ test signals at $P=3,123$ and $224$ days, at all other $P$-values the
test orbits were not recovered. This low detectability is due to the
smaller number of observations ($20$ $RV$ measurements) and the shorter time span of monitoring ($1068$ d).
The results for HR 8323 are not displayed.

The numerous gaps in the CES detectability of companions of the giant HR 8883 (G4III) are a direct
result of the large $RV$ scatter ($65.2~{\rm m\,s}^{-1}$) and the irregular monitoring of this star (see
Fig.~\ref{limfig7}).

In the case of GJ 433 (M2V) the small number of $RV$ measurements ($15$) and the short duration of monitoring
($337$ days) prohibit determination of companion limits. No simulated planetary signal could be detected at
$all$ orbital phases at the trial periods of $3,10,40,100,150,175,250$ and $333$ days. Only at the trial
period of $200$ days a signal with a $K$ amplitude of $600~{\rm m\,s}^{-1}$ was detected, corresponding to a
companion of $m\sin i=9.8~{\rm M}_{\rm Jup}$ with a semi-major axis of $a=0.5$ AU.

Barnard's star (M4V) constitutes another special case of the companion limit analysis.
The limits determined from the CES $RV$ data can be combined with the astrometric companion limits based on
the HST Fine Guidance Sensor (Benedict et al.~\cite{benedict}). This combination is very effective since
both methods are complementary to each other (the $RV$ method is more sensitive to close-by companions while
for astrometry the detectability of more distant companions is better).
Fig.~\ref{limfig7} displays the companion limits derived from the
CES data and combined with HST astrometric results. 
These combined limits demonstrate that - except for an $aliasing$-window near $a=0.08$ AU
(corresponding to $P= 20$ to $27$ d) - $all$ planets with $m\sin i > 1.2~{\rm M}_{\rm Jup}$ can be
exluded.

\subsection{Orbital eccentricity}
        
One limitation of this method is the restriction to sinusoidal signals ($e = 0$), i.e. planetary
system similar to our Solar System. The huge parameter
space for eccentric orbits (including the two additional parameters of orbital eccentricity $e$ and
anomaly $\omega$) simply makes it impossible to sample all possible orbits at a given period $P$.
The main motivation of examining the validity of the derived limits also for eccentric orbits is 
of course the fact that many extrasolar planets with longer periods known today have eccentric
orbits. 

\subsubsection{The test case: HR 4979}

In order to check the validity of the derived $e = 0$ limits also for eccentric orbits
we select the special test case of HR 4979 and compare the limits for $e = 0$ orbits for certain sets
of $K$ and $P$ values with $e > 0$ orbits. HR 4979 was chosen because the companion limits for this
star are a smooth function without gaps (see Fig.~\ref{limfig5}).
We want to emphasize that the following tests can only serve as an example and 
the results cannot be taken as generally valid for the complete survey.
Furthermore, due to the above mentioned feasibility limitations, the sampling rate of the parameter
space has to be kept low (e.g. the parameter $\omega$ is sampled only every $90^{\circ}$). The test
consists of the two following steps:
\begin{enumerate}
\item Simulations are performed, where the orbital phase $\phi$ is kept fixed and the parameters $P,e,$
and $\omega$ are varied, to obtain a rough estimate of the $e > 0$ limits validity.
\item For the smallest and largest $P$-value, the orbital phase parameter $\phi$ is also varied and the
detectability is determined for these cases.
\end{enumerate}

\begin{figure}
\centering{
  \vbox{\psfig{figure=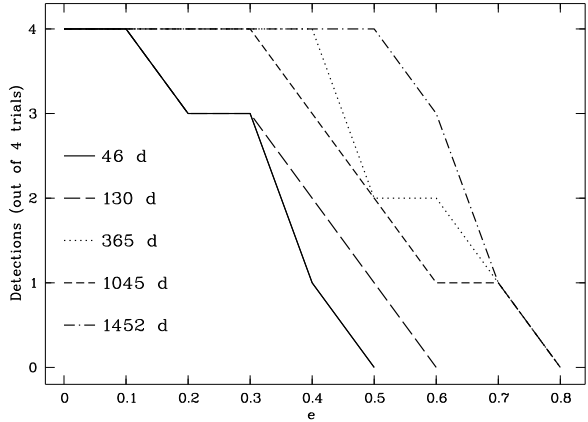,width=9.0cm,height=6.0cm,angle=270}}
   \par
        }
  \caption[]{Eccentricity-test part 1:
        Detectability of $e>0$ orbits in the case of HR 4979 and fixed orbital
        phase. For each trial period
        and eccentricity 4 different orbits ($\omega$ shifted by $90^{\circ}$) are created with
        the $K$-value obtained from the sinusoidal simulations.
        The graph shows how many of the 4 test signals are detected. The detectability
        increases with orbital period. For the shortest period $P = 46$ days the
        detectability ends at $e = 0.5$ (no test signal is recovered) and for the longer
        periods at $e = 0.8$.
        }
 \label{eccentric1}
\end{figure}

For step 1 we create 110 simulated orbits with different $e$ and $\omega$ values
at the trial periods $P = 46,130,365,1045$, and $1452$ days (with a fixed value of orbital
phase angle $\phi$), and the $K$ semi-amplitude of $22,20,35,26$, and $30~{\rm m\,s}^{-1}$ (these values
represent the limits obtained for
the $e = 0$ case). For each parameter-set of $P$,$K$
and $e$ we simulate $4$ different orbits with $\omega$ shifted by $90^{\circ}$. Again, we perform for
each simulated signal 1000 bootstrap runs to obtain the FAP level. A signal is classified as
{\it non-detected} if the FAP exceeds $1\%$ or the periodogram shows maximum power at a different period
than the input value $P$. Fig.~\ref{eccentric1} displays the detectability of these test
signals as a function of orbital eccentricity. 
For smaller values of $P$ these signals get undetectable
at $e = 0.5$ and $0.6$, while longer periodic signals are detectable until $e = 0.8$. 
The lower detectability at shorter periods might be
explained by the fact that with increasing eccentricity the
duration of the $RV$ maxima (minima) of the orbits is getting smaller than typical
observational time intervals. For longer periods these maxima or
minima are better sampled than for short periods and hence better detectable for the periodogram.

\begin{figure}
\centering{
  \vbox{\psfig{figure=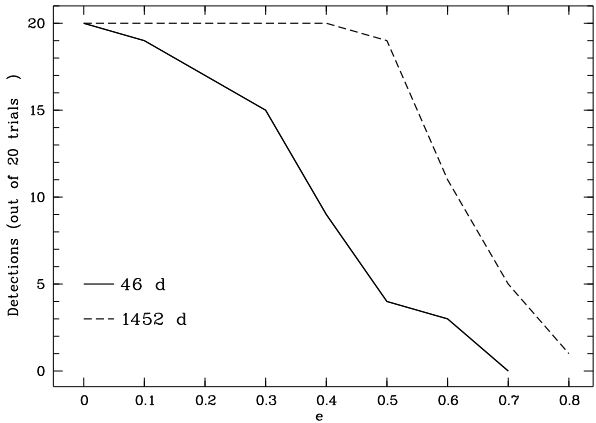,width=9.0cm,height=6.0cm,angle=270}}
  \par}
        \caption[]{Eccentricity-test part 2:
        Detectability of $e>0$ orbits for HR 4979 at $P = 46$ and $1452$ d with
        the orbital phase $\phi$ varied. For each pair of $P$ and $e$ 20 synthetic
        orbits at the 4 different $\omega$-values, each with 5 different $\phi$-values,
        are created. The plot displays how many of the 20 signals are recovered
        successfully. The general form of Fig.~\ref{eccentric1} is preserved.
        }
  \label{eccentric2}
\end{figure}

In step 2 the orbital phase $\phi$ is introduced as additional parameter for two selected $P$-values
(the minimum and maximum value of $P$: $46$ \& $1452$ d). For each set of $P,K,e$ and $\omega$ values
the parameter space of $\phi$ is additionally sampled $5$ times (equidistant). This means that for each
pair of $P$ and $e$ parameter $20$ synthetic planetary signals are generated (at $4$ different $\omega$
and $5$ different $\phi$ values). Fig.~\ref{eccentric2} shows the result of this second test. The
general form of the results from the first test is preserved.

These tests illustrate - at least in the case of HR 4979 - that the $e=0$ limits appear to be valid
for longer periods and eccentricities of $e<0.6$, while for smaller $P$ values the validity is
constrained to low eccentricities.  


\subsection{Average detection threshold}
        
By computing the ratio $F(P)$:
\begin{equation}
F(P) = \frac {K_{\rm detected}[{\rm m\,s}^{-1}]} {{\rm RV~scatter}[{\rm m\,s}^{-1}]}
\end{equation}
with $P$ the trial period,
$K_{\rm detected}$ the semi-amplitude of the detected $RV$ signal and the RV-scatter as the
individual noise-term of the star and by taking the mean
ratio \={F}$_{\rm Star}$ for all $P$-values where an upper mass limit was determined we obtain
an average detection threshold for each star.
For the $30$ stars, where we determined companion limits, and at all $P$-values we found $F(P) > 1.$, 
except in the case of $\beta$~Hyi where $F(3~{\rm d})=0.8$ 
(which also demonstrates the detectability of signals with smaller
amplitude than the measurement precision by a sufficient number of datapoints). The main bulk of values of
$F(P)$ lies in the range from $1.2$ to $4$, with an average value of $2.75$ (\={F}$_{\rm CES}$) for all
$30$ stars and $P$-parameter values where companion limits were derived.
 
The value of \={F}$_{\rm CES}$ corresponds to the average of the $RV$ signals detected and does not represent
the exact average detection threshold of the CES survey due to the presence of the 
non-detectability gaps, which cannot be taken into account in this
statistic. Table~\ref{threshold} gives the average detection factor for each star (\={F}$_{\rm Star}$) and
the resulting mean \={K}-amplitude of detectable planetary signals (the presence of non-detectability
gaps is also indicated).

The minimum value of \={F}$_{\rm Star}$ was found for HR~2667 (\={F}$_{\rm Star}=1.54$) and the
highest value for HR~448 (\={F}$_{\rm Star}=16.4$). In terms of average detectable $K$-amplitude the
minimum for the CES Long Camera survey is $19.9~{\rm m\,s}^{-1}$ for HR~5568, which is not surprising since
this is the star with the smallest RV-scatter (rms = $7.7~{\rm m\,s}^{-1}$), and the maximum at $369.4~{\rm
ms}^{-1}$ for the highly variable star HR 8883. 

\begin{table}
\begin{center}
\begin{tabular}{lll|lll}
 \hline
 \hline
 Star & \={F}$_{\rm Star}$ & \={K} & Star & \={F}$_{\rm Star}$ & \={K} \\
 & &[${\rm m\,s}^{-1}$] & & &[${\rm m\,s}^{-1}$] \\
 \hline
 $\zeta$ Tuc$^{*}$ & $3.3$ & $71.3$ & HR 3259$^{*}$ & $2.26$ & $36.7$ \\
 $\beta$~Hyi$^{*}$ & $1.76$ & $35.2$ & HR 3677 & $2.44$ & $56.8$ \\
 HR 209$^{*}$ & $5.37$ & $124$ & HR 4523$^{*}$ & $3.57$ & $53.6$ \\
 $\nu$ Phe & $2.16$ & $38.6$ & HR 4979 & $1.66$ & $23.2$ \\
 HR 448$^{*}$ & $16.4$ & $280.3$ & $\alpha$ Cen A & $1.88$ & $22.2$ \\
 HR 506$^{*}$ & $3.79$ & $90.5$ & $\alpha$ Cen B & $2.14$ & $26.9$  \\
 $\tau$ Cet$^{*}$ & $1.95$ & $22.6$ & HR 5568$^{*}$ & $2.58$ & $19.9$ \\
 $\kappa$ For & $2.42$ & $34.5$ & HR 6416 & $1.77$ & $34.3$  \\
 $\alpha$ For & $1.85$ & $94.2$ & HR 6998 & $2.67$ & $52.4$  \\
 $\zeta^{2}$ Ret & $2.04$ & $44.4$ & HR 7703 & $2.8$ & $37.3$  \\
 $\epsilon$ Eri$^{*}$ & $2.67$ & $40.0$ & $\phi^{2}$ Pav & $2.25$ & $79.5$ \\
 $\delta$ Eri & $2.12$ & $32.9$ & $\epsilon$~Ind & $3.35$ & $45.2$ \\
 $\alpha$ Men$^{*}$ & $2.51$ & $24.6$ & HR 8501$^{*}$ & $1.73$ & $40.4$ \\
 HR 2400 & $1.69$ & $42.0$ & HR 8883$^{*}$ & $5.67$ & $369.4$ \\
 HR 2667 & $1.54$ & $25.4$ & Barnard$^{*}$ & $4.58$ & $170.5$ \\
 \hline
 \hline
\end{tabular}
\end{center}
\caption[]{Average detection threshold \={F}$_{\rm Star}$ and resulting mean \={K}-amplitudes of
        detectable planetary signals for each star. Stars indicated with an asterix ($*$)
        possess windows of non-detectability.}
\label{threshold}
\end{table}

\begin{figure}
 \centering{
  \vbox{\psfig{figure=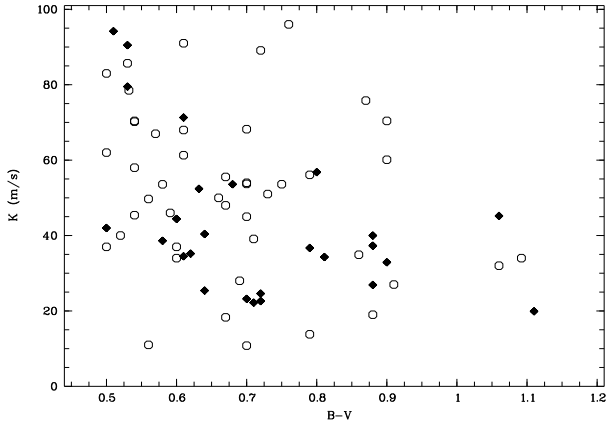,width=9.0cm,height=6.0cm,angle=270}}
   \par}
    \caption[]{
		 Average detectable K-amplitudes of the CES-survey (diamonds) as a function of $B-V$,
	compared to K-amplitudes of known extrasolar planets (cirlces). Only K-amlitudes less than
	$100~{\rm m\,s}^{-1}$ are displayed (excluding HR~448, HR~8883 and Barnard's star).
	The list of amplitudes of the known extrasolar planets was compiled using the information on the websites
	of the Geneva observatory program (http://obswww.unige.ch/$\sim$udry/planet/planet.html) and the
	California \& Carnegie planet search (http://exoplanets.org/) and the colors were derived
	from the SIMBAD database. Except for the blue part of the diagram (stars with $B-V<0.6$) 
	most of the known extrasolar planets are located within the detection range of the CES Long Camera
	survey.   
	}
	\label{knownplanets}
\end{figure}

Fig.~\ref{knownplanets} compares these CES-survey mean \={K}-amplitudes of detectable planetary signals 
with the K-amplitudes of known extrasolar planets as a function of $B-V$. For this purpose we selected 
only K-amplitudes less than $100~{\rm m\,s}^{-1}$ (which excludes three CES stars: HR~448, HR~8883 and
Barnard's star and 27 known extrasolar planets). 
Except for CES stars with $B-V<0.6$, most of the $RV$-signals
of the known extrasolar planets are within the detection range of the CES Long Camera survey.     

\begin{figure}
 \centering{
  \vbox{\psfig{figure=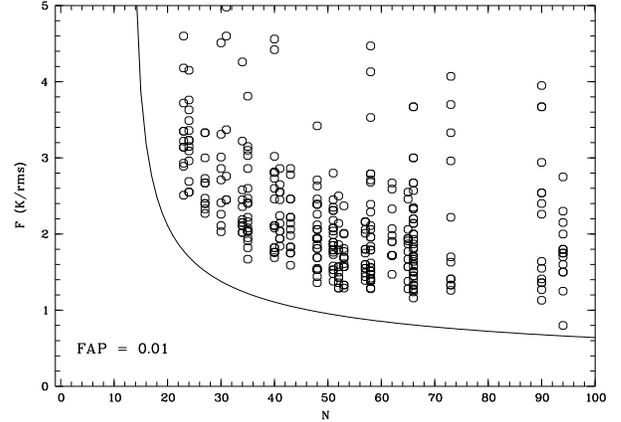,width=9.0cm,height=6.0cm,angle=270}}
   \par}
    \caption[]{Comparison of detected simulated signals (circles) with an analytically derived
	detection threshold (solid line) based on the Lomb-Scargle periodogram and a False Alarm
	Probability (FAP) of 0.01 (using Eg.(15) in Cochran \& Hatzes~\cite{bill96}). $F$ denotes the
	$S/N$-ratio of the signal in the power spectrum and $N$ the total number of measurements
	per star.   
	}
	\label{signal}
\end{figure}

Fig.~\ref{signal} shows a comparison of our numerical simulations with an analytic detection
threshold for a False Alarm Probability of 0.01 by using Eq.(15) from Cochran \& Hatzes (\cite{bill96}).
The theoratical curve (solid line in Fig.~\ref{signal}) is calculated on the basis of the Lomb-Scargle 
periodogram and gives the $S/N$-ratio ($F=K/\sigma$)
of signals detected with FAP $\le 0.01$ as a function of number of measurements $N$.
Clearly, the
plot shows that virtually $no$ signals can be detected with $N<20$, consistent with our results, and
that the curve constitutes a clear lower limit to our ``real'' detectability (circles in Fig.~\ref{signal}).
It is not surprising that the detected signals do not get closer to the theoratical limit, 
since they had to be recovered at $all$ phase angles. In many cases the 
CES Long Camera survey successfully recovered signals at certain phase angles at lower $F$-values, 
which are not included in Fig.~\ref{signal} due to the above mentioned criterion. 
Fig.~\ref{signal} demonstrates again that 
the sensitivity of $RV$ planet search programs can also be improved 
by increasing $N$, i.e. by taking more data, beside raising the $RV$ precision.

\subsection{Mass range}
 
\begin{figure}
 \centering{
  \vbox{\psfig{figure=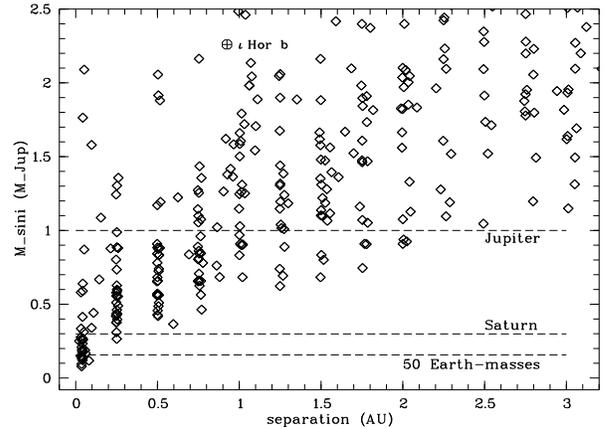,width=9.0cm,height=6.0cm,angle=270}}
   \par}
    \caption[]{Lower section of all detectable $m\sin i$-values (diamonds) of the CES Long Camera survey.
        The horizontal dashed lines
        show the mass values of Jupiter, Saturn and a hypothetical $50$ Earth-mass planet. The location of
	$\iota$ Hor b ($m\sin i = 2.26~{\rm M}_{\rm Jup}$, $a=0.925$ AU)
        is also indicated. $\iota$ Hor b lies well above the main bulk of detectable $m\sin i$-values.
        At orbital separations larger than $2$ AU no planets with $m\sin i < 1~{\rm M}_{\rm Jup}$ are
        detectable.
        For $22$ stars the detectable $m\sin i$-value at $a \approx 0.04$ AU reach down into the sub-Saturn
        mass regime ($m\sin i < 0.3~{\rm M}_{\rm Jup}$).
        }
        \label{massrange}
\end{figure}

The lowest companion mass which was detected during our simulations corresponds to
an $m\sin i = 0.079~{\rm M}_{\rm Jup} \approx 0.27~{\rm M}_{\rm Saturn}$ or $25~{\rm M}_{\rm Earth}$ 
planet in a $P=3$ d and $a=0.036$ AU orbit around HR 5568. 
Fig.~\ref{massrange} shows the lower section of the limiting $m\sin i$-values of all simulated planetary
signals which were determined as detectable by the CES survey.
Planets with $m\sin i = 1~{\rm M}_{\rm Jup}$ were found undetectable for the $30$ CES survey stars at
separations larger than $2$ AU. However, at small orbital separations of $a \approx 0.04$ AU
the CES survey would have detected ``51 Peg''-type planets in $all$ cases and for $22$ survey stars
even the presence of sub-Saturn-mass planets.
The bias of $RV$ planet searches towards short-period planets is obvious in Fig.~\ref{massrange}.
Also indicated in Fig.~\ref{massrange} is the location of the planet discovered around $\iota$ Hor.
With $m\sin i = 2.26~{\rm M}_{\rm Jup}$ and $a=0.925$ AU $\iota$ Hor b clearly lies above the main part of the
limiting detectable $m\sin i$-values and was thus ``easily'' detectable by the CES Long Camera survey.

\section{Discussion}

The planet around $\iota$ Hor remains the single clear detection of an extrasolar planet by the  
CES Long Camera survey. This corresponds to a detection rate of $\approx$ 3\%, a value similar
to other precise Doppler searches. The discovery of $\iota$~Hor b demonstrated for the first time
the feasibility of the $RV$ technique in planet detection in the case of young and
thus moderately active stars.  

Seven stars (19\%) of the CES sample show minor signs of variability (minor in the sense that they
pass one but fail at other tests): $\zeta$~Tuc, HR~506, $\zeta^{2}$~Ret, $\epsilon$~Eri,
$\phi^{2}$~Pav and HR~8883.
While for $\epsilon$~Eri this is an indication for the presence of the highly eccentric RV
signature of the planet, the cause of variability for the other stars remains unknown since no
convincing periodicity or trends were found. For HR~8883 the $RV$ variations can be explained by the high
instrinsic activity of this star (high X-ray luminosity). 

Low amplitude linear $RV$ trends were found for the following 5 targets (14\%): $\beta$~Hyi,
$\alpha$~For, HR~6416, $\epsilon$~Ind and HR~8501. For the known binaries $\alpha$~For, HR~6416
and HR~8501 these trends agree well with the expected acceleration by the stellar secondary.
The large scatter around the linear trend of $\alpha$~For is probably also due to high stellar activity
(again a high X-ray luminosity).

$\beta$~Hyi and $\epsilon$~Ind are identified as candidates for having long-period and probably stellar
companions. But could these trends also be caused by planets? If we assume that the minimum orbital
period would be 4 times the monitoring time span (i.e. $P\approx20$ years) and that the $RV$
semi-amplitude is of the order of the $RV$ shift over the 5 years ($\beta$~Hyi: $38~{\rm m\,s}^{-1}$,
$\epsilon$~Ind: $21~{\rm m\,s}^{-1}$) and $e=0$, the observed trends could be caused in the case of 
$\beta$~Hyi by a planet with $m\sin i \approx 4~{\rm M}_{\rm Jup}$ at $a\approx 7.6$ AU and for 
$\epsilon$~Ind by an $m\sin i \approx 1.6~{\rm M}_{\rm Jup}$ companion at $a\approx 6.5$ AU.  
Such planetary systems with a distant giant planet would resemble our Solar System more closely than
the extrasolar planetary systems found so far. For $e>0$ orbits the period can even be much shorter
than 20 years and we therefore conclude that although the linearity of the $RV$ trends points towards
distant and previously unknown stellar companions, both stars constitute prime targets for follow-up
observations by the CES planet search program.

$\phi^{2}$ Pav has been earlier announced by our team as a possible candidate for having a planetary
companion with an orbital period of about 43 days and $m\sin i = 0.7~{\rm M}_{\rm Jup}$
(K\"urster et al.~\cite{martin99}). This signal was found with a low confidence level and based
on a preliminary analysis of a subset of the Long Camera data, using an early version of the 
$Radial$ code (Cochran \& Hatzes~\cite{radial}) to obtain the $RV$ measurements. The analysis
of the complete data set of $\phi^{2}$ Pav using the $Austral$ software did not confirm the 
presence of this companion. The total rms scatter over the entire 5 1/2 years is $35.4~{\rm m\,s}^{-1}$, 
slightly larger than the mean internal error of $31.3~{\rm m\,s}^{-1}$ for this star. No 
apparent Keplerian signal is present. This is consistent with results coming from the Anglo-Australian
planet search (Butler et al.~\cite{butler01}), who collected 7 measurements of $\phi^{2}$ Pav over the course
of 1 year, which reveal a total rms scatter of only $5~{\rm m\,s}^{-1}$ (the Anglo-Australian 
planet search uses the UCLES echelle spectrometer which covers a much larger spectral region and the entire
$1000~{\rm \AA}$ of the I$_2$-reference spectrum, hence the higher precision of their results).
However, we have identified in our much longer and higher sampled data a periodic signal of
$\approx 7$ days, again with low confidence (the FAP of this signal is still higher than $0.001$ 
but it appears in both the unbinned original $RV$ data as well as in the nightly averaged results). 
If the periodic signal is indeed real what could produce such an $RV$ signature?
$\phi^{2}$~Pav belongs to the $\zeta$~Ret stellar kinematic group, a group of metal deficient stars
with an age of $\approx 5$ Gyr (del Peloso et al.~\cite{peloso}).  
The iron abundance was determined as $[$Fe/H$] = -0.37$ by Porto de Mello \& da Silva (\cite{gustavo})
and as $[$Fe/H$] = -0.44$ by Edvardsson et al. (\cite{edvardsson}).  
This low metallicity can account for the observed large $RV$ scatter, since fewer and shallower 
absorption lines in the small CES bandpass degrade our measurement precision. In fact $\phi^{2}$ Pav 
has the second largest internal $RV$ error of the F-type stars in the CES sample.
Based on H$\alpha$ emission, $\phi^{2}$~Pav appears to be slightly more active than the Sun 
(del Peloso et al.~\cite{peloso}) and the star is already evolving into the subgiant phase (Porto de Mello 
\& da Silva~\cite{gustavo}). With this higher level of activity we suspect that the
$P=7$ day $RV$ variation is in fact the stellar rotation period and that our $RV$ measurements 
are affected by cool spots in the photosphere of $\phi^{2}$ Pav. These spots would modulate
the $RV$ measurements with a typical timescale of $P_{\rm Rot}$. Since these
spots appear and disappear on short timescales compared to the monitoring duration and 
the overall activity level might change over 5 1/2 years, the amplitude as well as the phase of
this modulation varies with time. Such a signal is therefore difficult to detect significantly, 
which is exactly what we observe here. 
The expected size of the subsurface convection zone for a low-metallicity F-type star is smaller than
for a star of solar metallicity. Even with $P_{\rm Rot}=7$ days such a star would not
display a much larger activity level than $\phi^{2}$~Pav due to the inefficiency of the dynamo.
From the $v\sin i = 6.7~{\rm kms}^{-1}$ and $R_{*}=1.86~R_{\odot}$ 
(Porto de Mello priv. comm.) and the $P_{\rm Rot}$ value of $7$ days we derive a viewing angle 
of $\approx 30^{\circ}$. The continued monitoring of $\phi^{2}$ Pav will demonstrate 
whether the $P=7$ day is robust and can be recovered with a higher confidence level.      

Roughly 50\% of the targets (18 stars) of the CES Long Camera survey show absolutely no sign
of variability or trends in their $RV$ data. Within the given $RV$ precision of the CES 
Long Camera survey the following stars were found to be $RV$-constant: HR 209, HR 448, HR 753,
$\zeta^{1}$Ret, $\delta$ Eri, HR 2667, HR 4523, HR 4979, HR 6998, HR 7373, HR 7703, HR 8323 and GJ 433. 

In the cases of the binaries $\kappa$~For, HR~2400, HR~3677 and $\alpha$~Cen A \& B (see Endl et
al.~\cite{michl01}) no sign of significant periodic signals were found in the $RV$ residuals after
subtraction of the binary orbit. Interestingly, $\kappa$~For does not show any excess scatter 
although based on its $L_{X}$-flux and H$\alpha$-emission (Porto de Mello priv. comm.) 
it is an active star. Still, the residuals after subtraction of the binary orbit are consistent 
with our measurement errors. 

The CES Long Camera survey is in $all$ cases sensitive to short-period (``51 Peg''-type) planets with
orbital separations of $a < 0.15$ AU. This result confirms the general bias of precise Doppler searches
towards short-period companions. For 22 stars of the CES Long Camera survey these mass-limits reach
down into the sub-Saturn mass regime at $a \approx 0.04$ AU.

For most stars the region where planets with $m \sin i < 1~{\rm M}_{\rm Jup}$ could have been detected 
is confined to orbital separations of less than 1 AU. Beyond 2 AUs no planets with $m \sin i \approx 
1~{\rm M}_{\rm Jup}$ were found to be detectable around any star of the survey. Subsequently, in order
to detect a Solar System analogue the time baseline and (if possible) the $RV$ precision of the CES
planet search has to be increased. 
Within the limitations of our numerical simulations ($e=0$, $P^{2/3}$ sampling)
we can rule out the presence of giant planets within 3 AU of the CES survey stars according to the
limits presented here (with the exceptions of HR 209, HR 8883 and periods inside the
non-detectability windows). 
   
Spectral leakage is the main cause for the windows of non-detectability. Even for well observed stars
like e.g. $\tau$~Cet or $\beta$~Hyi these windows exist close to the seasonal one year period. This
demonstrates how difficult the detection of $RV$ signals with a one year periodicity is.

The average detection threshold \={F}$_{\rm CES}$ for the examined 30 Long Camera survey stars is
$2.75$, meaning that on average detectable planetary signals have $K$ amplitudes which exceed the noise
level by a factor of $2.75$. 

\subsection{Outlook}

With the decommissioning of the Long Camera in April 1998, phase I of the CES 
planet search program came to an end. All results based on this homogeneous set of 
observations are included in this work or were already presented earlier. 

\begin{figure}
\centering{
  \vbox{\psfig{figure=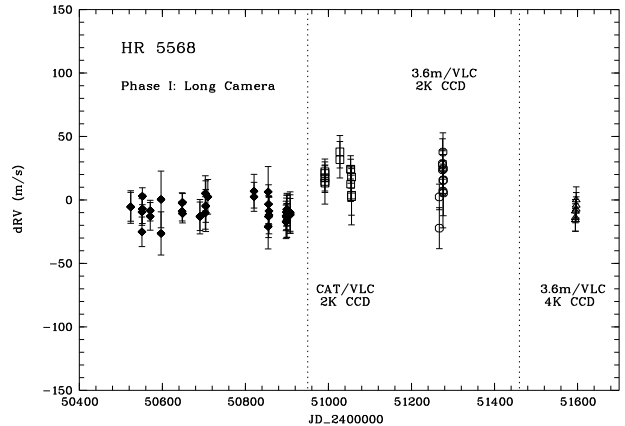,width=9.0cm,height=6.0cm,angle=270}}
   \par
        }
  \caption[]{$RV$ monitoring of HR 5668 during the refurbishment of the CES. A 
	comparison of the Long Camera results (full diamonds) with data
	collected with the new VLC and the 2K CCD (boxes and circles) show a
	slight offset. This offset disappears with the installation of the 4K
	CCD (triangles) which increased and equalized the spectral bandwidth (see text for details).
	The total rms scatter over the 3 years is $15.5~{\rm m\,s}^{-1}$, 
	and $7.5~{\rm m\,s}^{-1}$ without the intermediate data (between the vertical dotted lines).	
        }
 \label{gj570a_vlc}
\end{figure}

Although the CES was modified quite substantially after that, with the installation of the 
Very Long Camera (VLC) yielding a higher resolving power of $R\approx220,000$ and an optical fibre-link 
to the 3.6 m telescope being the most significant changes, the CES planet search program
was continued using the same I$_2$-cell for self-calibration. This ensures the 
capability to merge the $RV$ results from phase I with the newer phase II data set without
the need to compensate for velocity zero-point drifts as demonstrated for HR 5568 in
Fig.~\ref{gj570a_vlc}. The displayed $RV$ results now cover almost 3 years for this star (compared to 1
year of the Long Camera survey). For the intermediate time when the 1.4 m CAT and the 
3.6 m telescope were used in combination with the VLC and the 2K CCD (which meant a reduced bandwidth
of $\approx 18~{\rm \AA}$ due to the higher spectral dispersion) we observe a small $RV$ offset 
of $\approx 25~{\rm m\,s}^{-1}$. This offset can be explained by the difference in spectral regions 
which were analysed to obtain the $RV$s. However, after the VLC was equipped with a longer 
4K CCD the spectral bandwidth was increased to $36.5~{\rm \AA}$. To assure that the $RV$ results are
based on the same spectral regions we analyse both the Long Camera and the VLC data using a stellar
template spectrum obtained with the most current instrumental setup (i.e. VLC \& 4K CCD). A comparison
of the $RV$ results derived with the current CES and Long Camera results does no longer show any velocity 
offset (see Fig.~\ref{gj570a_vlc}). This demonstrates that the I$_2$-cell technique 
successfully compensates even for major instrumental setup changes. This guarantees a high
long-term $RV$ precision and allows a smooth continuation of the CES planet search program.    

The Very Long Camera at the CES is promising to increase the $RV$ precision of the CES planet search 
due to several reasons: the resolving power is doubled with respect to the Long Camera while the
spectral bandwidth is not reduced by a large amount ($36.5~{\rm \AA}$ instead of $48.5$), and the 
$S/N$-ratio of spectra is higher due to the usage of image-slicers and the larger aperture 
of the 3.6 m telescope.   
With the successful merging of the new Very Long Camera data with the Long Camera survey and an 
expected better $RV$ precision the CES planet search might become sensitive to Solar System 
analogues in the near future. 

\section{Summary}
%
\begin{enumerate}

   \item We present more than 5 years of high precision differential radial velocities 
	of 37 stars in the southern hemisphere observed with the 1.4 m CAT telescope and the 
	CES spectrograph in Long Camera configuration. 

   \item In this sample of stars we detected 1 extrasolar planet around the young and 
	active G-dwarf $\iota$ Hor. The results for $\epsilon$ Eri contribute to the
	evidence for a long-period planetary companion.

   \item $\kappa$~For, HR~2400 and HR~3677 were found to be spectroscopic binaries with orbital
	periods significantly longer than the duration of the Long Camera survey. For
	all three stars the $RV$ residuals after subtraction of the binary motion 	 	
	are consistent with our measurement errors. 

   \item We identified low-amplitude linear $RV$ trends for $\alpha$~For, HR~6416 and HR~8501 in agreement
	with their long-period binary orbtial motion. In the cases of $\beta$~Hyi and $\epsilon$~Ind
	we detected low-amplitude linear $RV$ slopes of unknown origin. 

   \item A period search within the complete $RV$ data set did not reveal another significant
	periodic signal beside the planetary signal of $\iota$~Hor b and in the case of HR~8501 
	a signal with $P>5000$ days indicating the found linear $RV$ trend for this star.   

   \item We set quantitative upper mass-limits for planets orbiting the CES Long Camera survey stars
	based on our long-term $RV$ data and numerical simulations of planetary signatures.  

   \item Based on these mass-limits we can exclude the presence of short-period Jupiter-type planets around
	all examined survey stars, while planets with $m\sin i < 1~{\rm M}_{\rm Jup}$ outside of
	2 AU were undetectable by the CES Long Camera survey. 

   \item Except for CES targets with $B-V<0.6$, a comparison of our simulation results with the $RV$-signals 
	of known extrasolar planets shows that most of these planets could have been detected 
	by the CES Long Camera survey.

   \item We demonstrate that the I$_2$-cell technique successfully compensates for the instrumental changes 
	at the CES after April 1998 and that a smooth continuation of the CES planet search program 
	is guaranteed. The increase in time baseline as well as a possible better $RV$ precision with the 
	Very Long Camera data will allow us to become sensitive to planetary systems more similar to our
	own Solar System.       
 
\end{enumerate}
%

\begin{acknowledgements}

We are thankful to the ESO OPC for generous allocation of observing time to the
CES planet search program and to the ESO night assistants and support staff at 
La Silla and Garching (during remote observing runs). 
Our referee Stephane Udry had many important comments which helped to improve this
article significantely.   
We would also like to thank Gustavo F. Porto de Mello, who contributed with many valuable
discussions on the stellar properties of the CES target stars. 
ME and SE both acknowledge support by the ESO science office, ME was also supported by
the Austrian Fond zur F\"orderung der wissenschaftlichen Forschung Nr.~S7302 and 
SE under E.U. Marie Curie Fellowship contract HPMD-CT-2000-00005.  
ME, APH and WDC acknowledge support from NSF Grant AST-9808980 and NASA Grant
NAG5-9227. During most of the time for this work MK was employed
by the European Southern Observatory whose support is
gratefully acknowledged. This work made us of the online SIMBAD database. 

\end{acknowledgements}

%

%

\clearpage
\newpage

\appendix
\section{Radial velocity results}

$RV$ results for all CES Long Camera survey stars are displayed for comparison in the same 
time frame (JD 2,448,800 to JD 2,451,000), and with the y-axis adjusted 
according to the individual velocity dispersion.

\begin{figure}[b]
\centering{
        \vbox{\psfig{figure=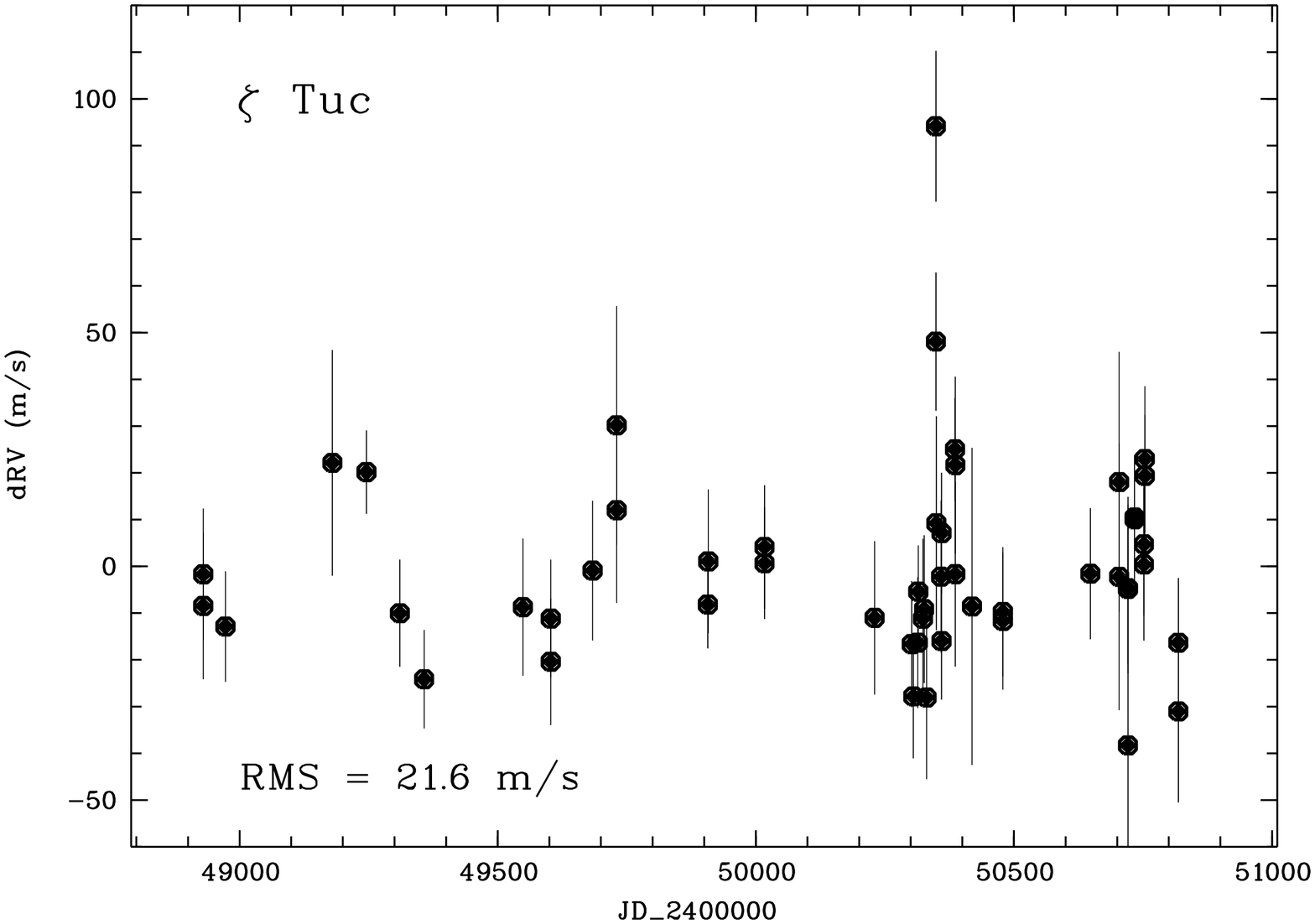,width=9.0cm,height=5.5cm,angle=270}}
        \vbox{\psfig{figure=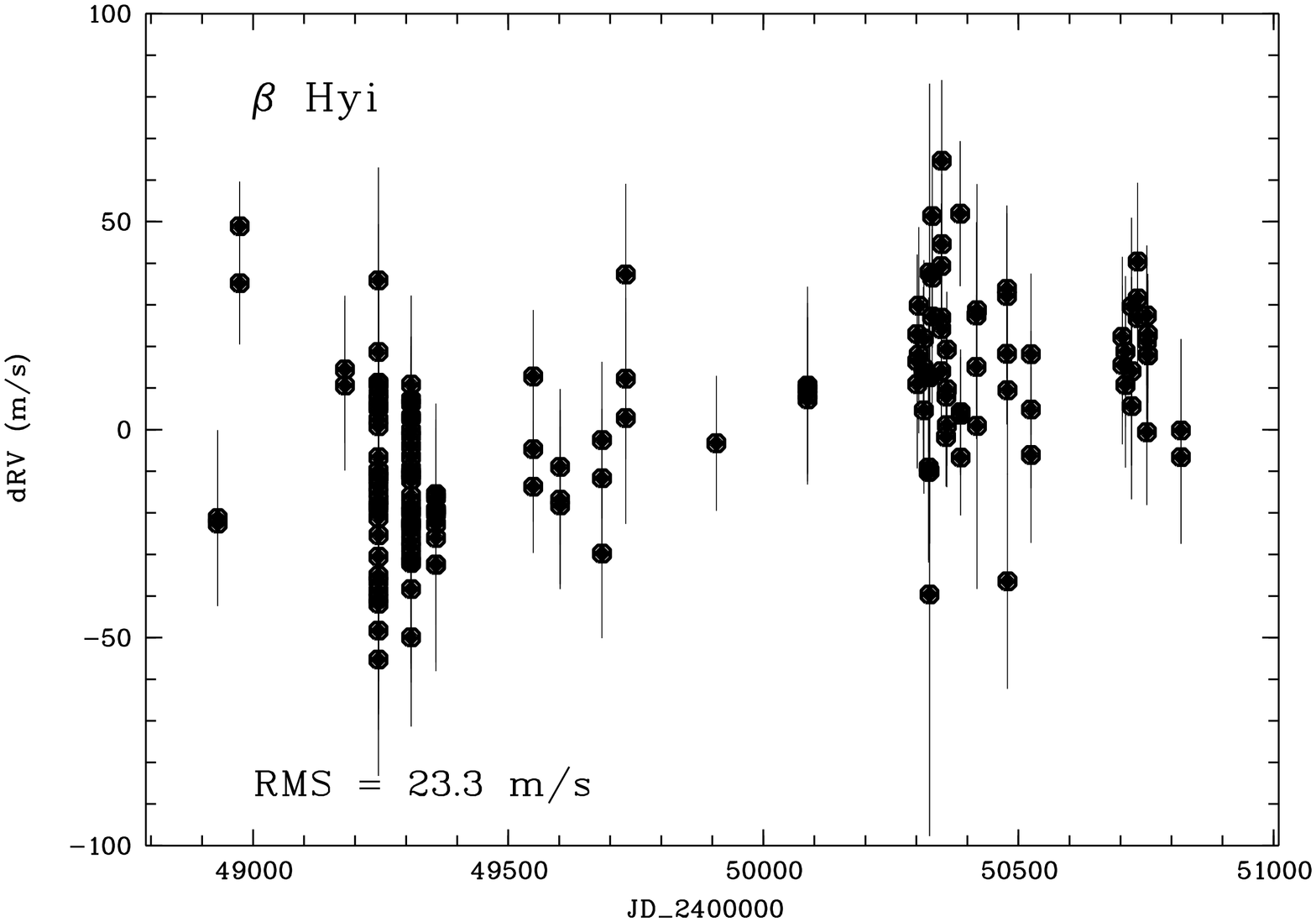,width=9.0cm,height=5.5cm,angle=270}}
        \vbox{\psfig{figure=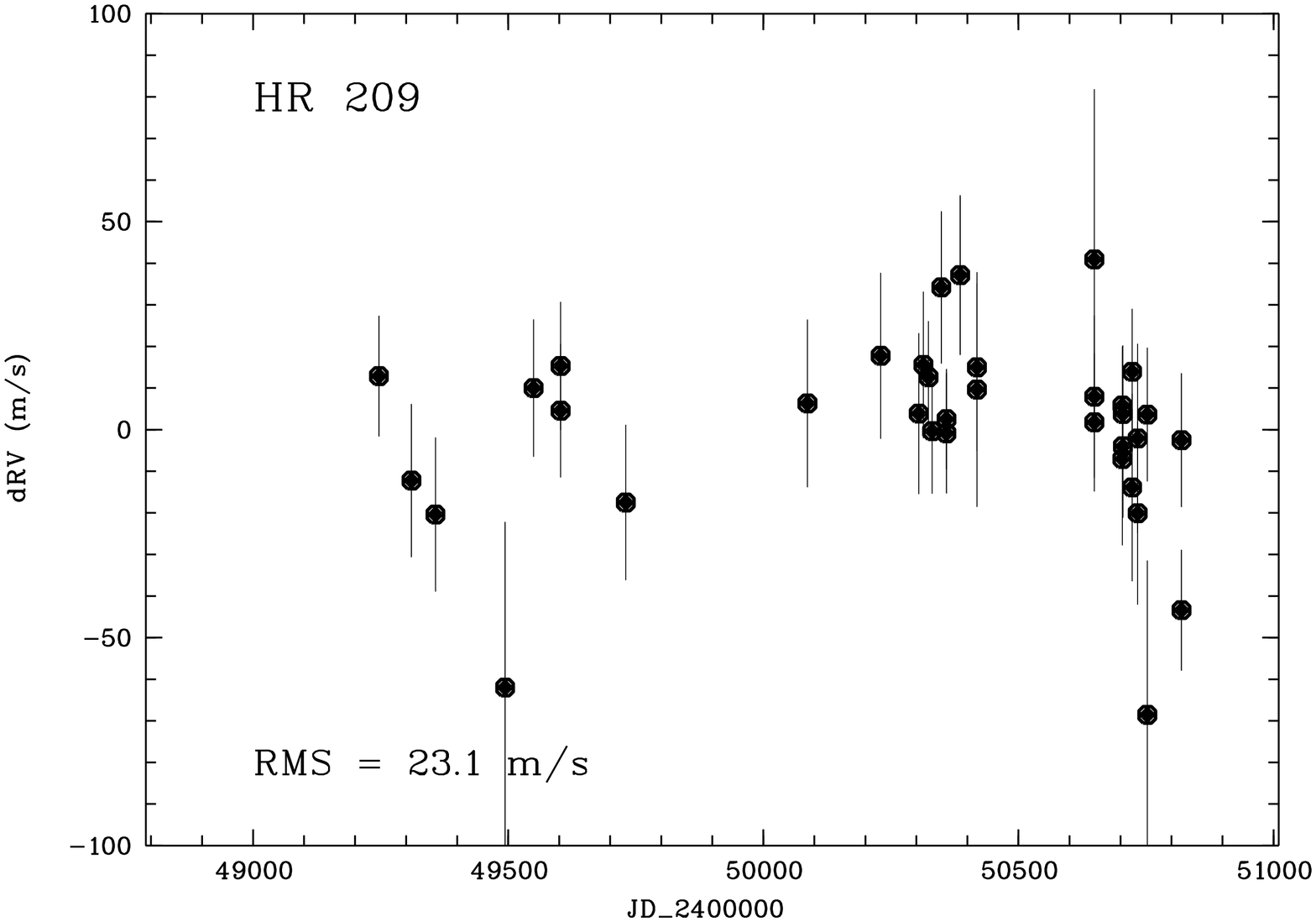,width=9.0cm,height=5.5cm,angle=270}}
        \vbox{\psfig{figure=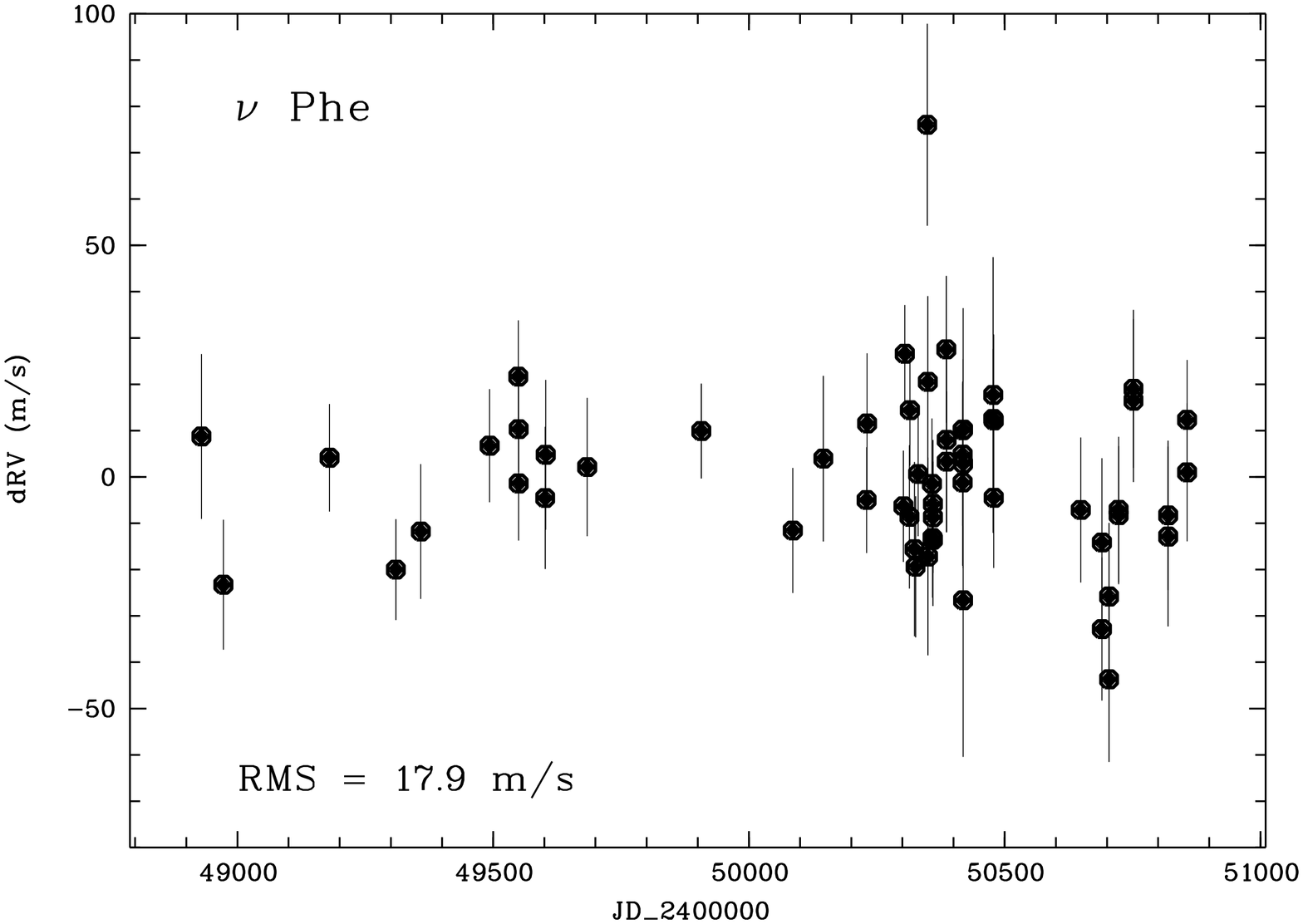,width=9.0cm,height=5.5cm,angle=270}}
   \par 
        }
  \caption[]{
        Radial velocity results for $\zeta$~Tuc, $\beta$~Hyi, HR~209 and $\nu$~Phe.}
 \label{rvsfig1}
\end{figure}
\begin{figure}
 \centering{
        \vbox{\psfig{figure=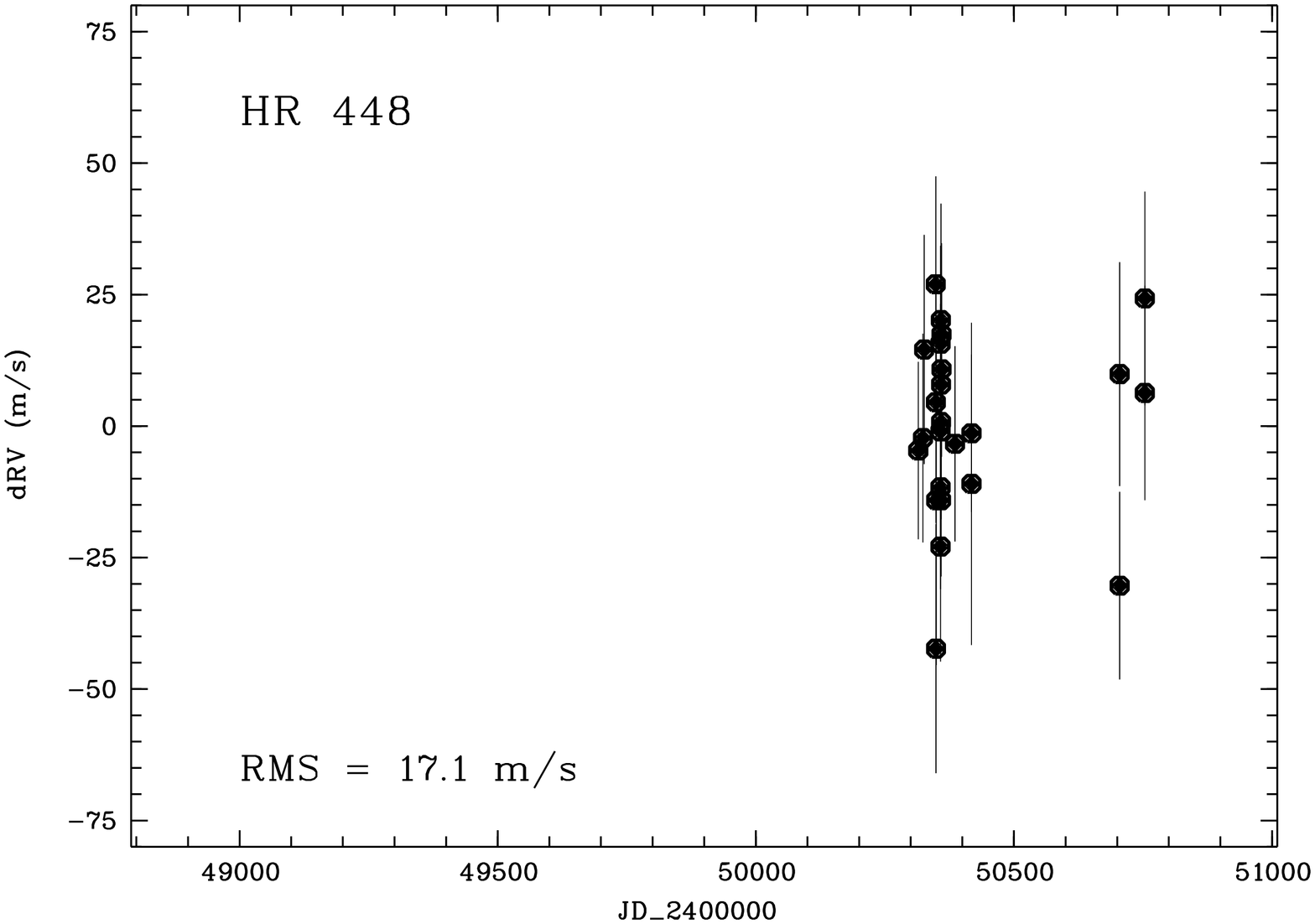,width=9.0cm,height=5.5cm,angle=270}}
        \vbox{\psfig{figure=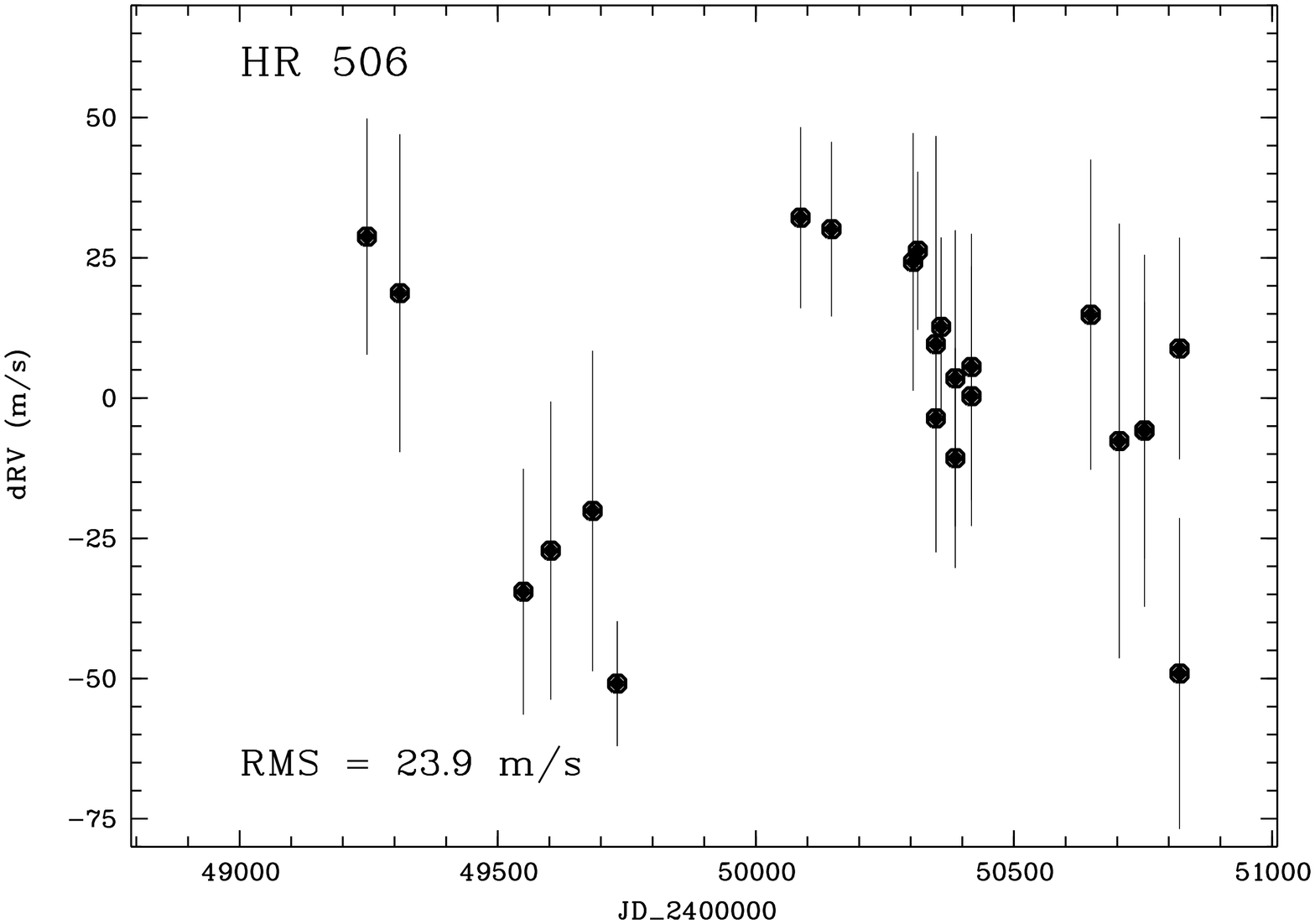,width=9.0cm,height=5.5cm,angle=270}}
        \vbox{\psfig{figure=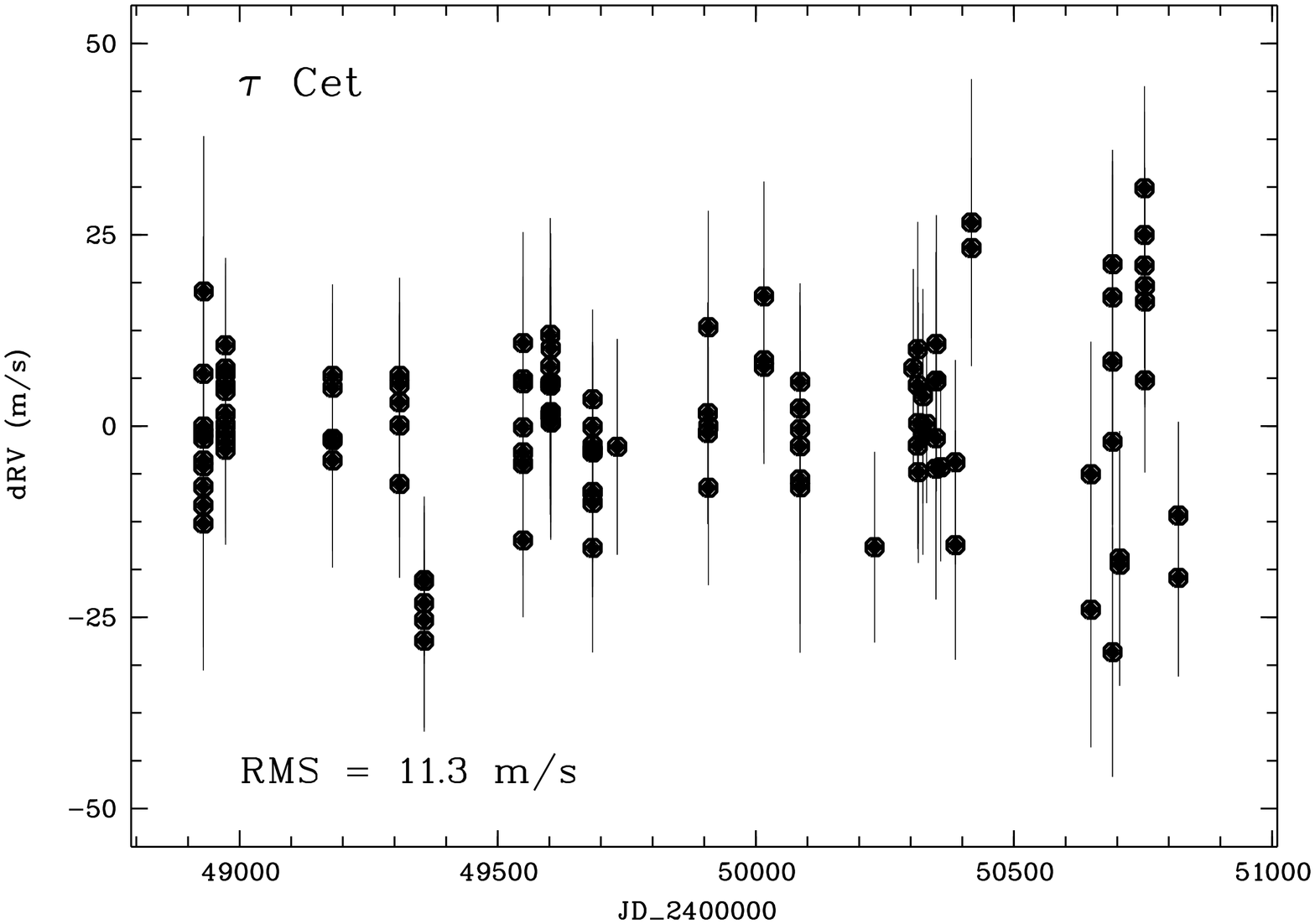,width=9.0cm,height=5.5cm,angle=270}}
	\vbox{\psfig{figure=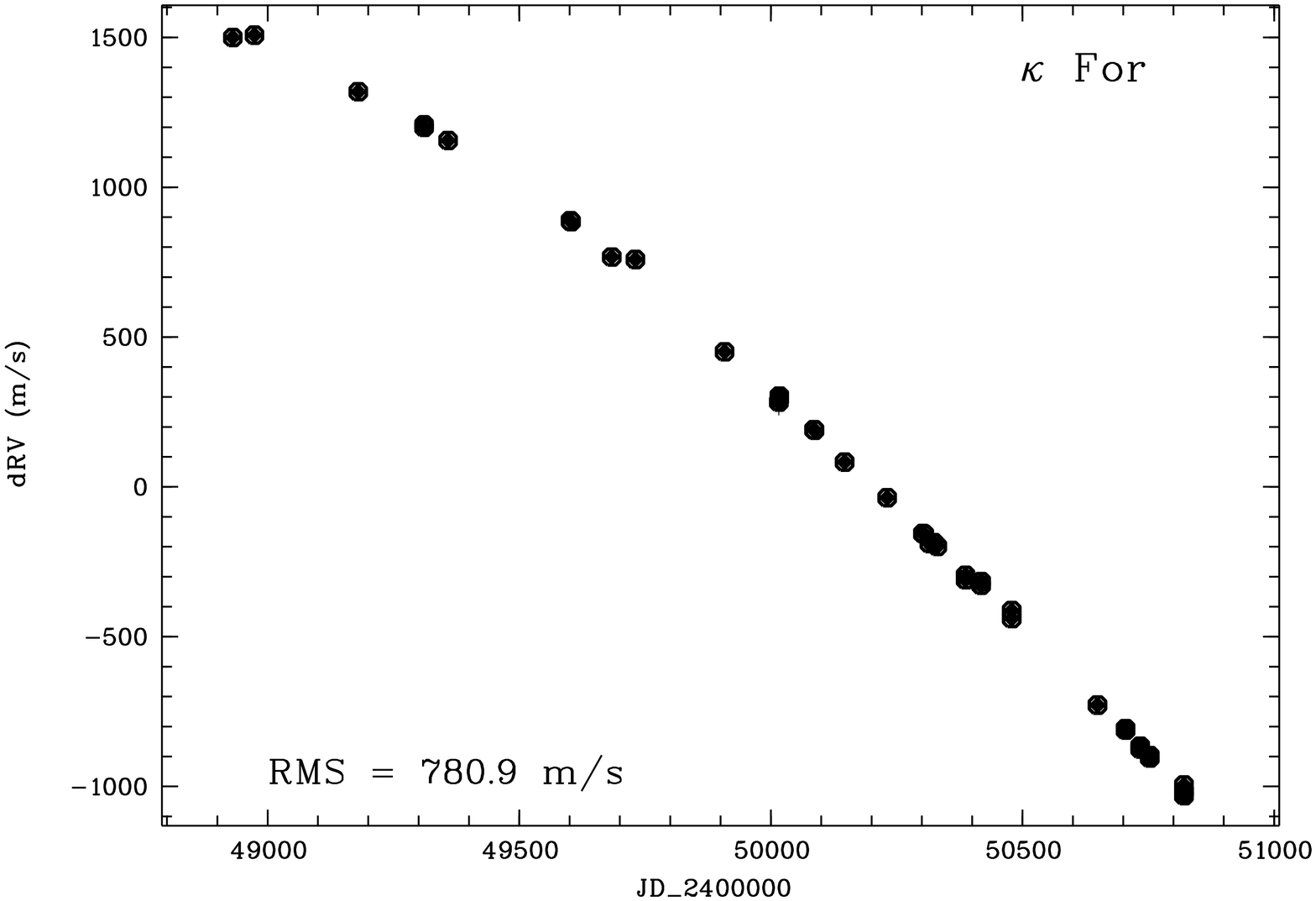,width=9.0cm,height=5.5cm,angle=270}}
   \par
        }
   \caption[]{
        Radial velocity results for HR~448, HR~506, $\tau$~Cet and $\kappa$~For.}
  \label{rvsfig2}
\end{figure}
\begin{figure}
 \centering{
	\vbox{\psfig{figure=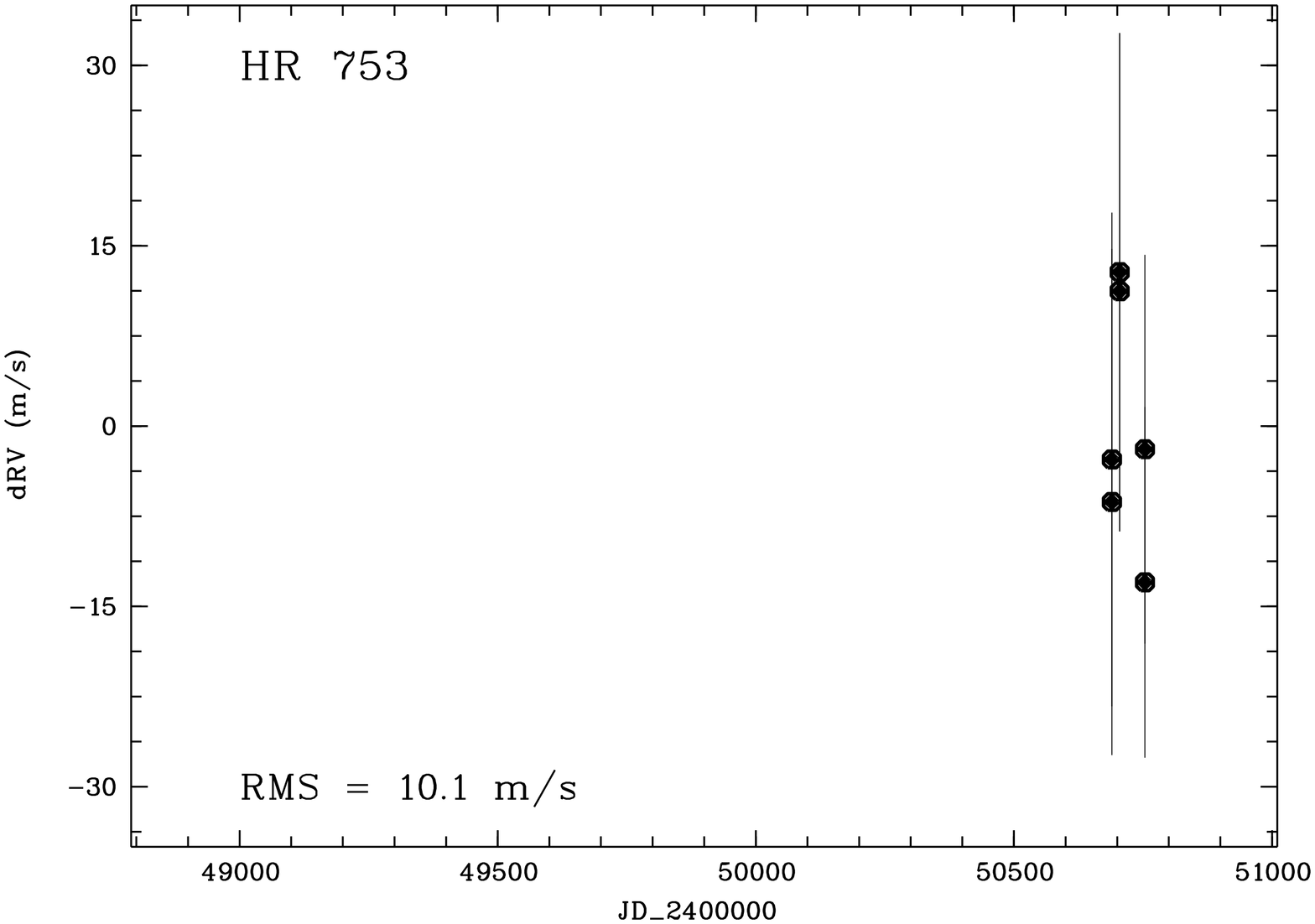,width=9.0cm,height=5.5cm,angle=270}}
        \vbox{\psfig{figure=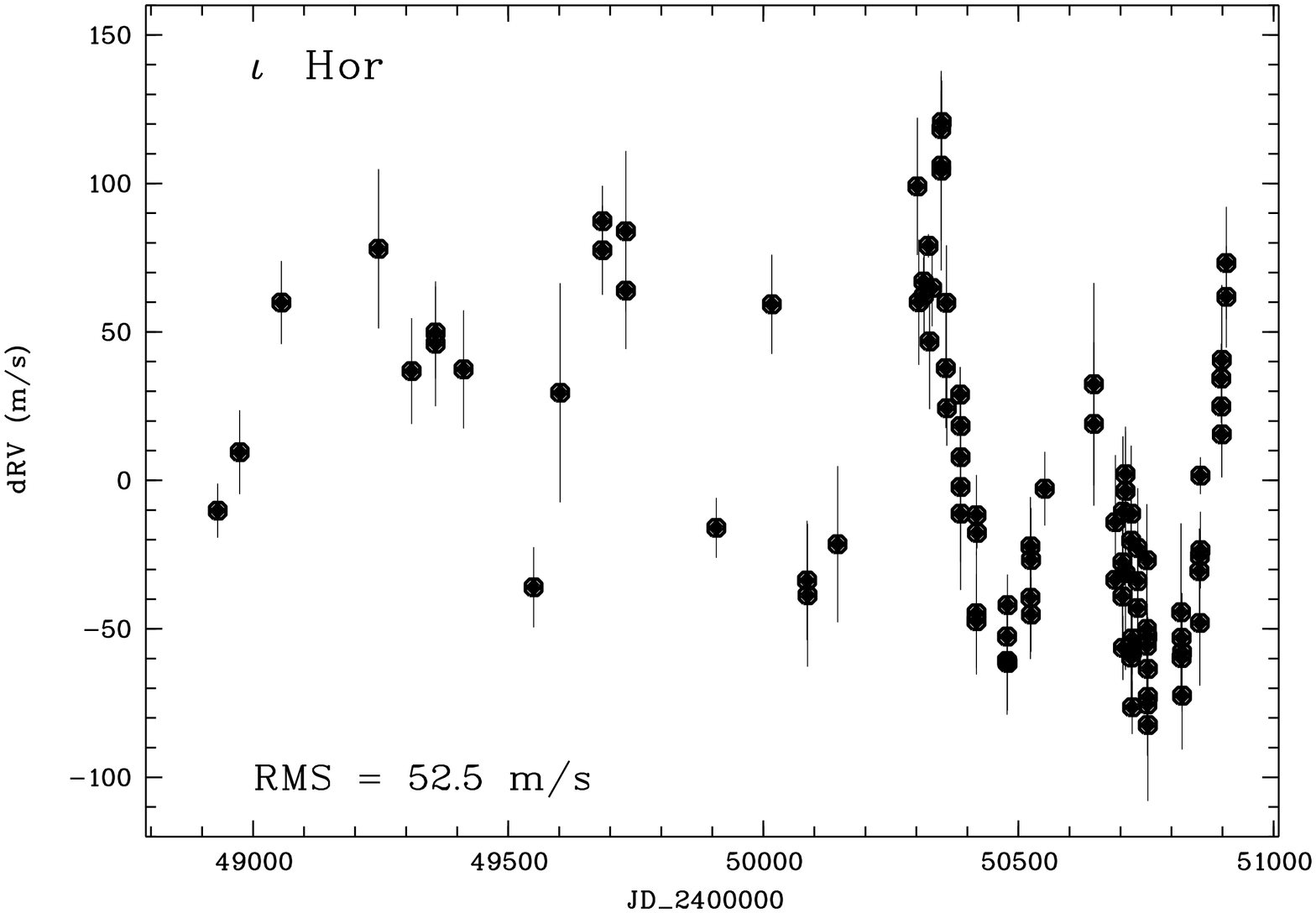,width=9.0cm,height=5.5cm,angle=270}}
	\vbox{\psfig{figure=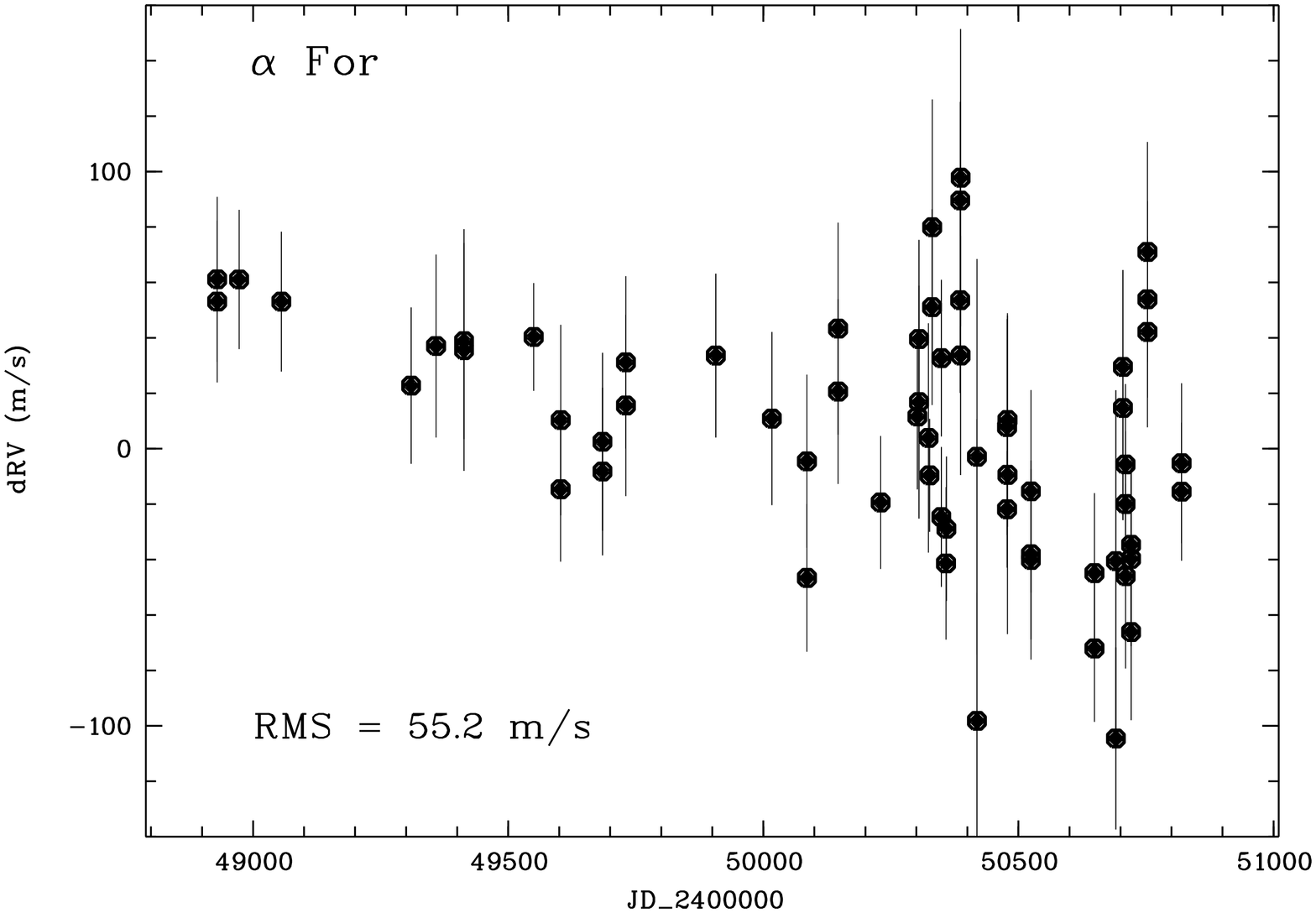,width=9.0cm,height=5.5cm,angle=270}}
        \vbox{\psfig{figure=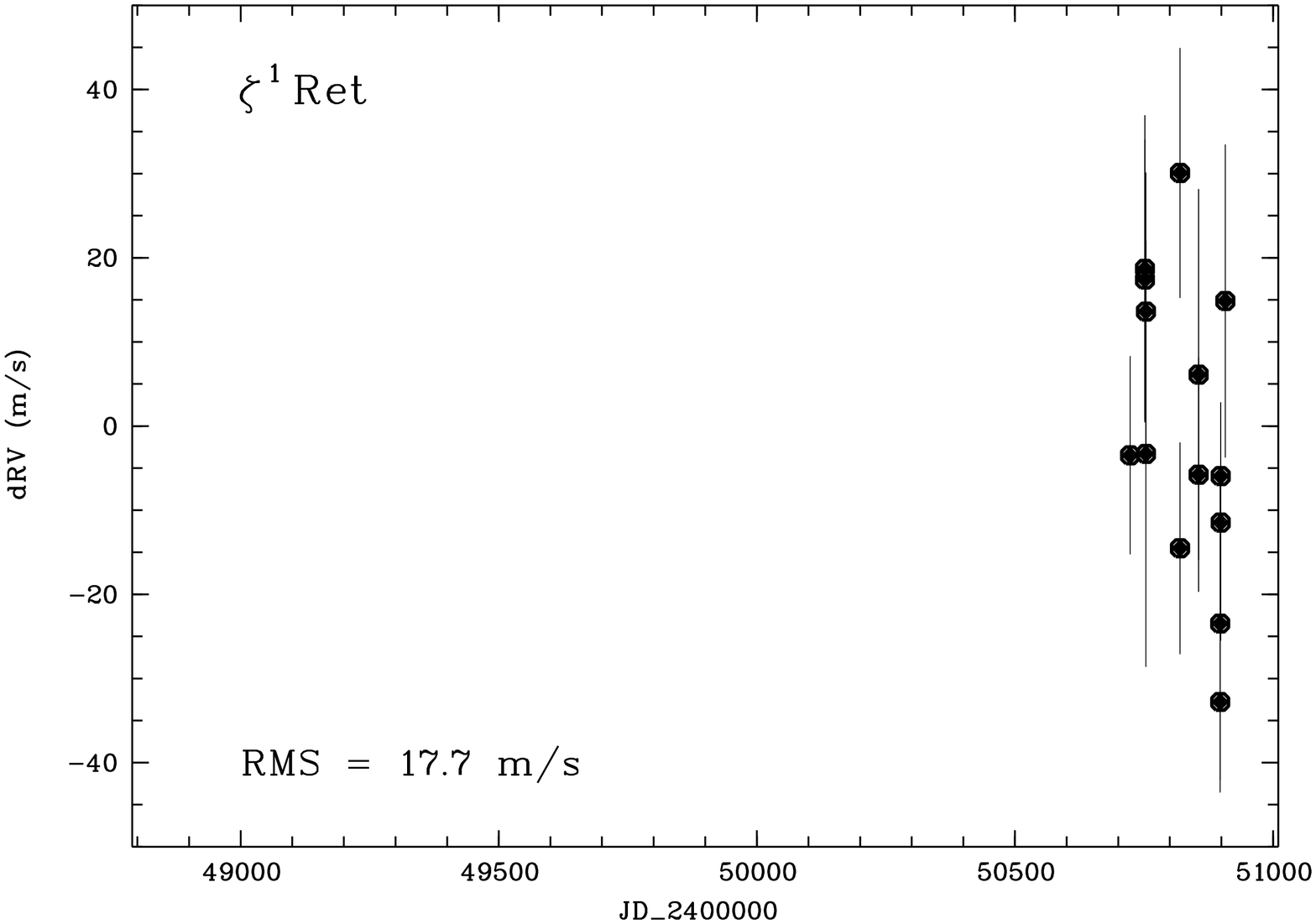,width=9.0cm,height=5.5cm,angle=270}}
   \par
        }
   \caption[]{Radial velocity results for HR~753, $\iota$~Hor, $\alpha$~For and 
	$\zeta^{1}$~Ret.} 
     \label{rvsfig3}
\end{figure}
\begin{figure}
 \centering{
	\vbox{\psfig{figure=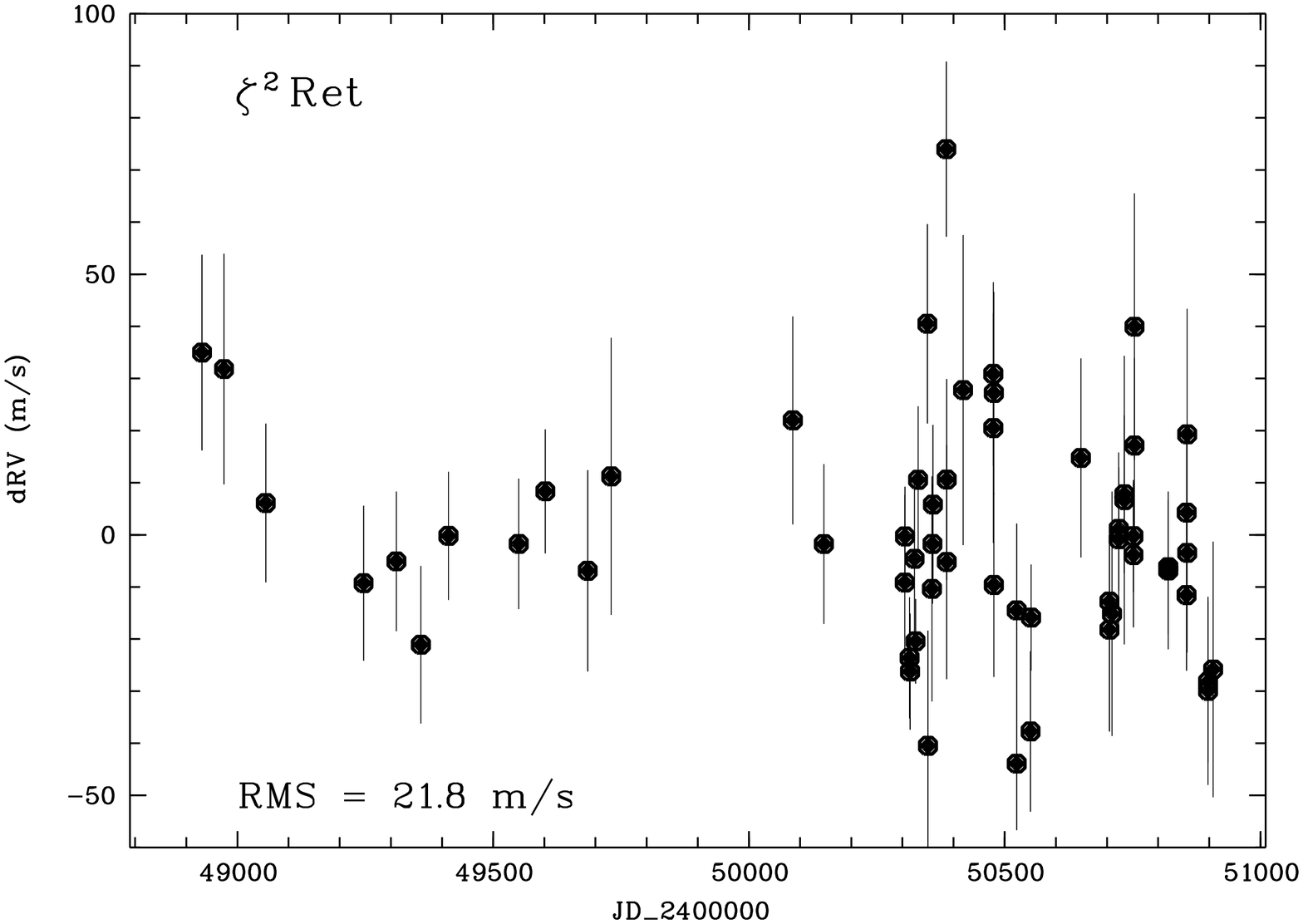,width=9.0cm,height=5.5cm,angle=270}}
	\vbox{\psfig{figure=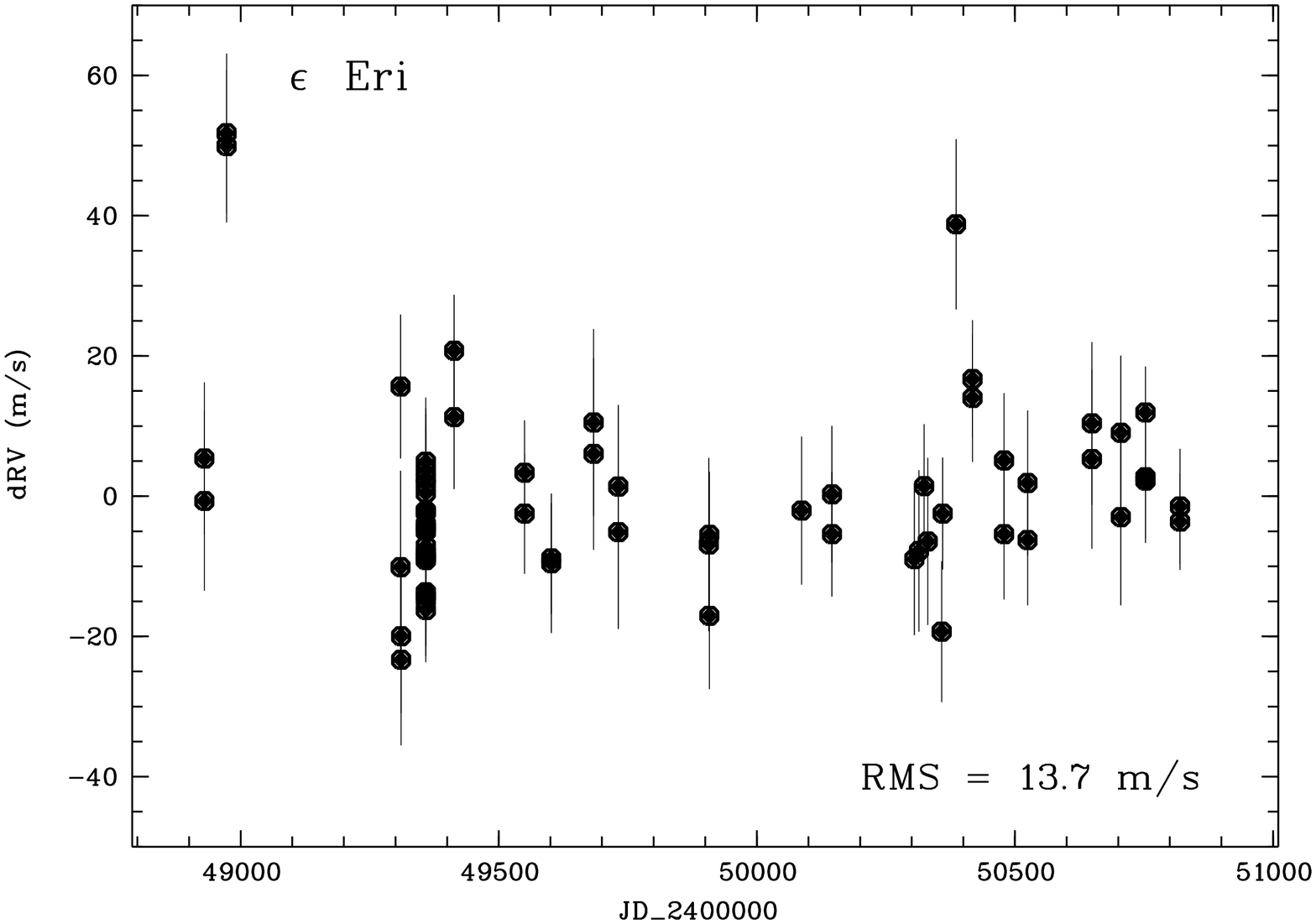,width=9.0cm,height=5.5cm,angle=270}}
        \vbox{\psfig{figure=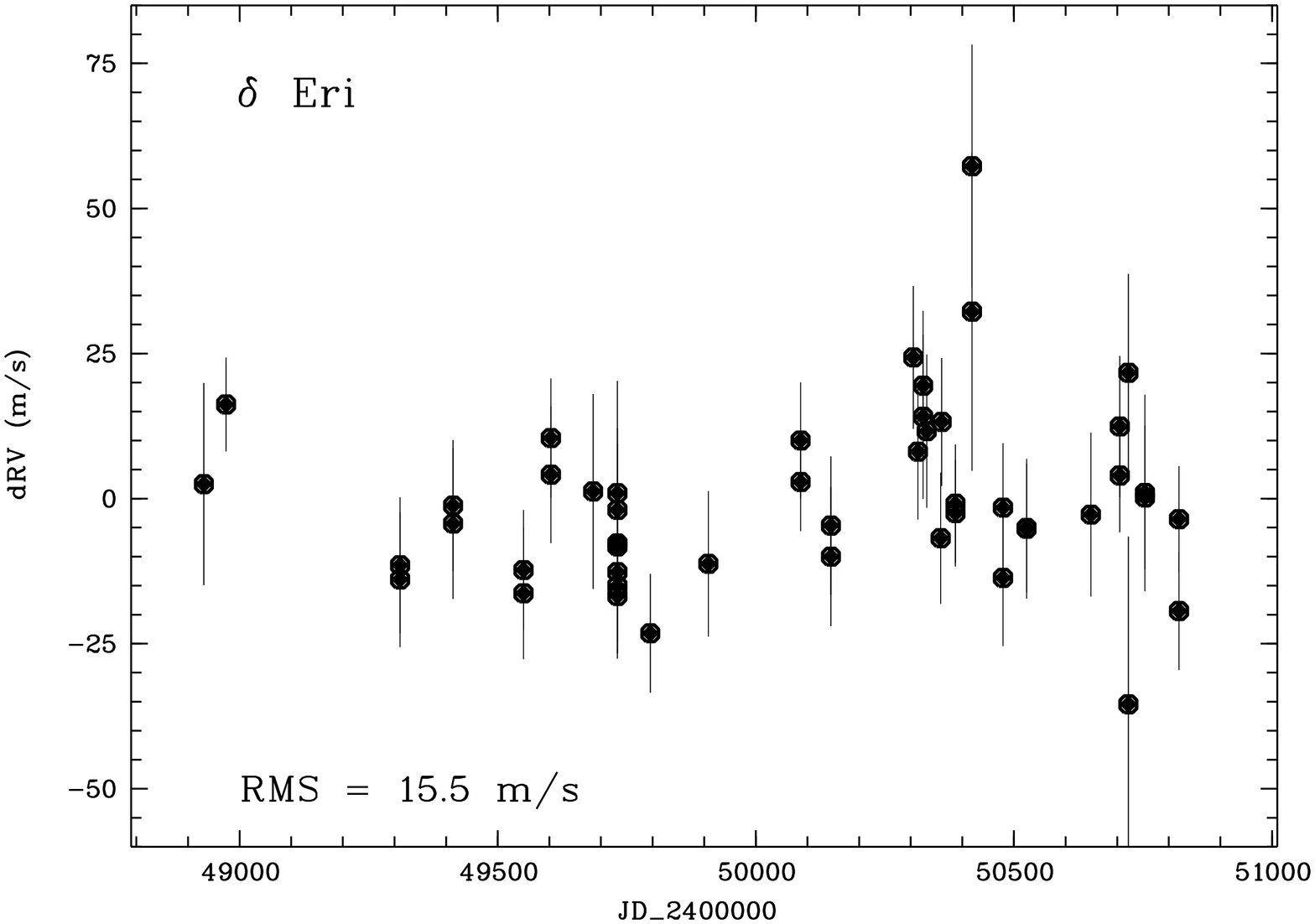,width=9.0cm,height=5.5cm,angle=270}}
        \vbox{\psfig{figure=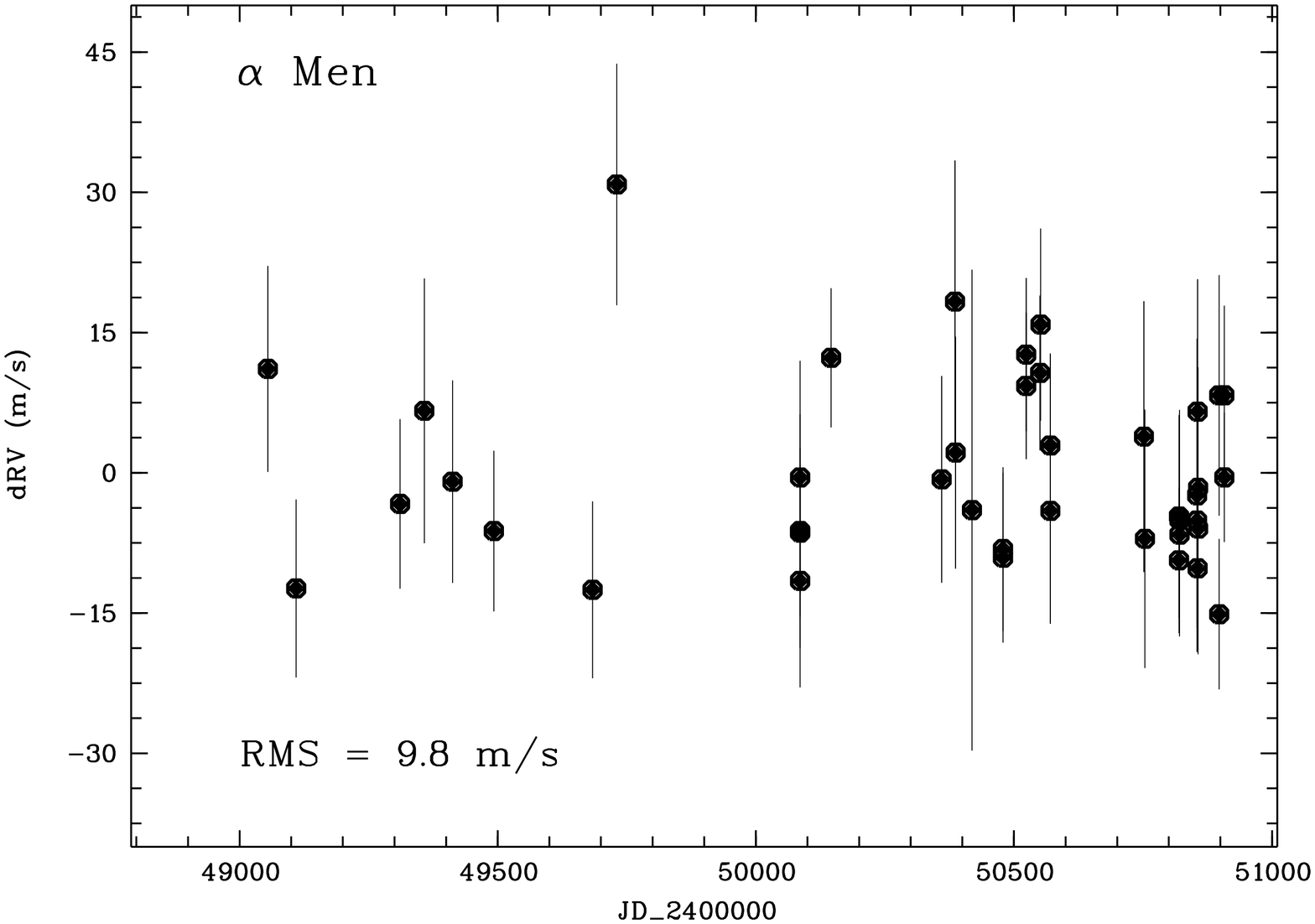,width=9.0cm,height=5.5cm,angle=270}}
    \par
        }
   \caption[]{Radial velocity results for $\zeta^{2}$~Ret, $\epsilon$~Eri, $\delta$~Eri
	and $\alpha$~Men.}
  \label{rvsfig4}
\end{figure}
\begin{figure}
 \centering{
	\vbox{\psfig{figure=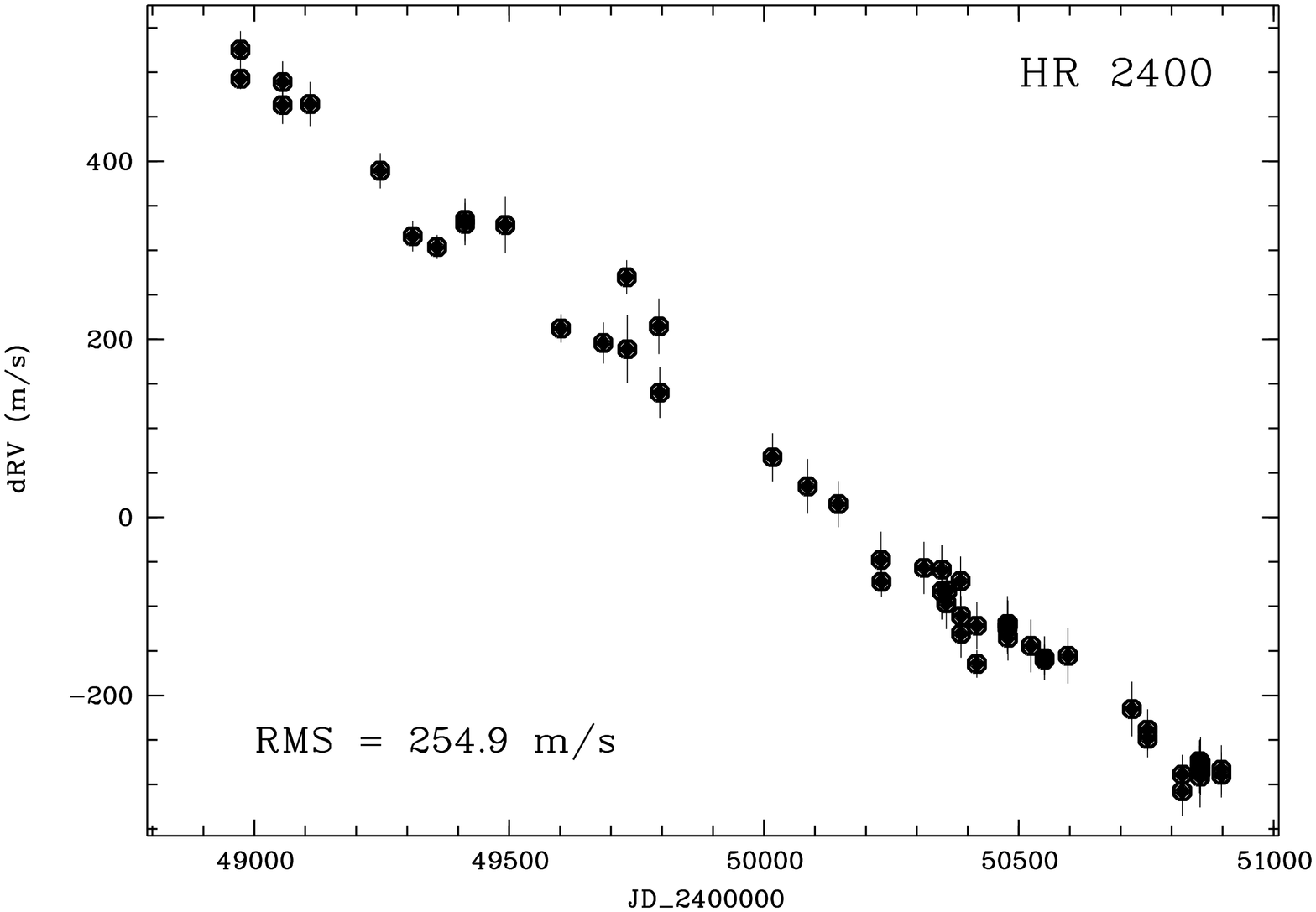,width=9.0cm,height=5.5cm,angle=270}}
	\vbox{\psfig{figure=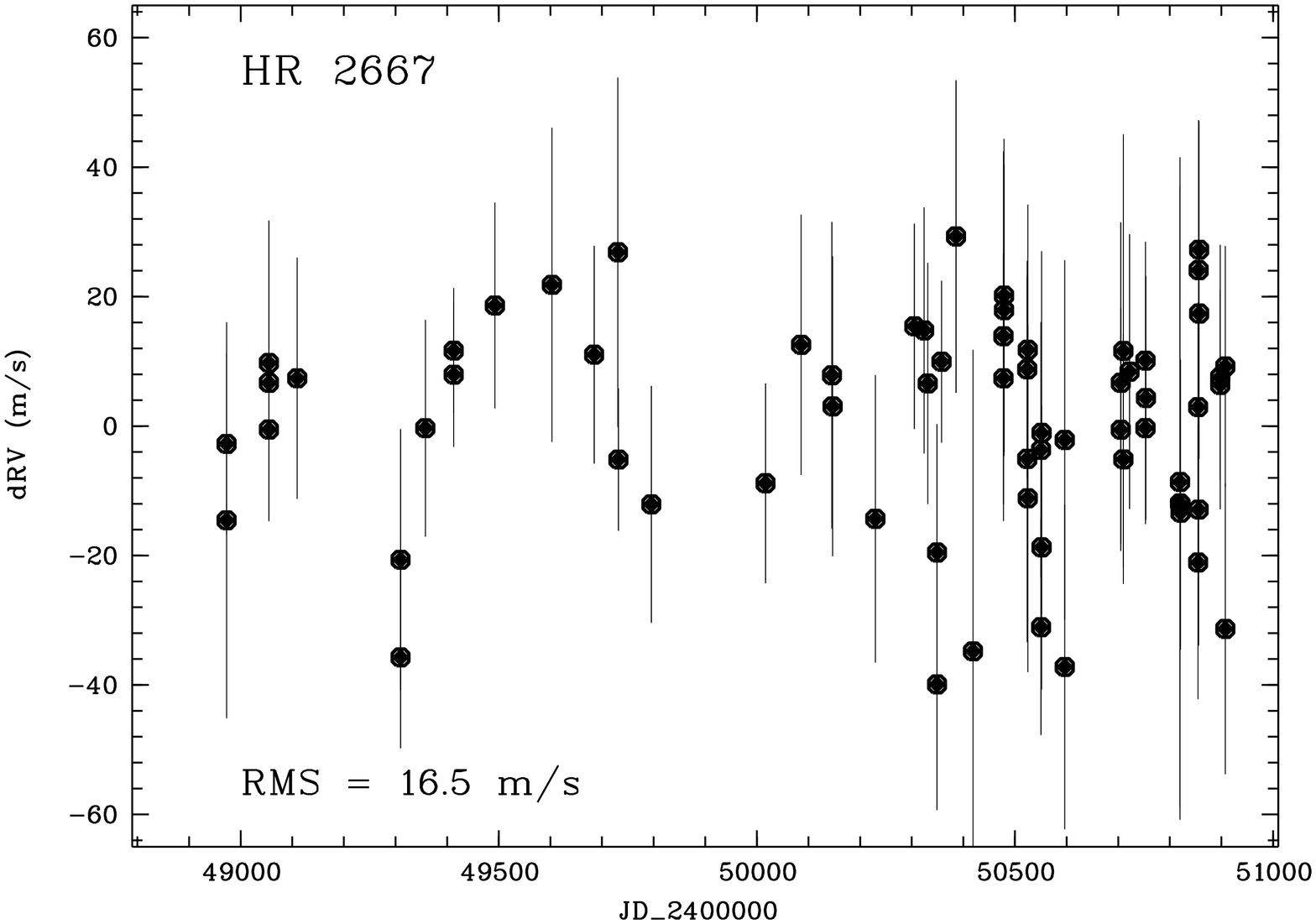,width=9.0cm,height=5.5cm,angle=270}}
	\vbox{\psfig{figure=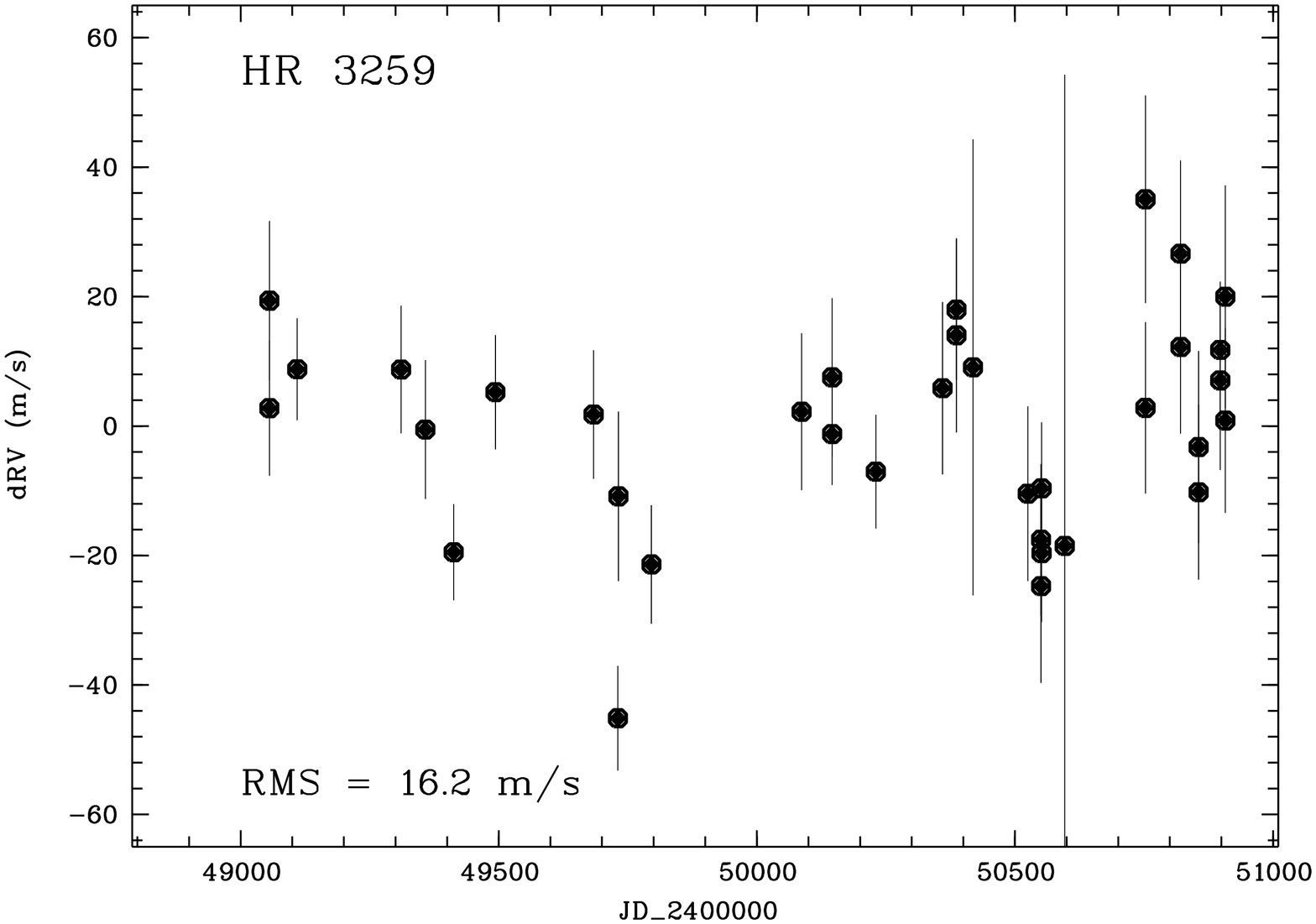,width=9.0cm,height=5.5cm,angle=270}}
	\vbox{\psfig{figure=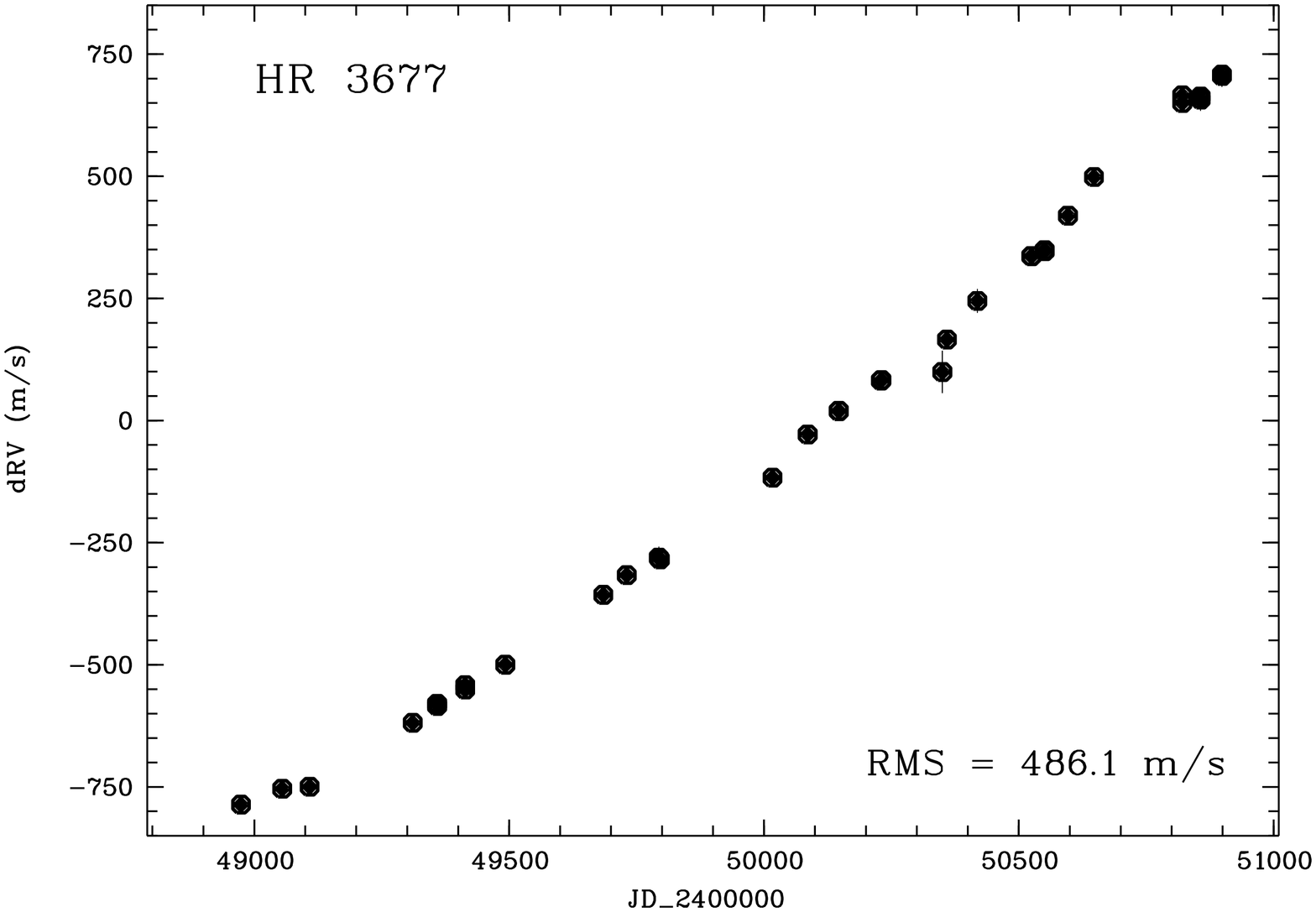,width=9.0cm,height=5.5cm,angle=270}}
   \par
        }
   \caption[]{ 
         Radial velocity results for HR~2400, HR~2667, HR~3259 and HR~3677.} 
  \label{rvsfig5}
\end{figure}
\begin{figure}
 \centering{
        \vbox{\psfig{figure=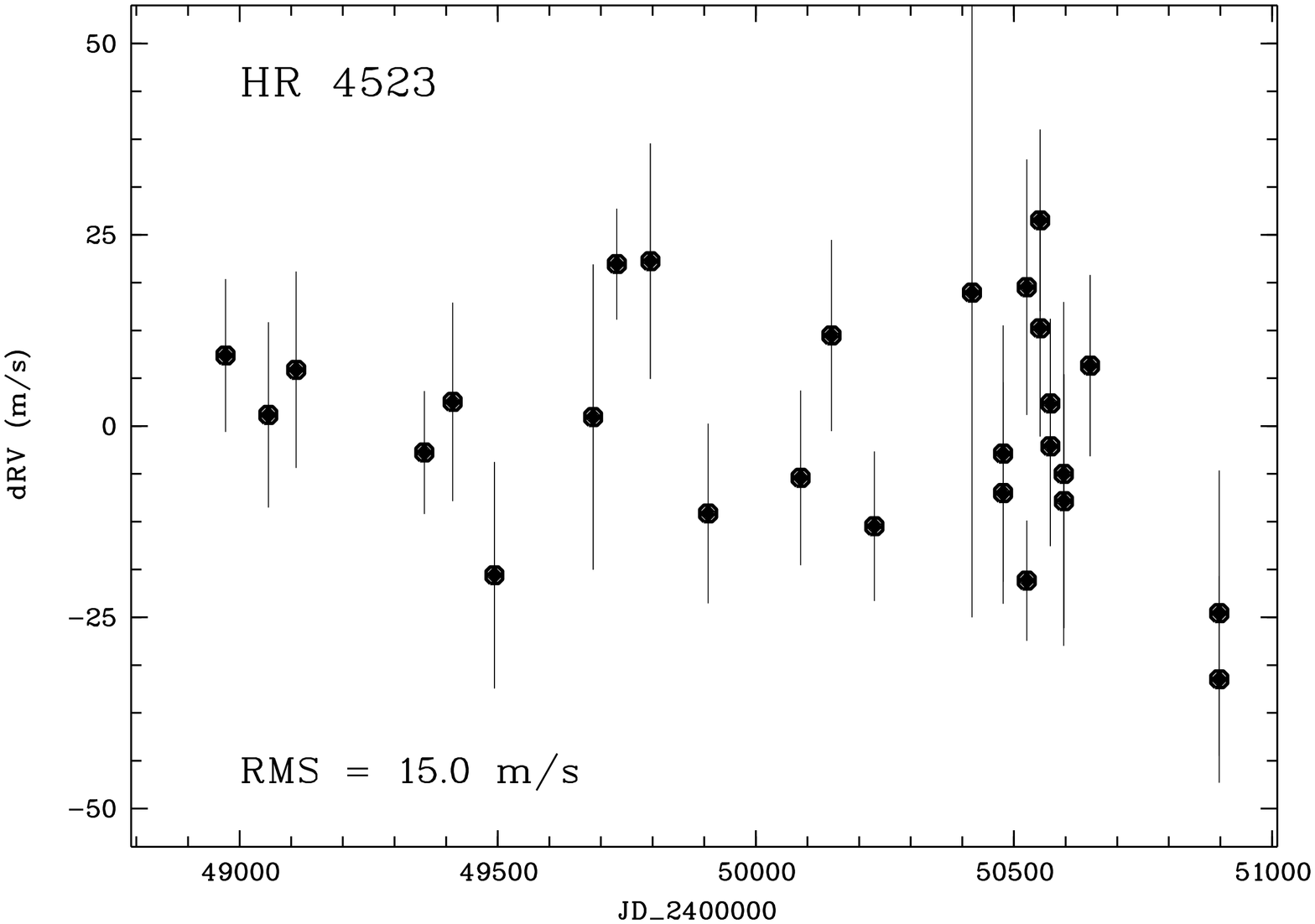,width=9.0cm,height=5.5cm,angle=270}} 
        \vbox{\psfig{figure=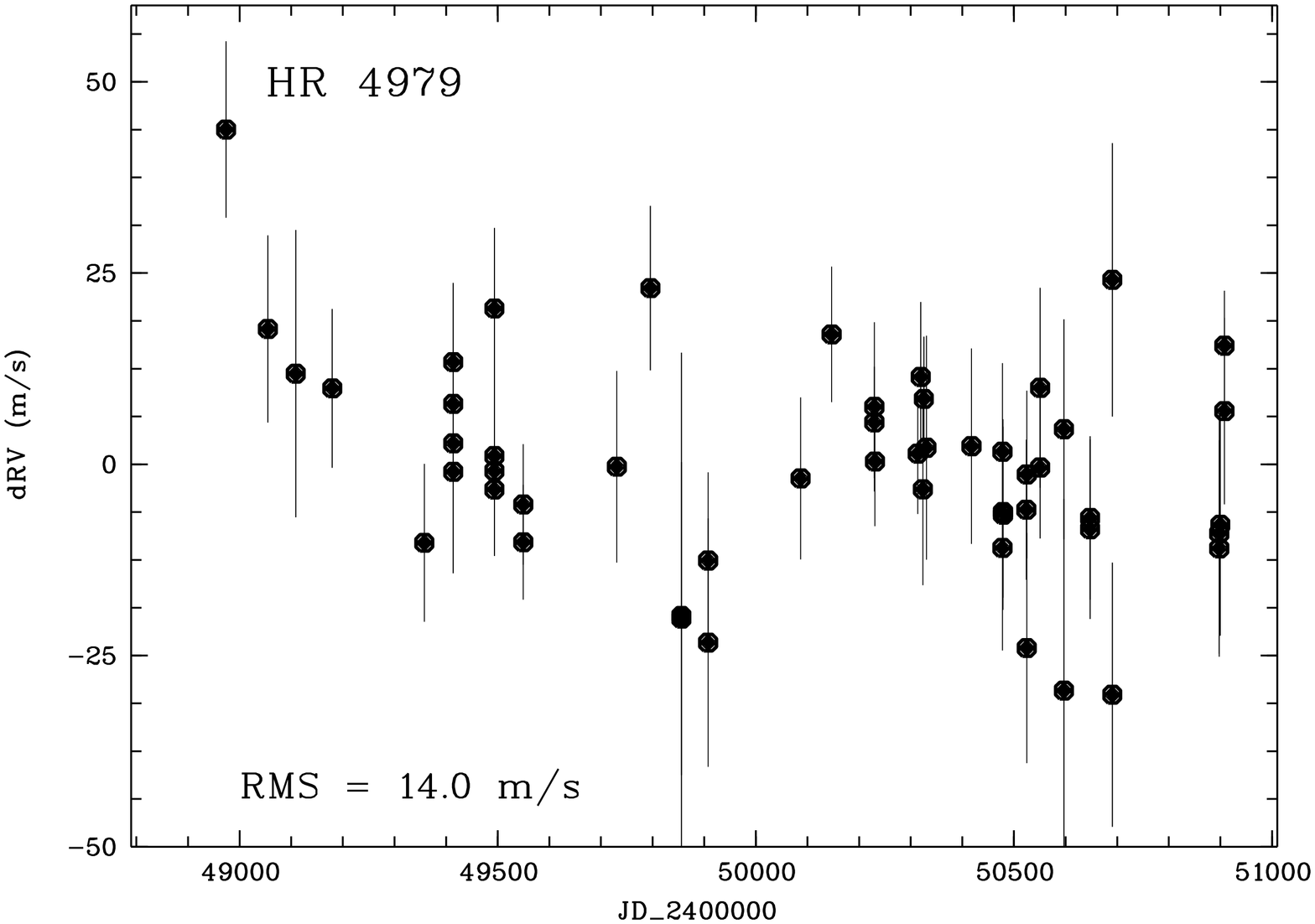,width=9.0cm,height=5.5cm,angle=270}}
	\vbox{\psfig{figure=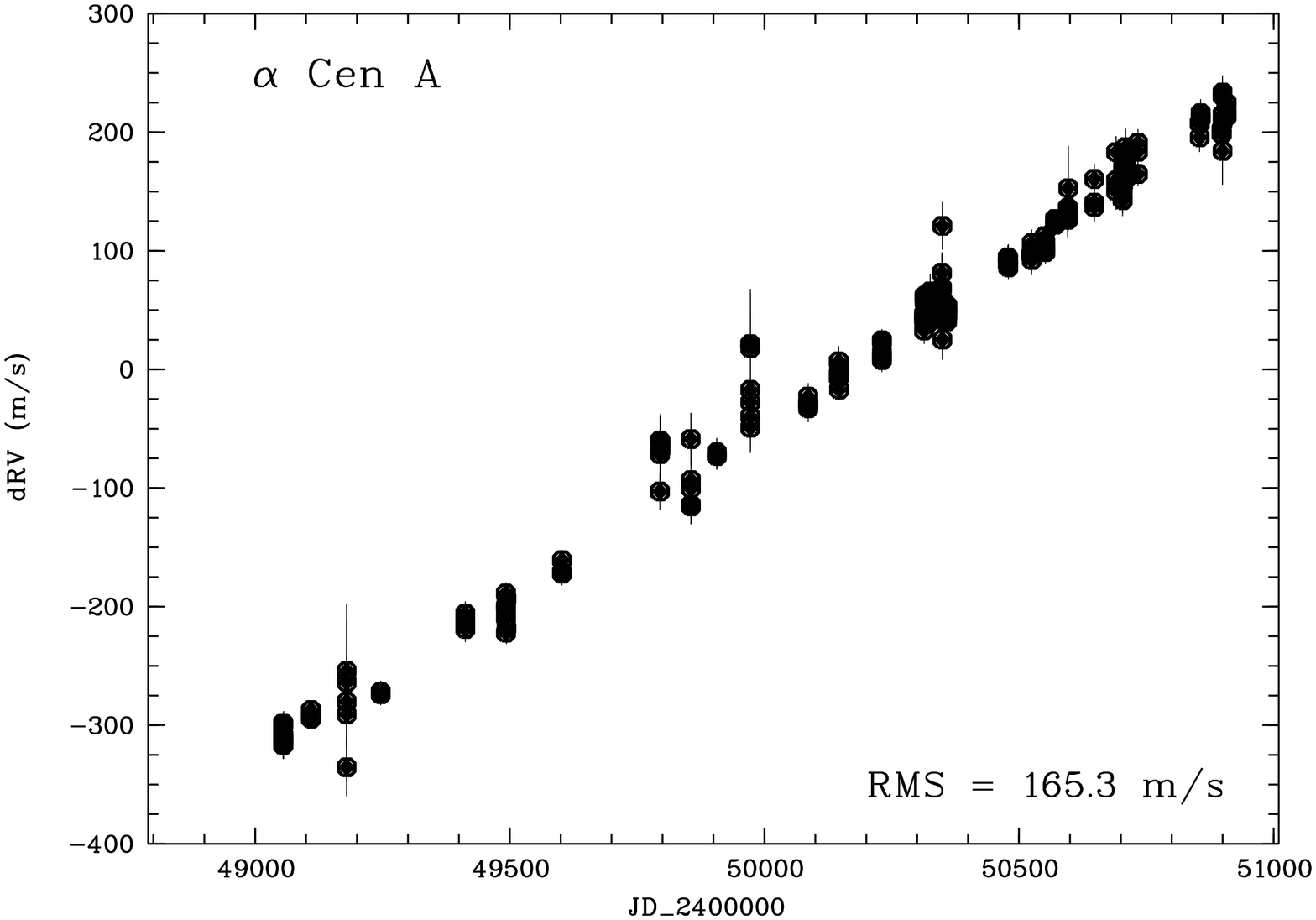,width=9.0cm,height=5.5cm,angle=270}}
	\vbox{\psfig{figure=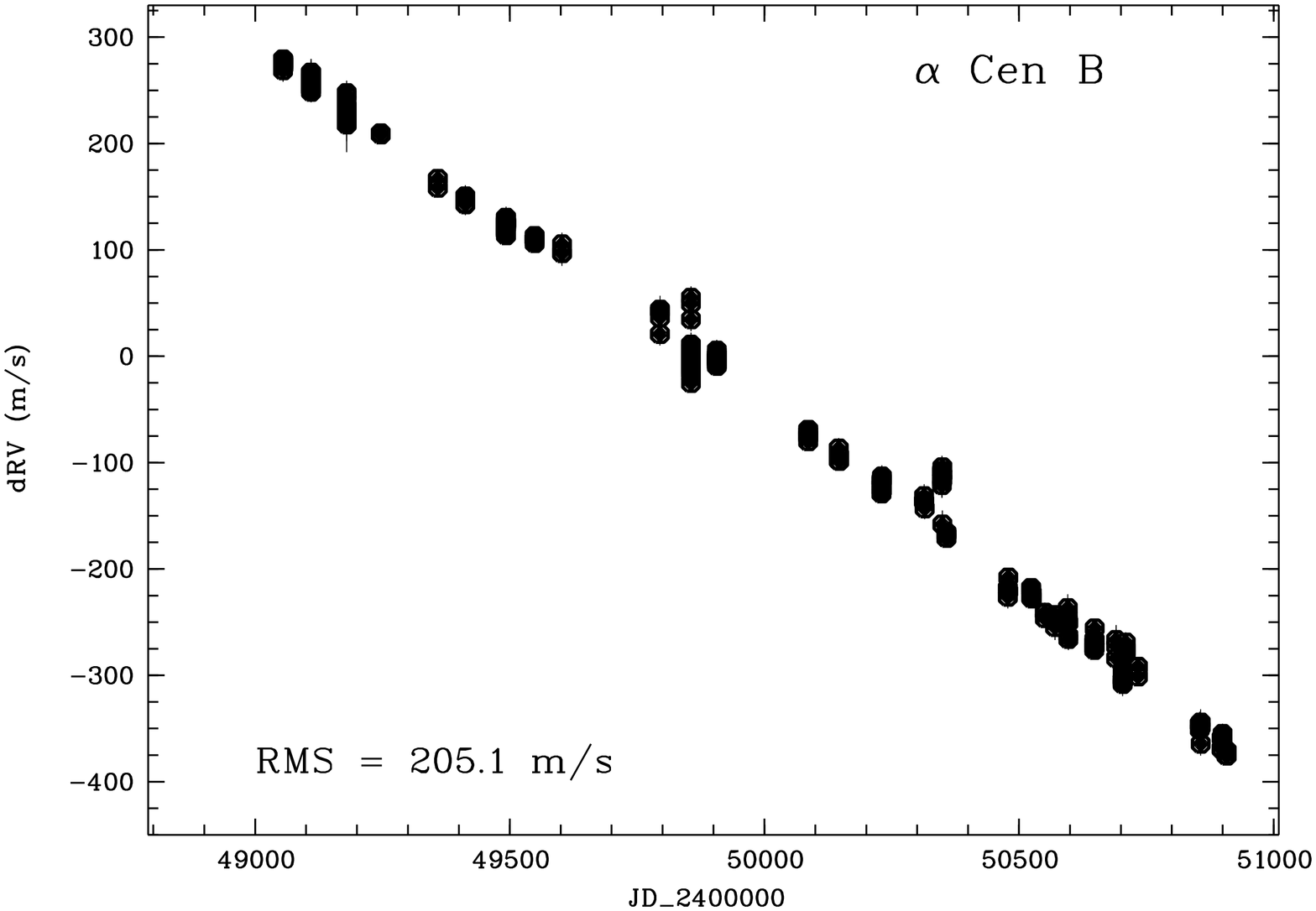,width=9.0cm,height=5.5cm,angle=270}}
	\par
	}
  \caption[]{Radial velocity results for HR~4523, HR~4979, $\alpha$~Cen~A and $\alpha$~Cen~B.}
  \label{rvsfig6}
\end{figure}  
\begin{figure}
 \centering{

	\vbox{\psfig{figure=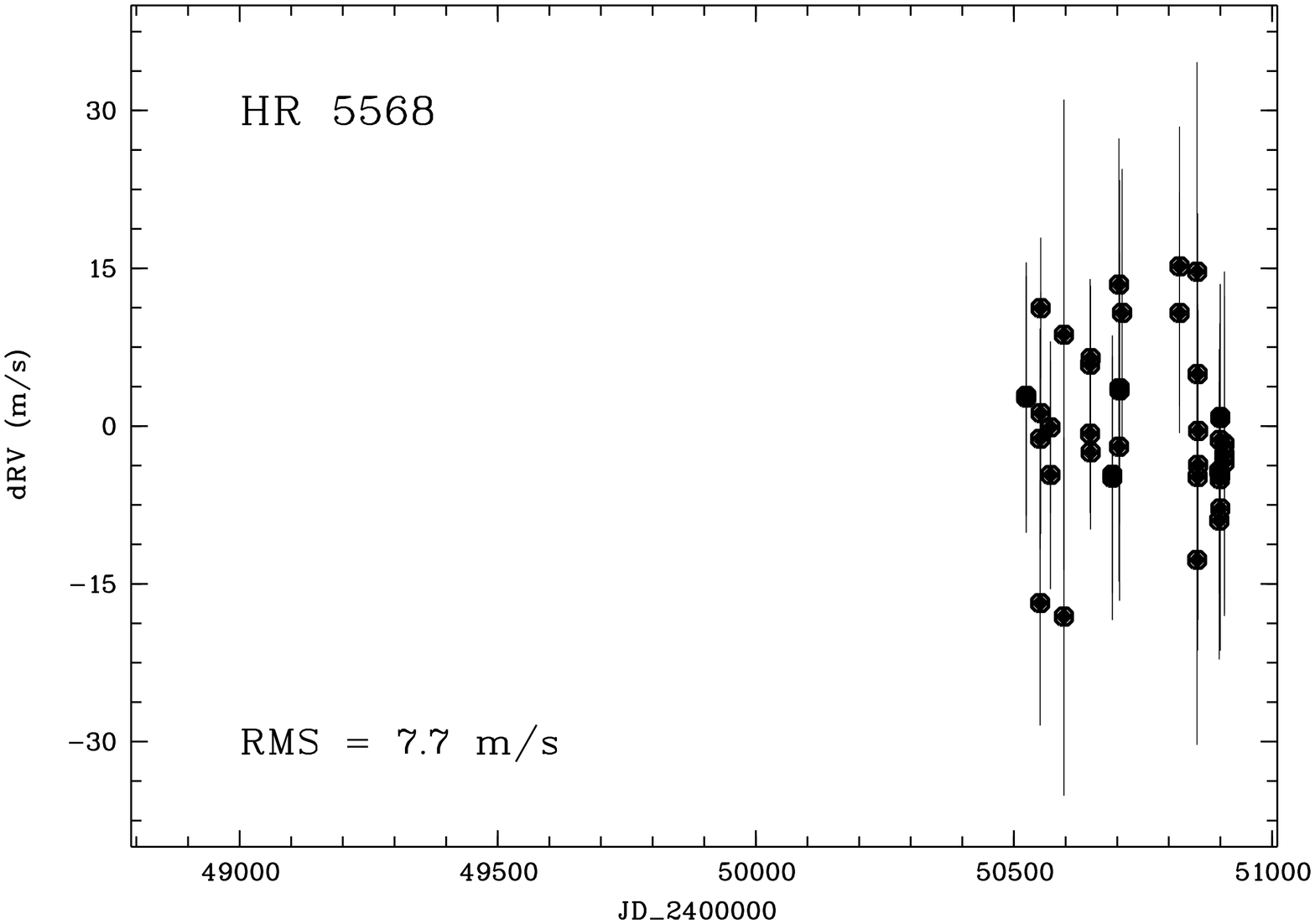,width=9.0cm,height=5.5cm,angle=270}}
	\vbox{\psfig{figure=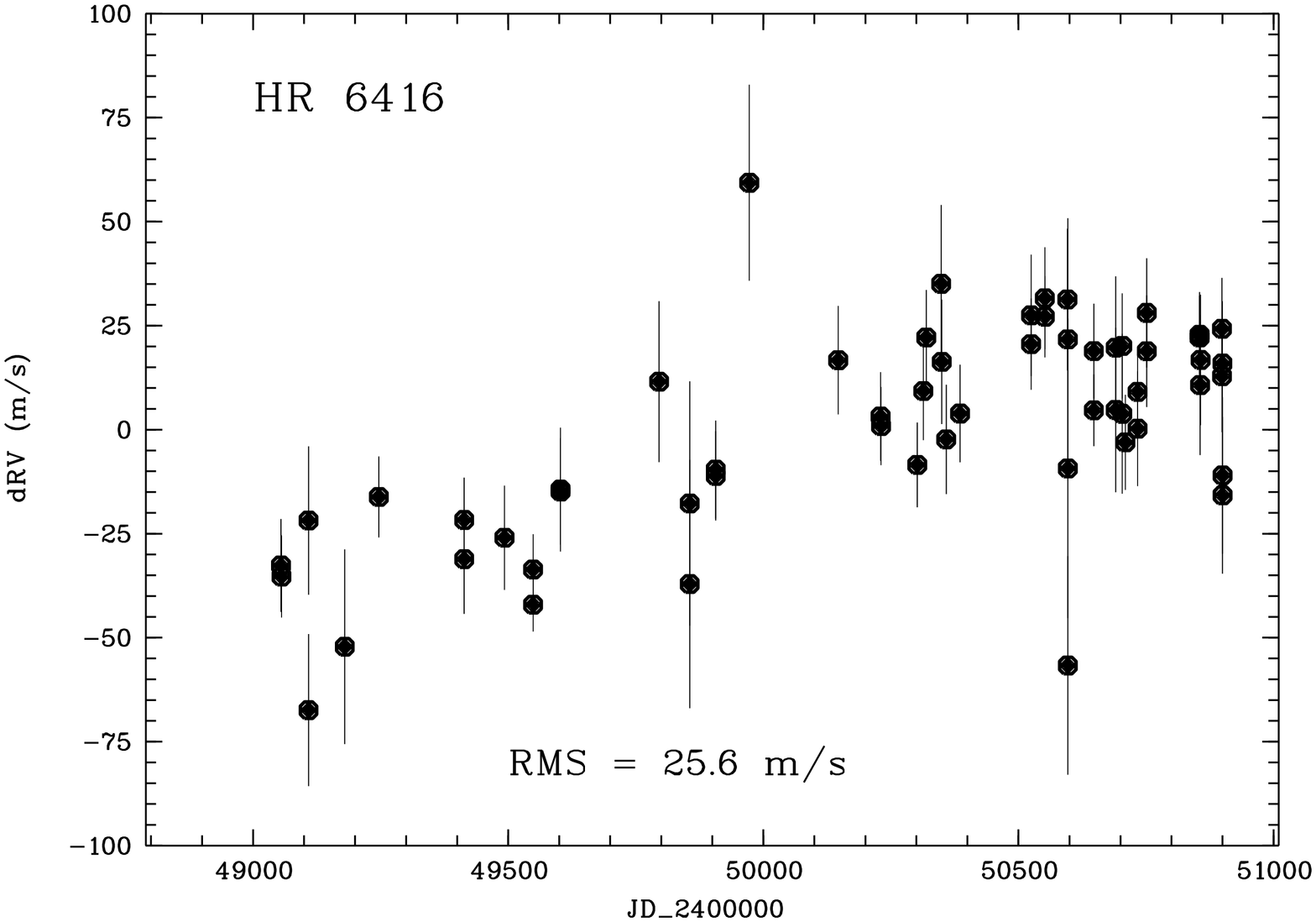,width=9.0cm,height=5.5cm,angle=270}}
	\vbox{\psfig{figure=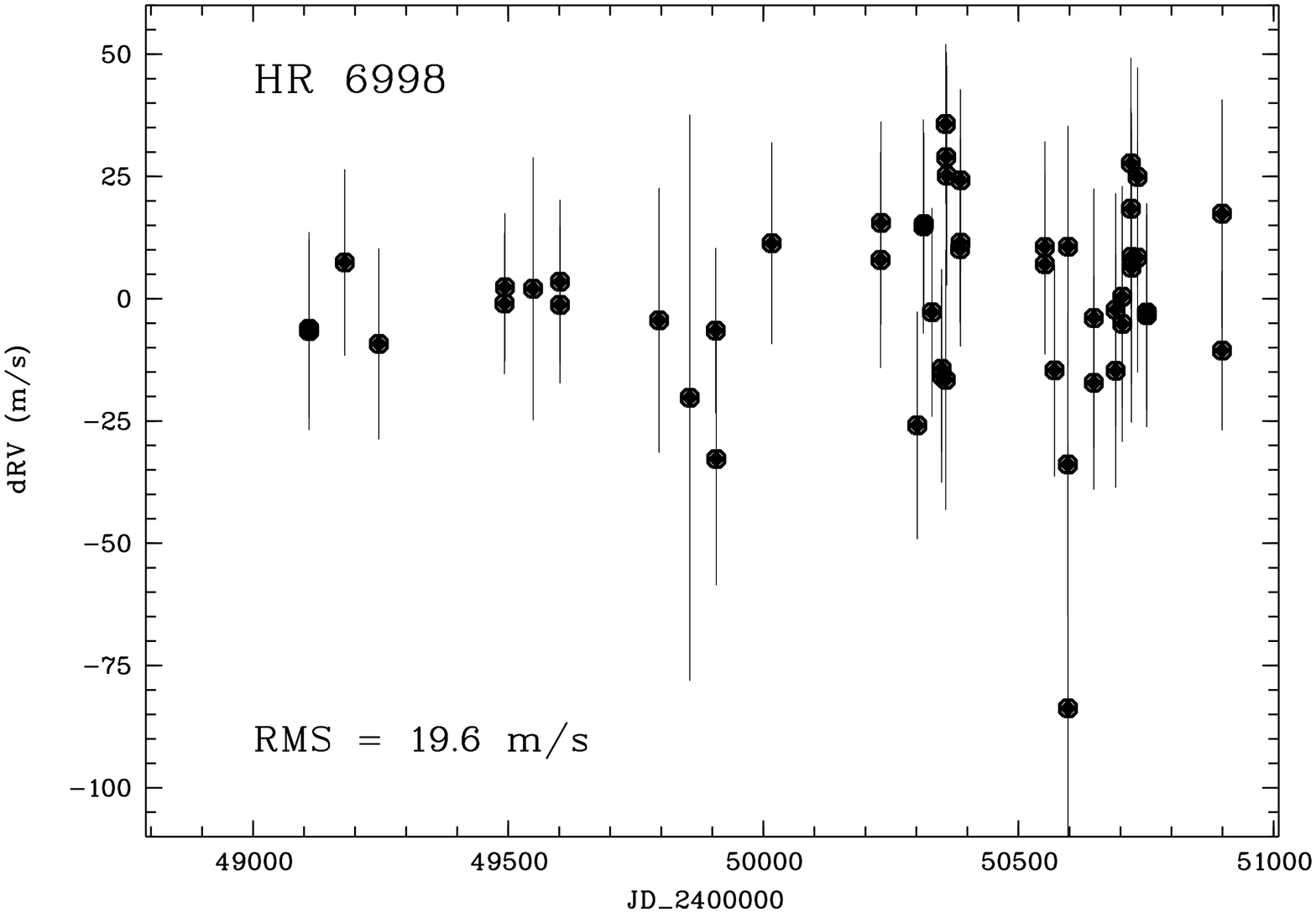,width=9.0cm,height=5.5cm,angle=270}}
	\vbox{\psfig{figure=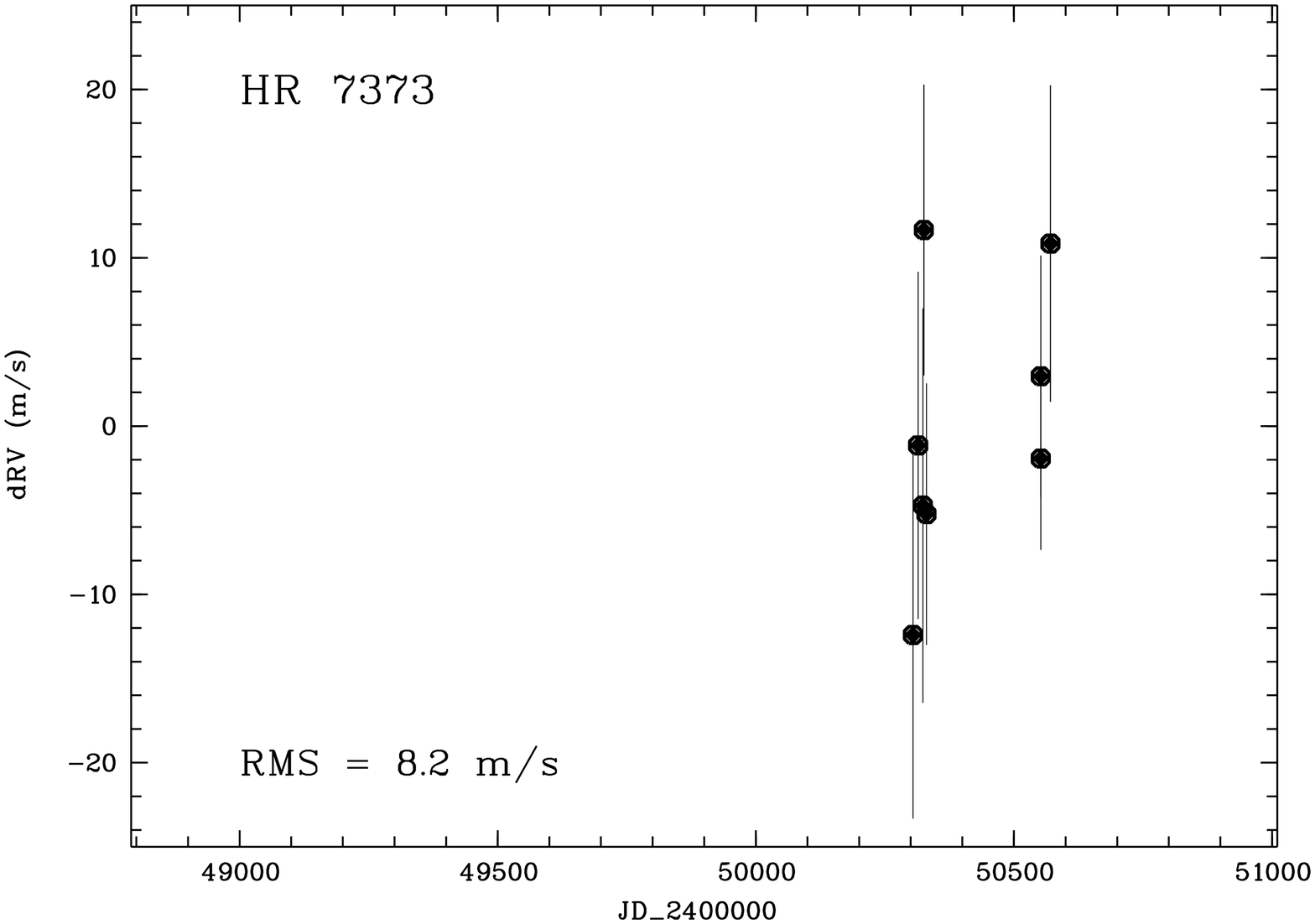,width=9.0cm,height=5.5cm,angle=270}}
   	\par
        }
   \caption[]{Radial velocity results for HR~5568, HR~6416, HR~6998 and HR~7373.}
  \label{rvsfig7}
\end{figure}  
\begin{figure}
 \centering{
        \vbox{\psfig{figure=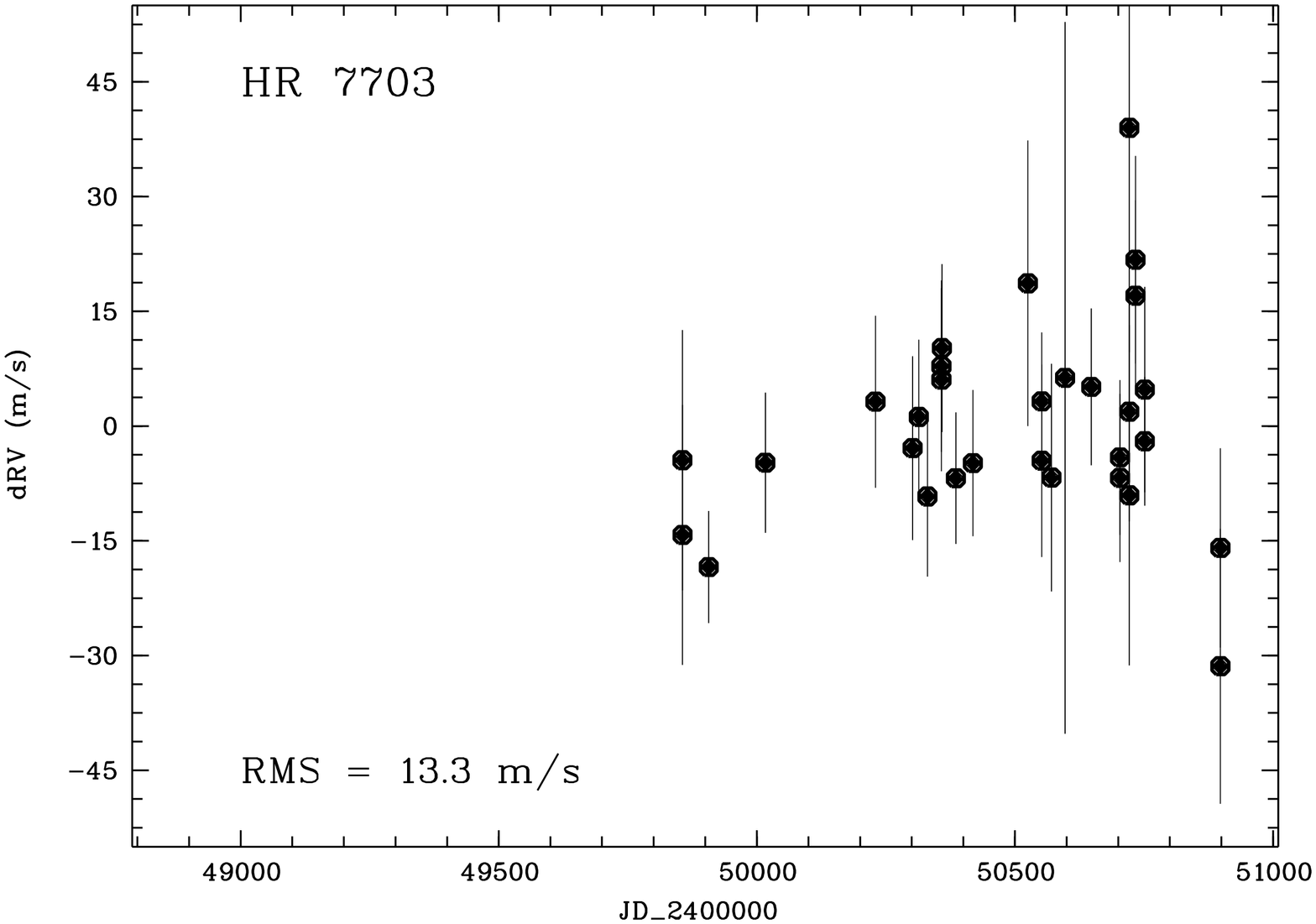,width=9.0cm,height=5.5cm,angle=270}}
        \vbox{\psfig{figure=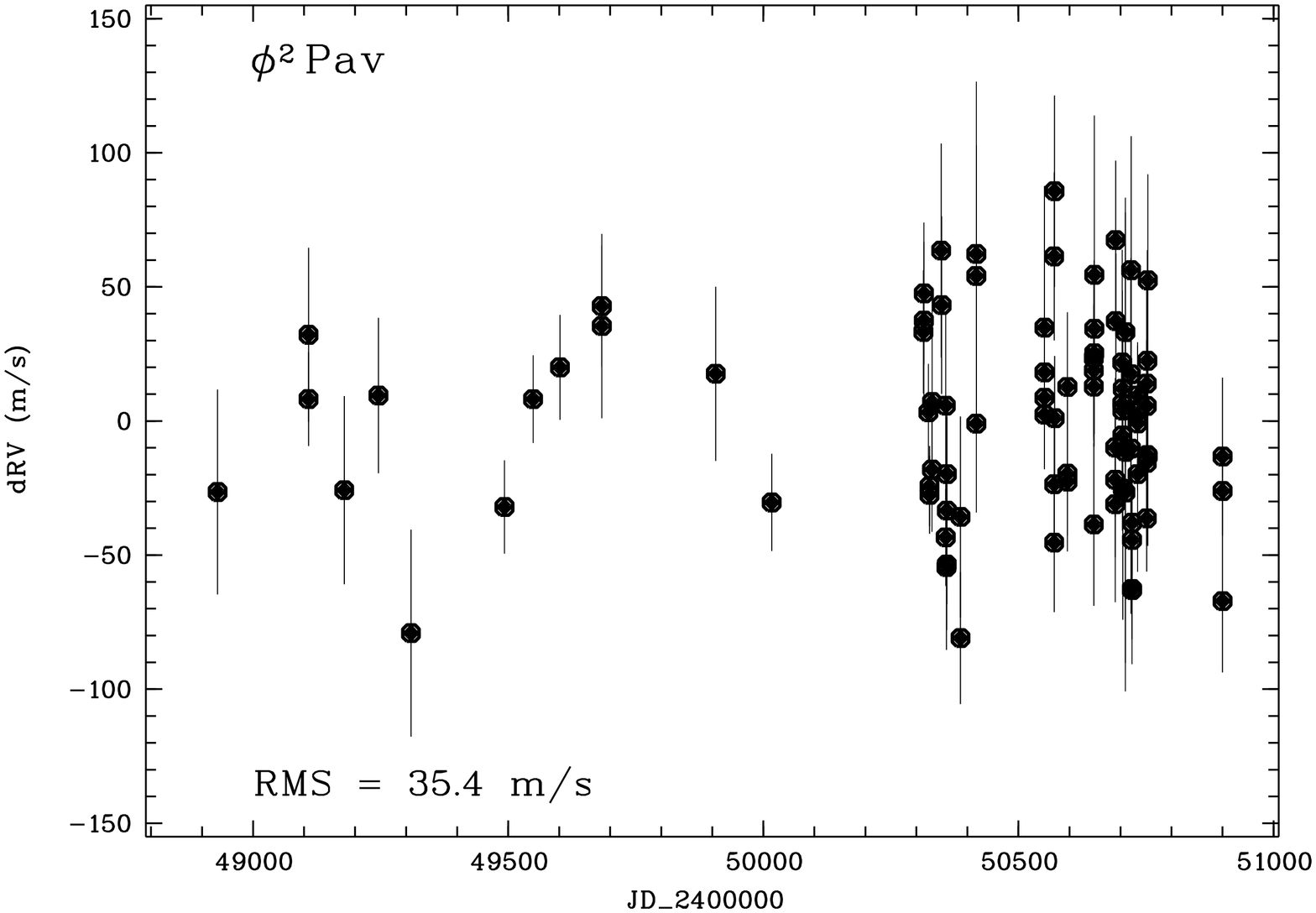,width=9.0cm,height=5.5cm,angle=270}}
	\vbox{\psfig{figure=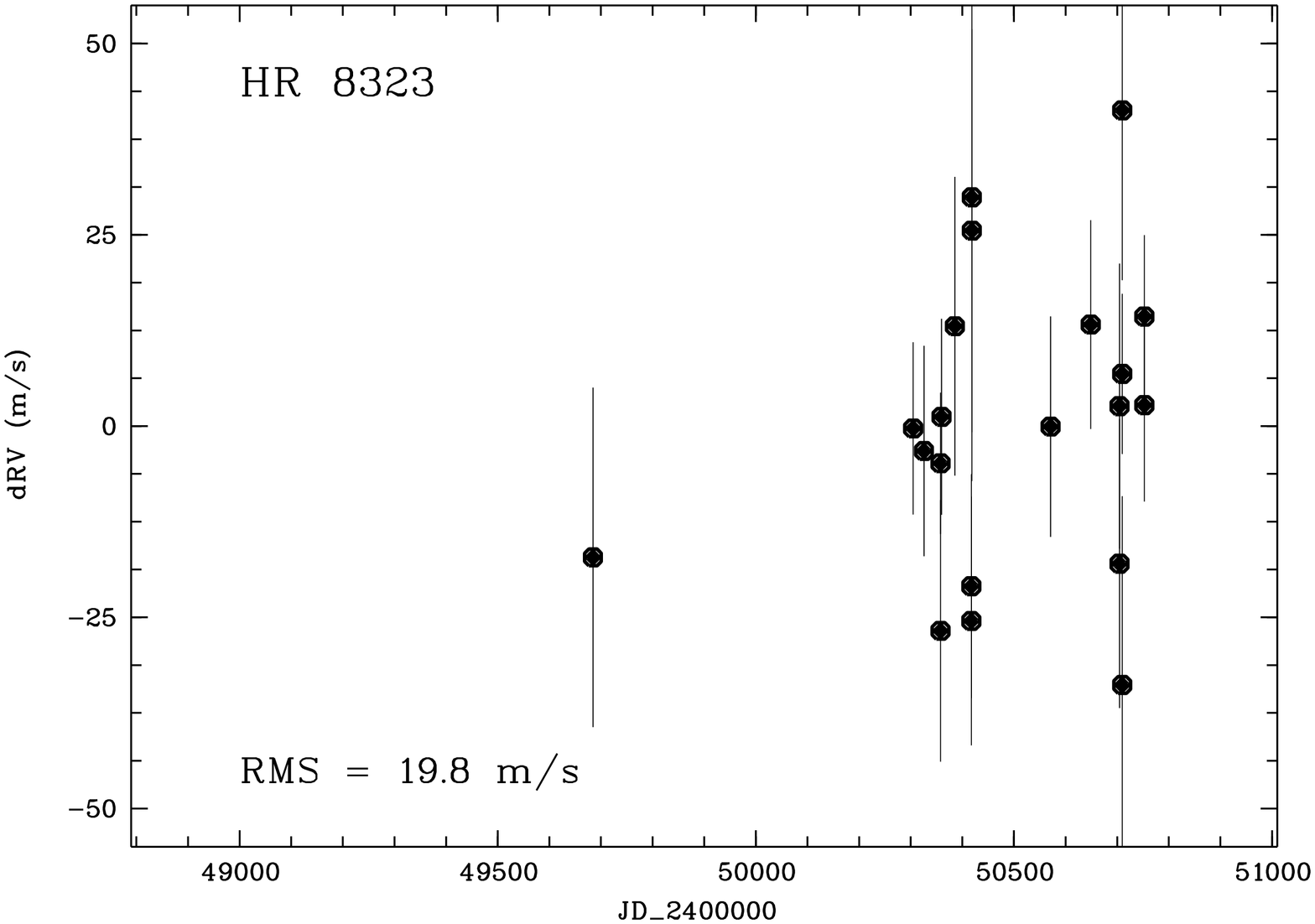,width=9.0cm,height=5.5cm,angle=270}}
        \vbox{\psfig{figure=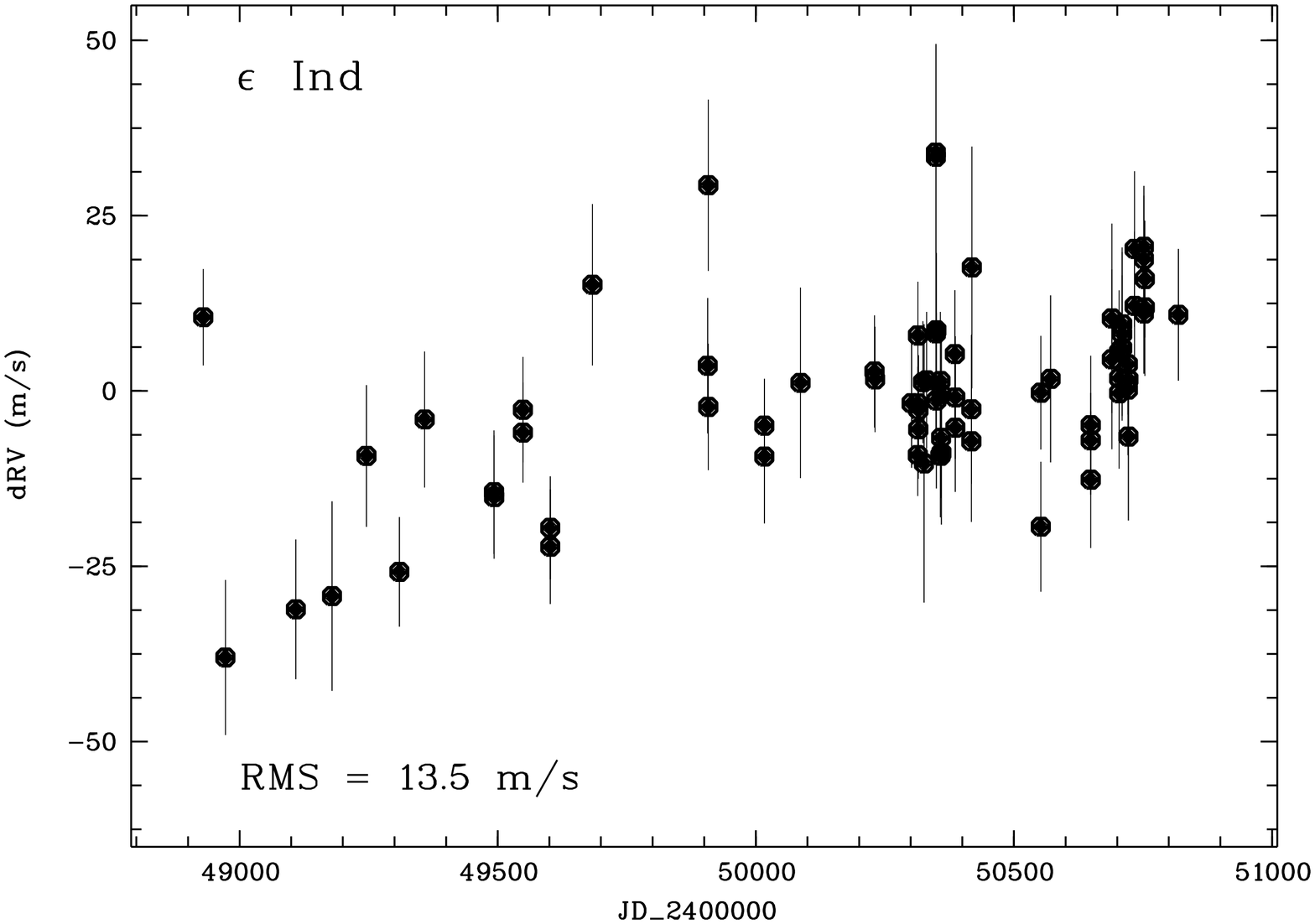,width=9.0cm,height=5.5cm,angle=270}}
   \par
        }
   \caption[]{   
        Radial velocity results for HR~7703, $\phi^{2}$~Pav, HR~8323 and $\epsilon$~Ind.}
  \label{rvsfig8}
\end{figure}  
\begin{figure}
 \centering{
        \vbox{\psfig{figure=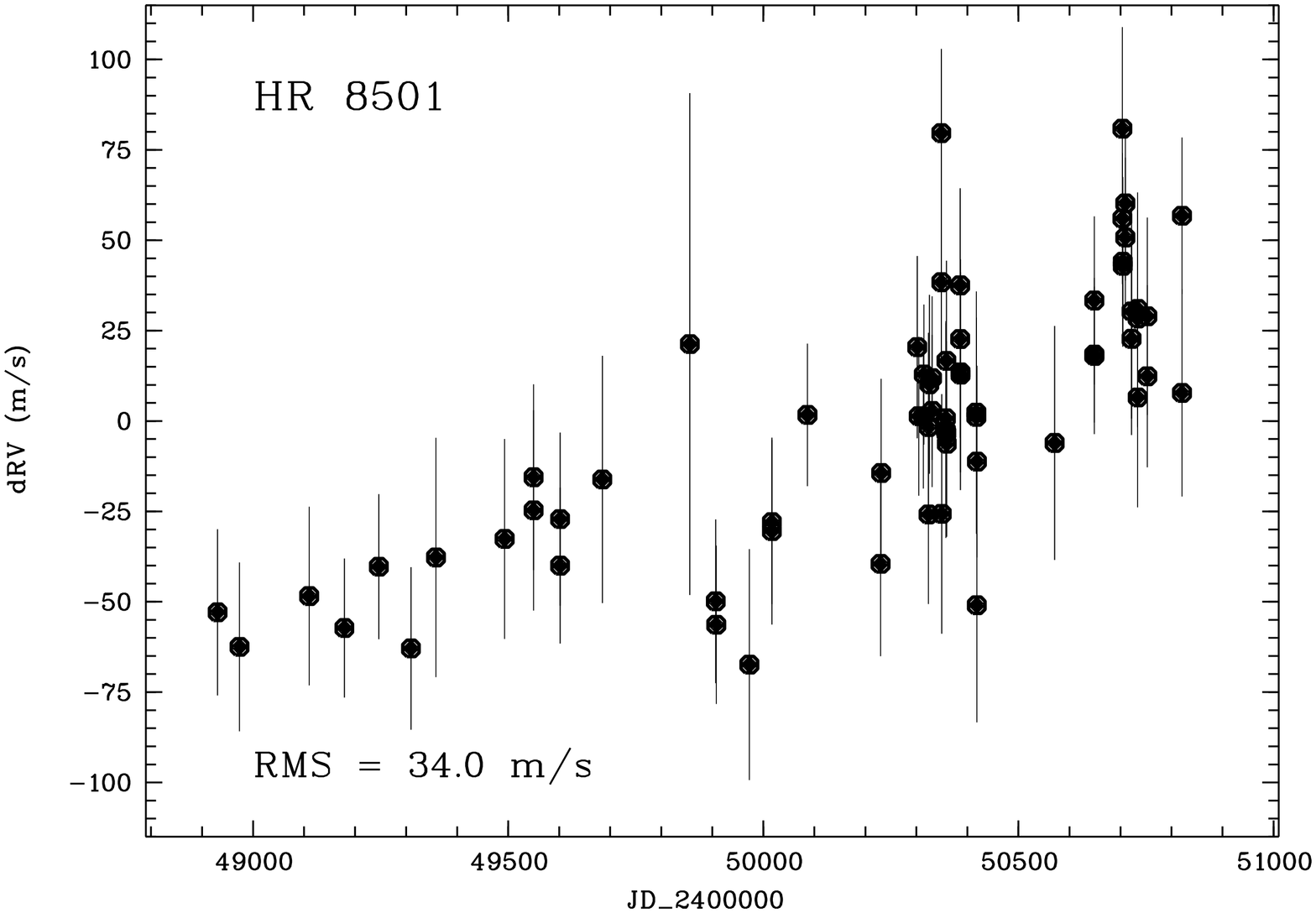,width=9.0cm,height=5.5cm,angle=270}}
        \vbox{\psfig{figure=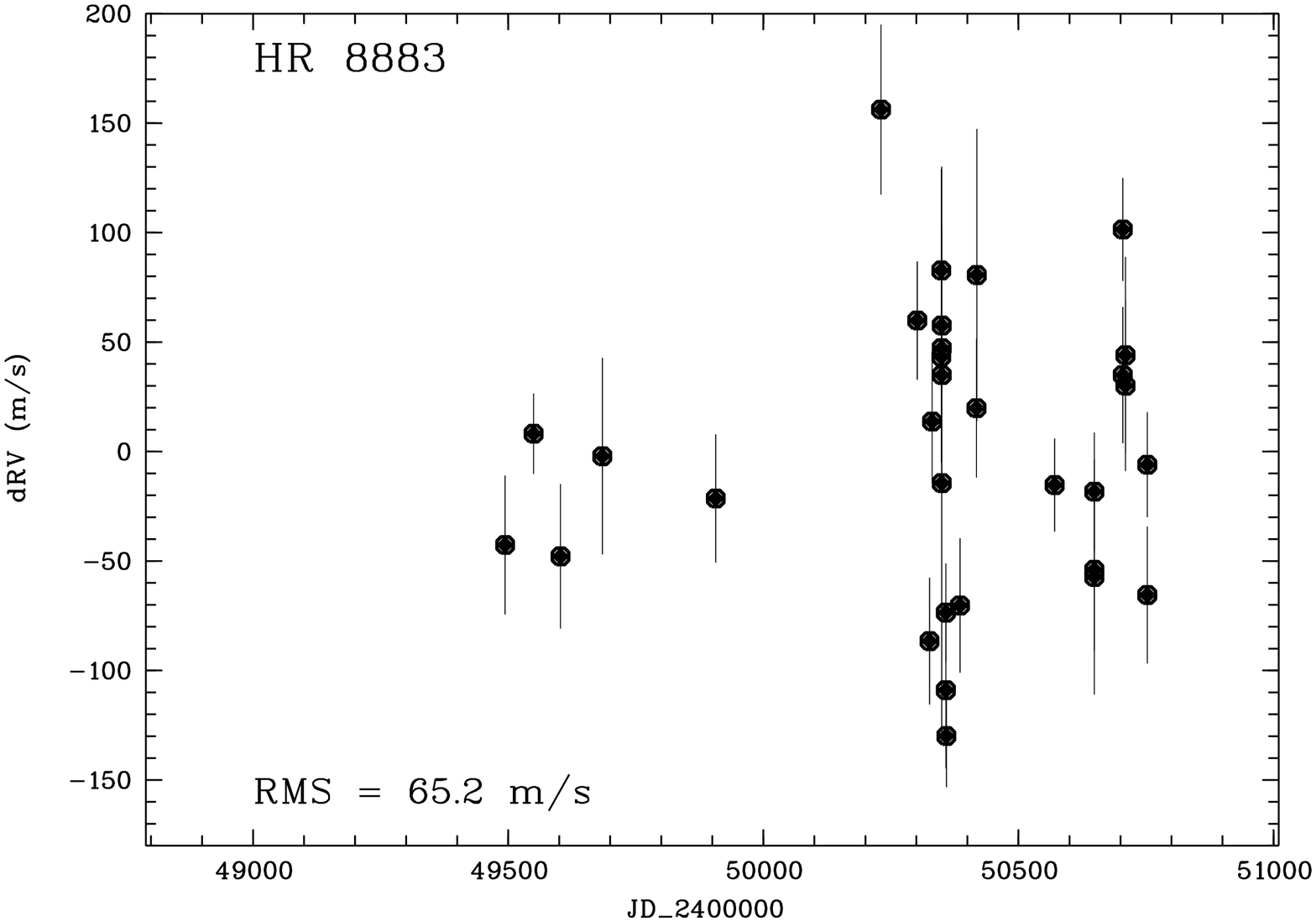,width=9.0cm,height=5.5cm,angle=270}}
	\vbox{\psfig{figure=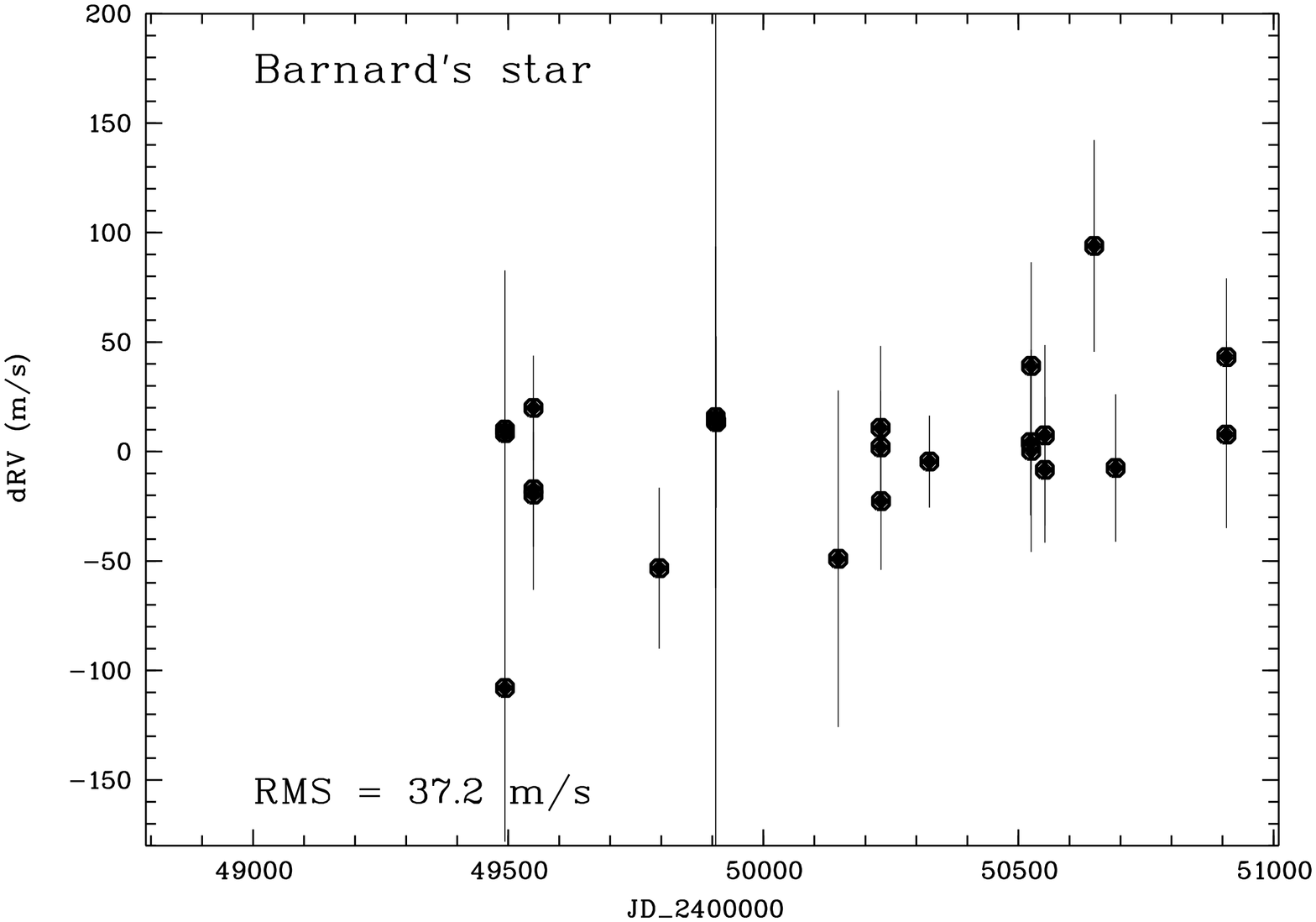,width=9.0cm,height=5.5cm,angle=270}}
        \vbox{\psfig{figure=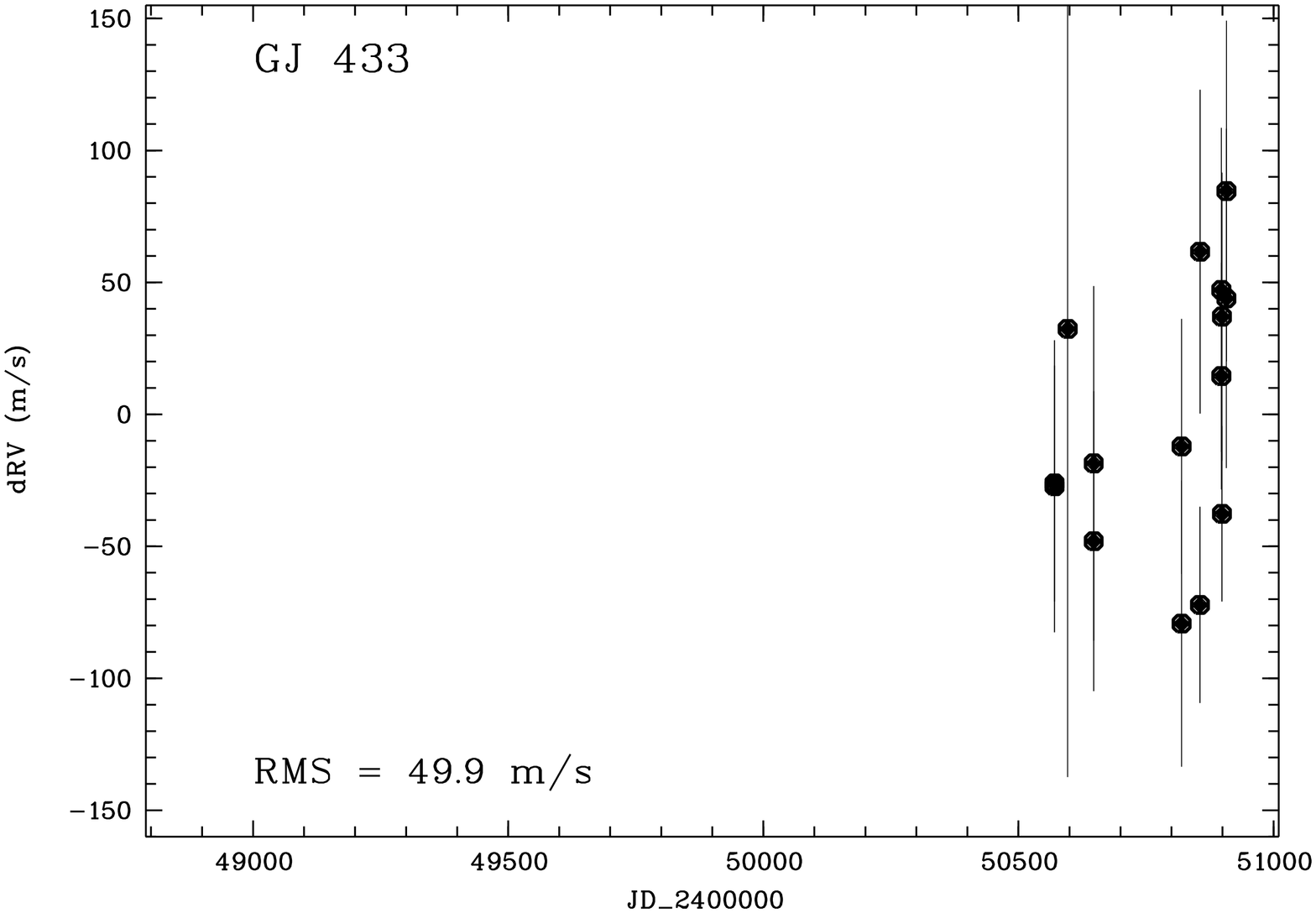,width=9.0cm,height=5.5cm,angle=270}}
   \par
	}
	\caption[]{   
        Radial velocity results for HR~8501, HR~8883, Barnard's star and GJ 433.}
	\label{rvsfig9}
\end{figure} 
\begin{figure}
 \centering{
        \vbox{\psfig{figure=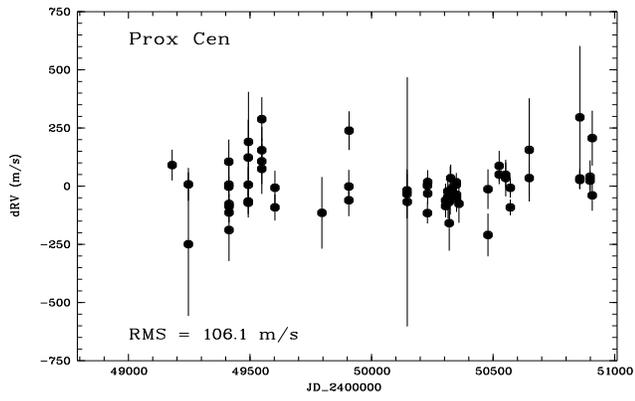,width=9.0cm,height=5.5cm,angle=270}}
       \par
        }
        \caption[]{Radial velocity results for Proxima Centauri.}
 \label{rvsfig10}
\end{figure}

\clearpage
\newpage

\section{Limits for planetary companions}

\begin{figure}
\centering{
        \vbox{\psfig{figure=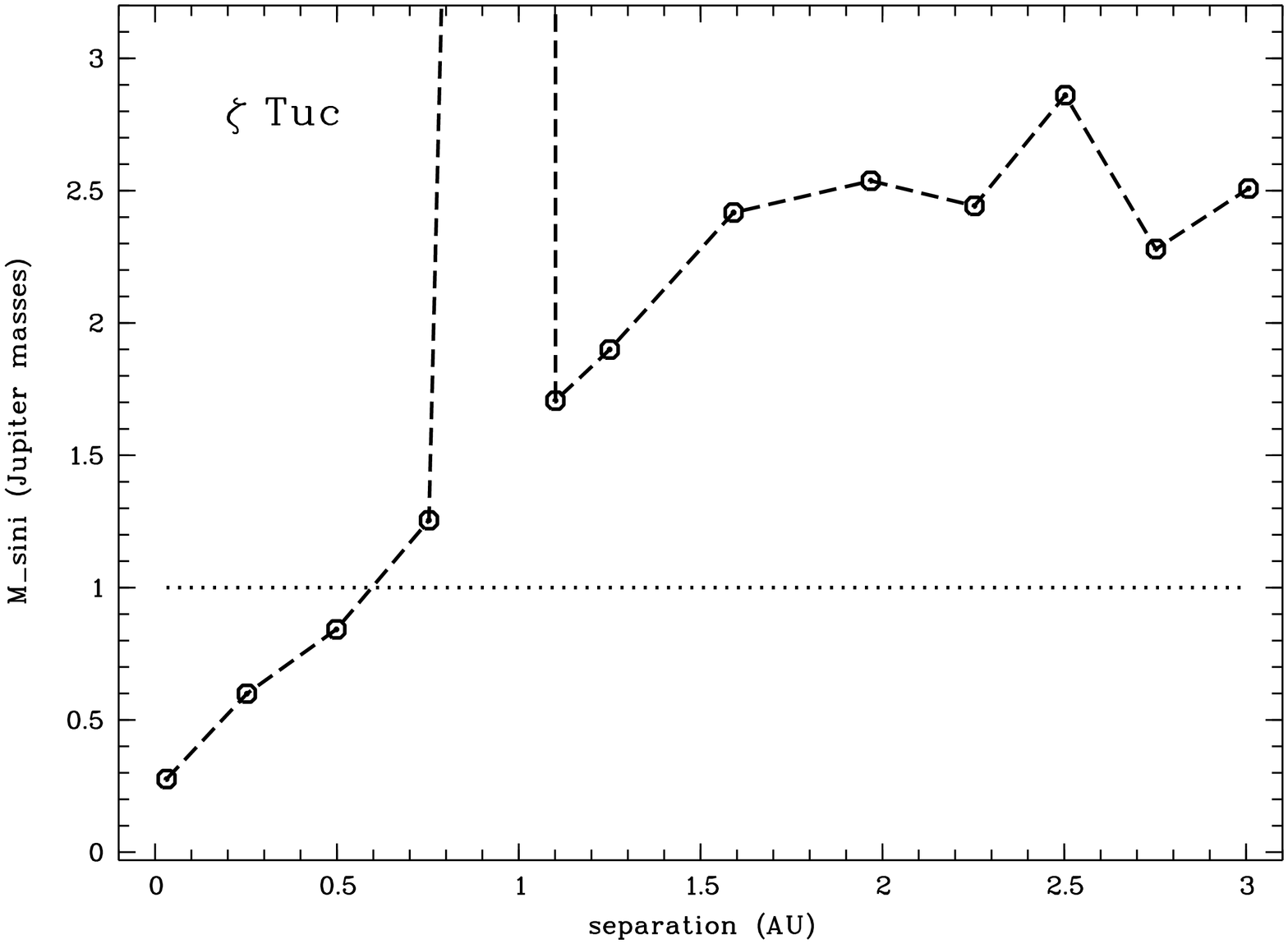,width=9.0cm,height=5.5cm,angle=270}}
        \vbox{\psfig{figure=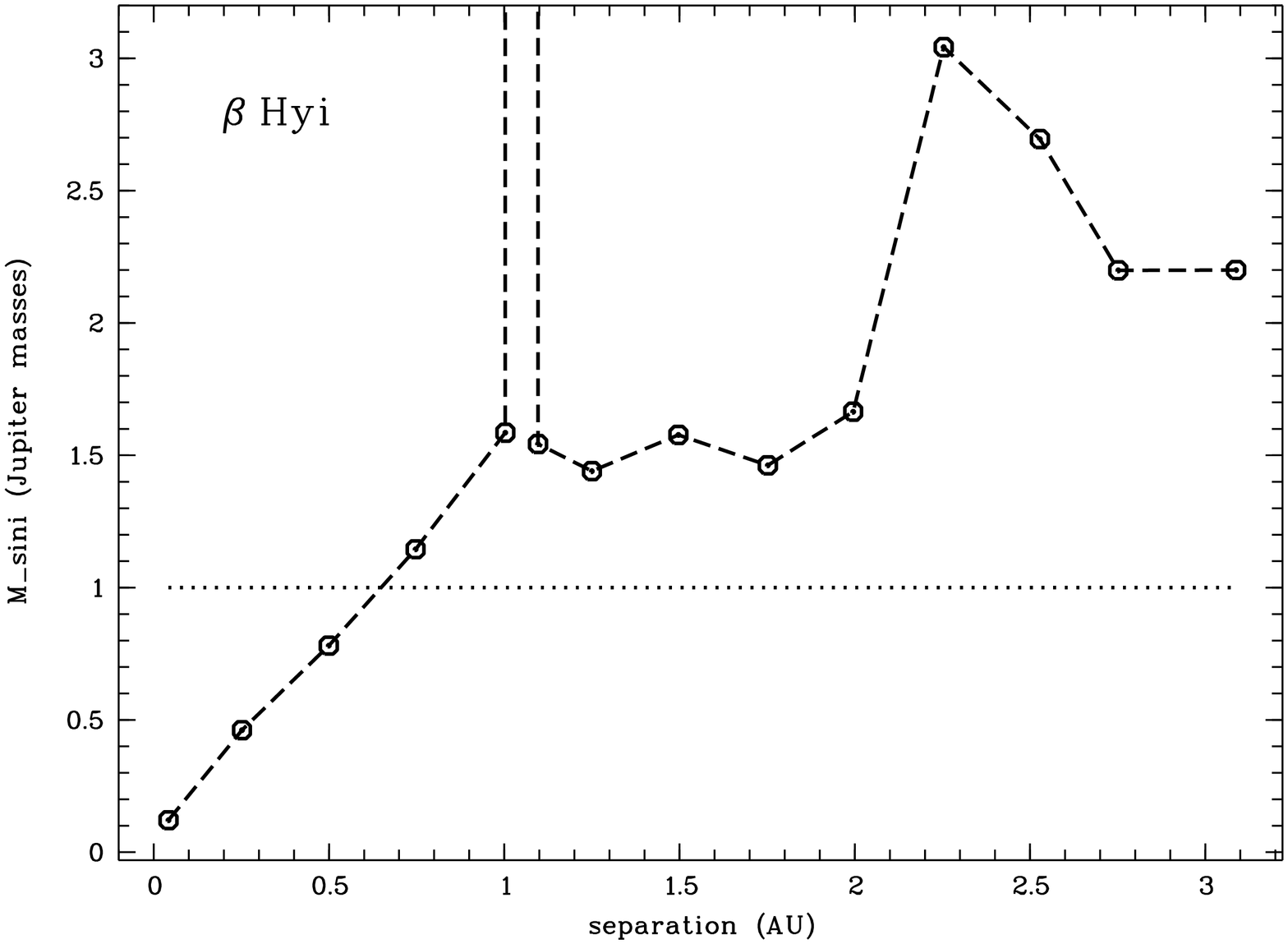,width=9.0cm,height=5.5cm,angle=270}}
        \vbox{\psfig{figure=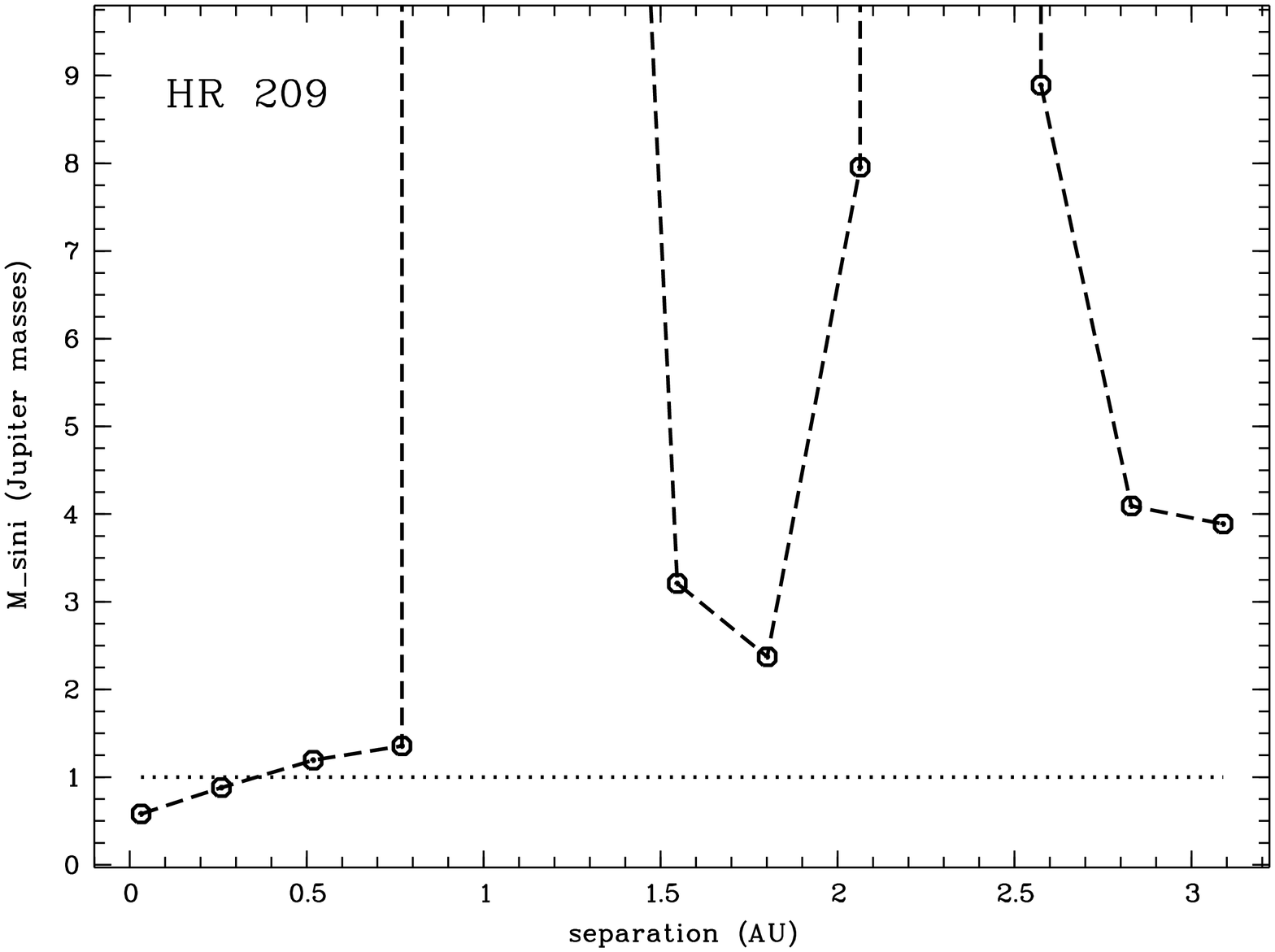,width=9.0cm,height=5.5cm,angle=270}}
        \vbox{\psfig{figure=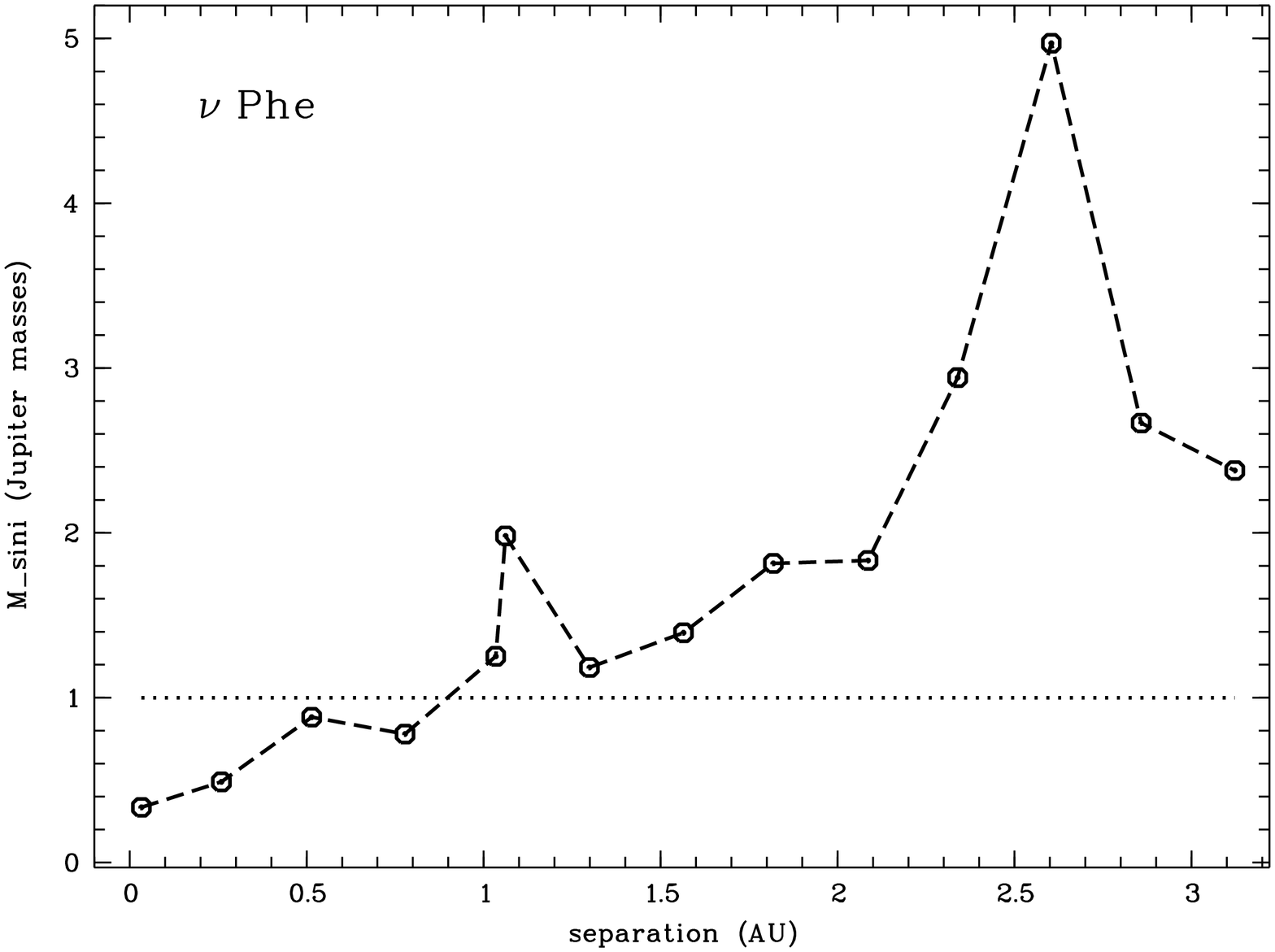,width=9.0cm,height=5.5cm,angle=270}}
   \par 
        }
  \caption[]{
	Planetary companion limits for $\zeta$ Tuc, $\beta$~Hyi, HR 209 and $\nu$ Phe.
	}
 \label{limfig1}
\end{figure}

\begin{figure}
 \centering{
        \vbox{\psfig{figure=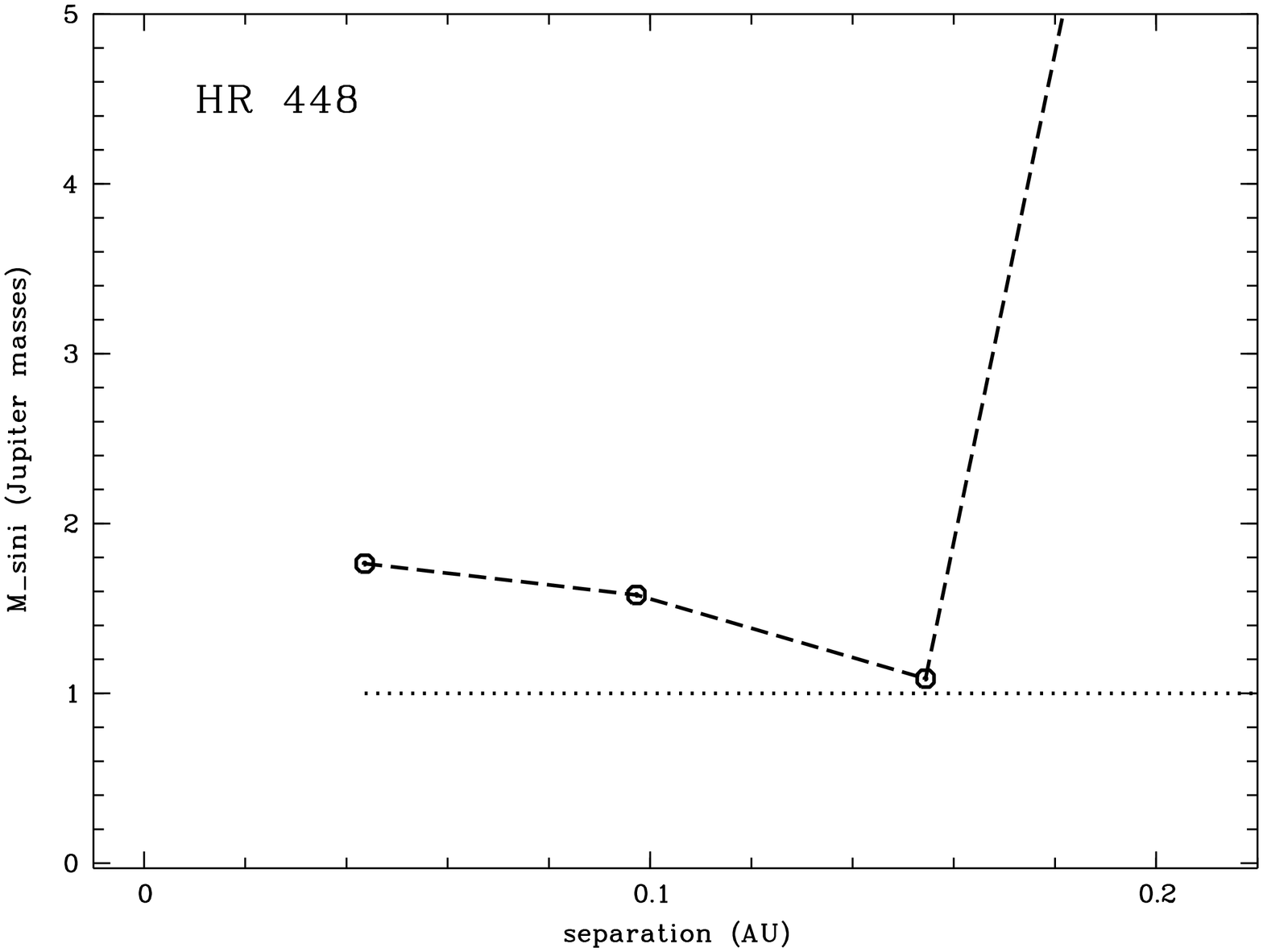,width=9.0cm,height=5.5cm,angle=270}}
        \vbox{\psfig{figure=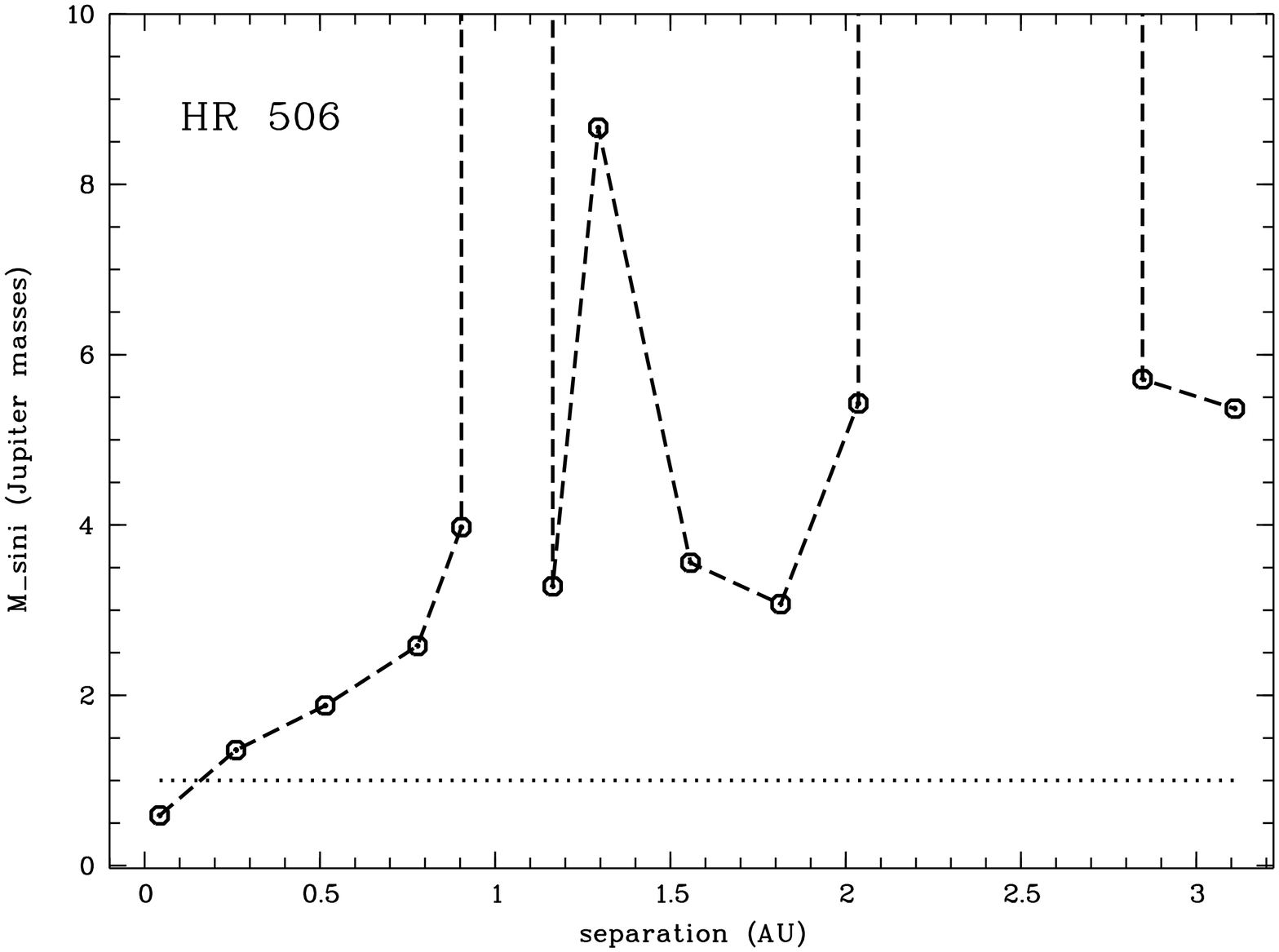,width=9.0cm,height=5.5cm,angle=270}}
        \vbox{\psfig{figure=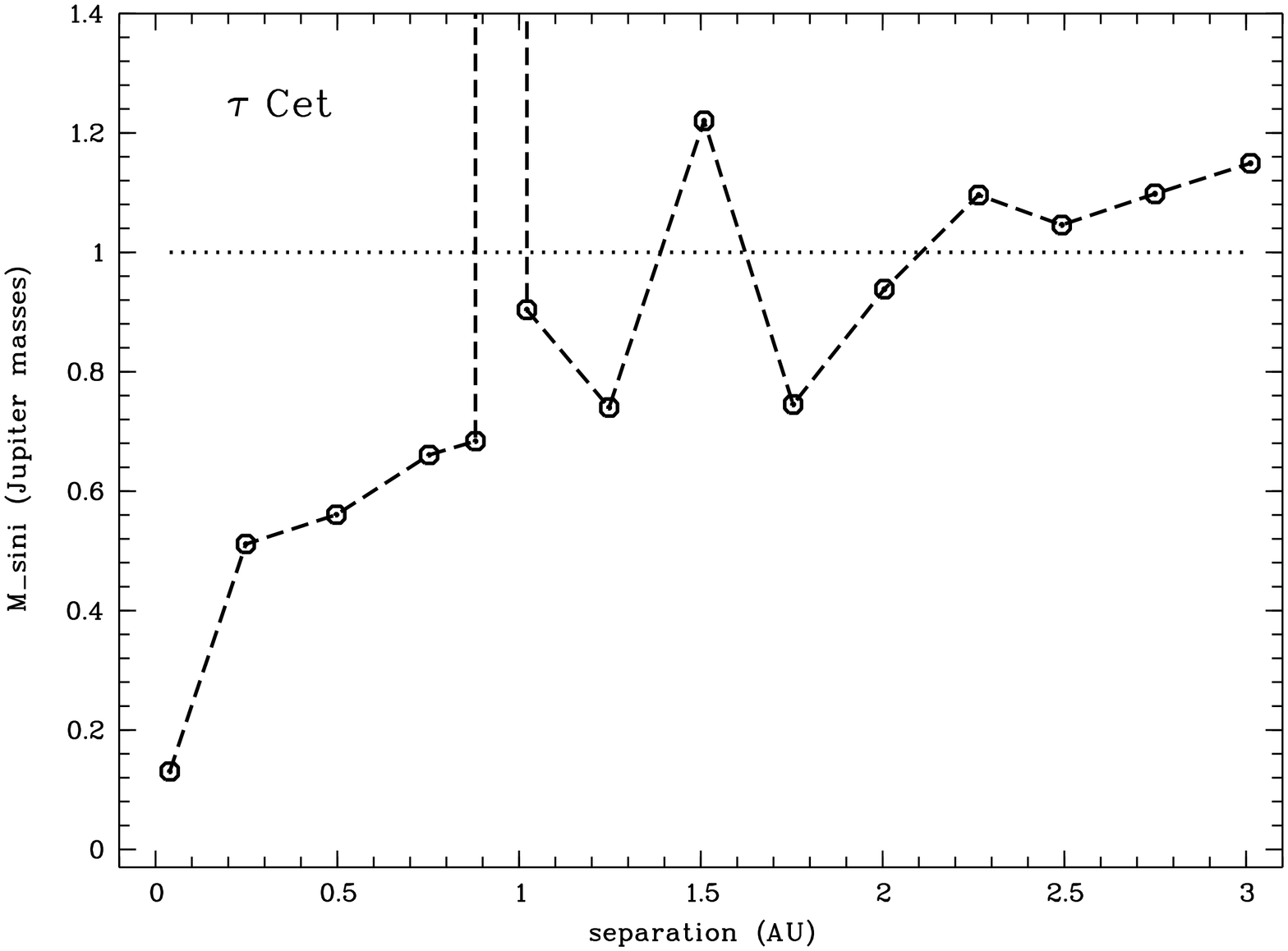,width=9.0cm,height=5.5cm,angle=270}}
        \vbox{\psfig{figure=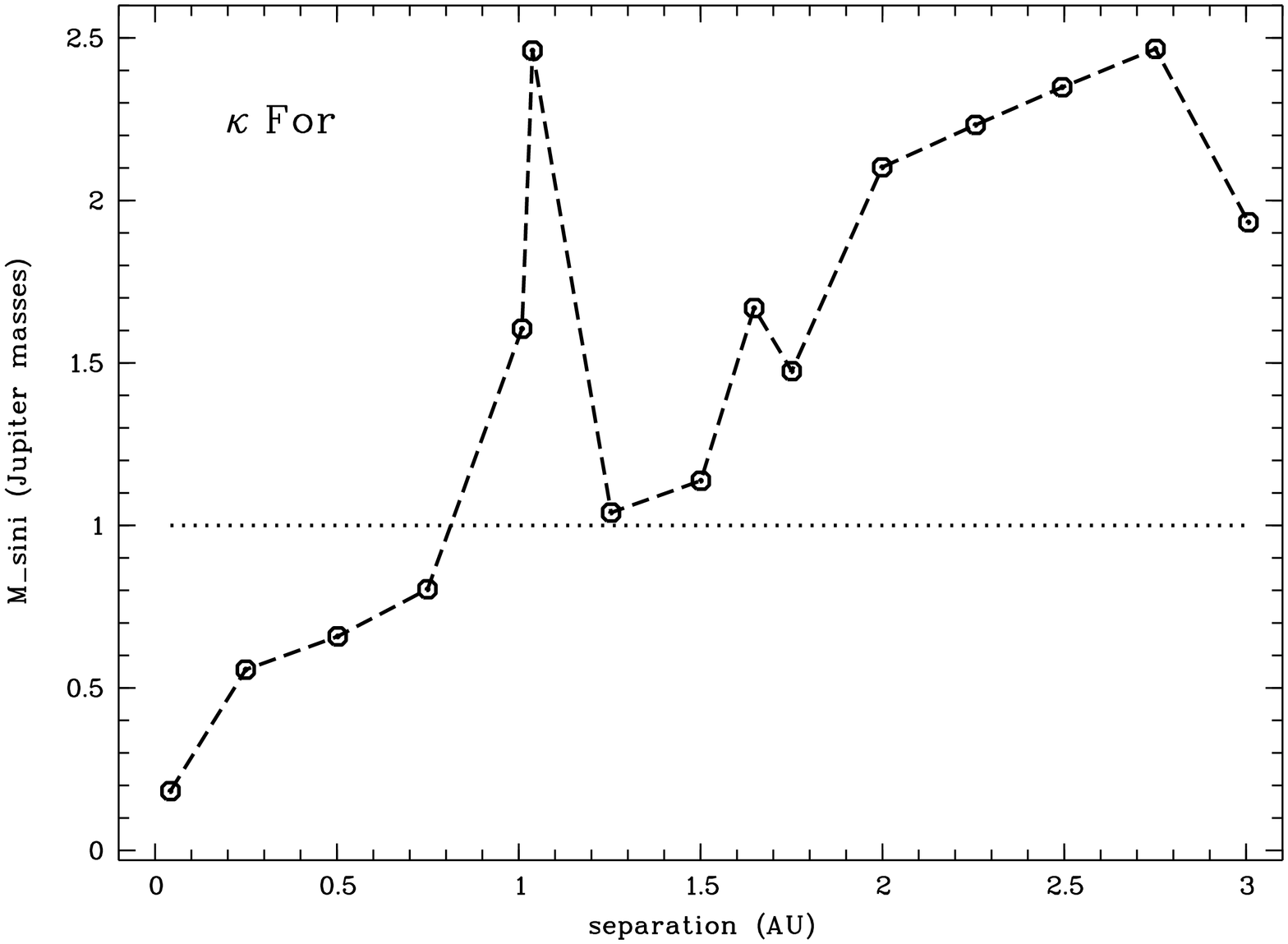,width=9.0cm,height=5.5cm,angle=270}}
   \par
        }
   \caption[]{
	Planetary companion limits for HR 448, HR 506, $\tau$ Cet and $\kappa$ For.
	}
  \label{limfig2}
\end{figure}

\begin{figure}
 \centering{
        \vbox{\psfig{figure=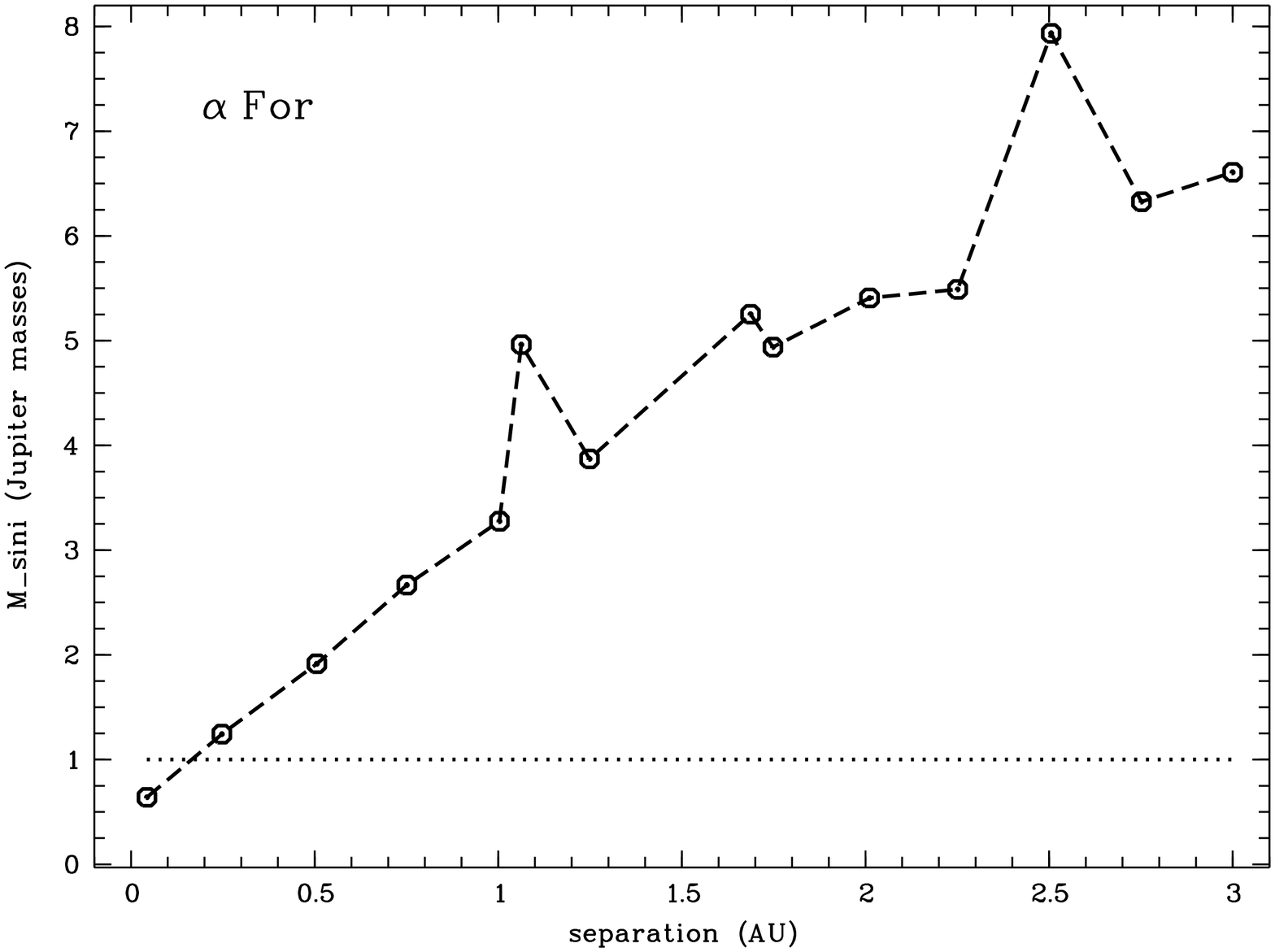,width=9.0cm,height=5.5cm,angle=270}}
        \vbox{\psfig{figure=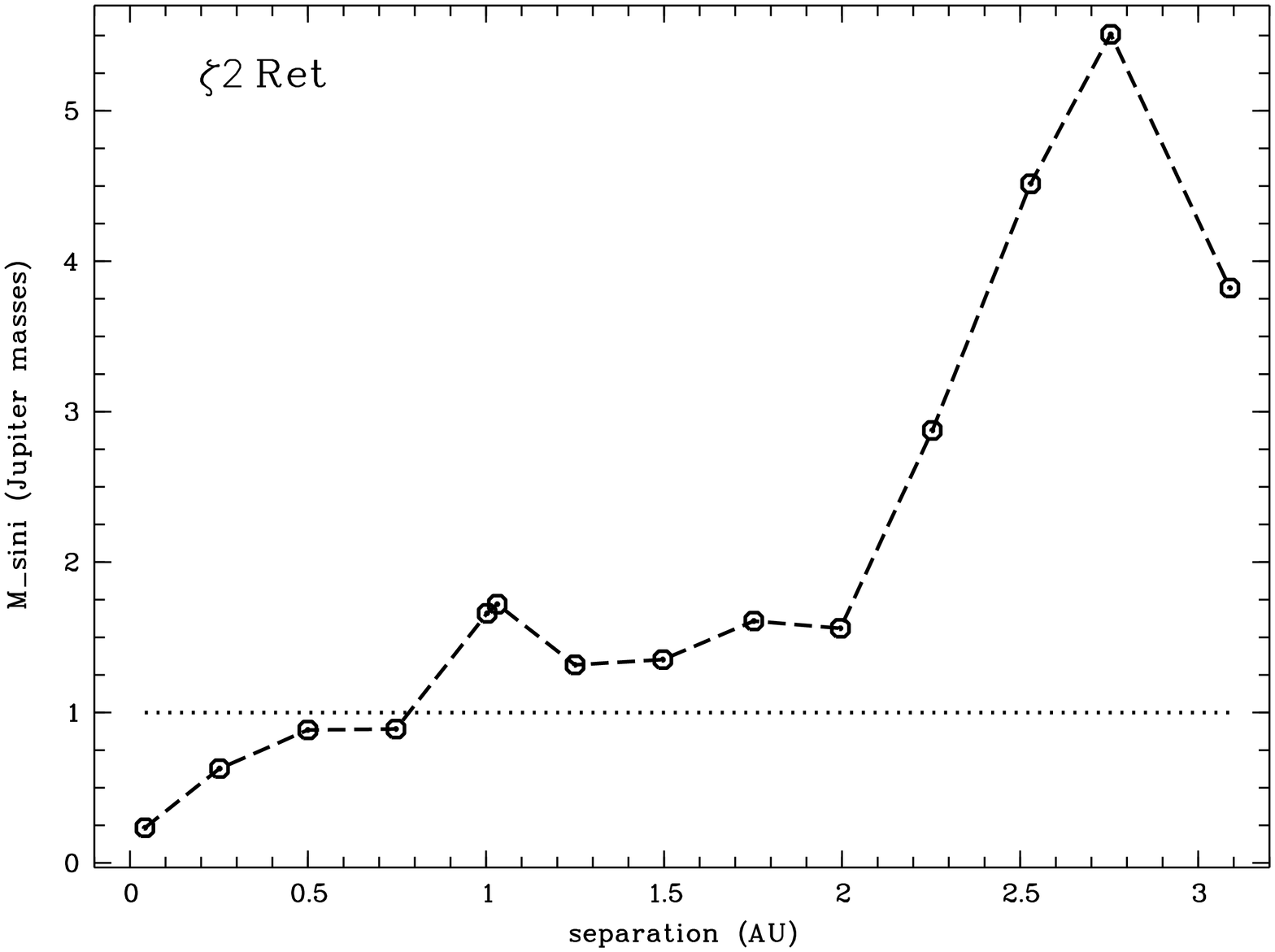,width=9.0cm,height=5.5cm,angle=270}}
        \vbox{\psfig{figure=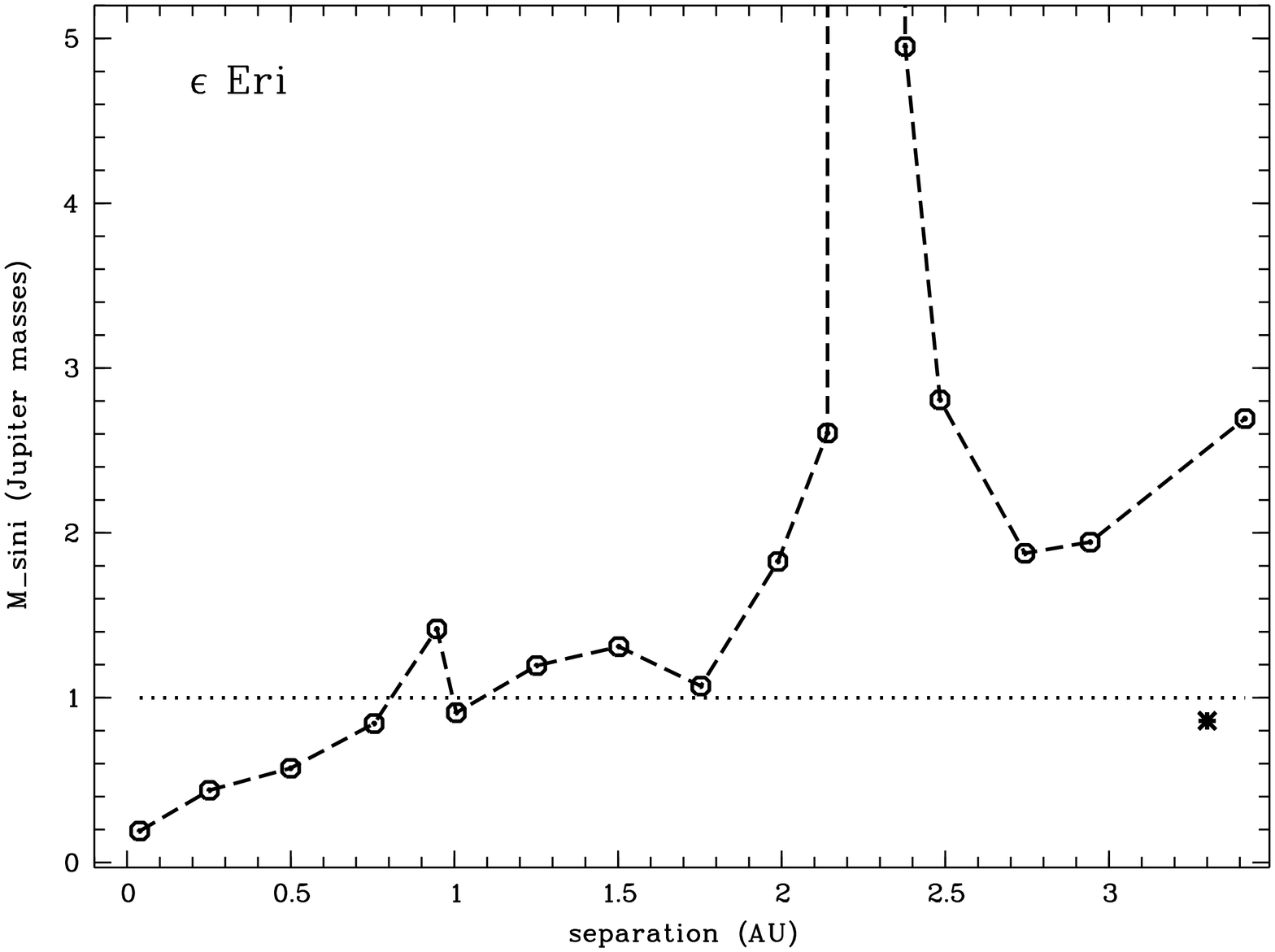,width=9.0cm,height=5.5cm,angle=270}}
        \vbox{\psfig{figure=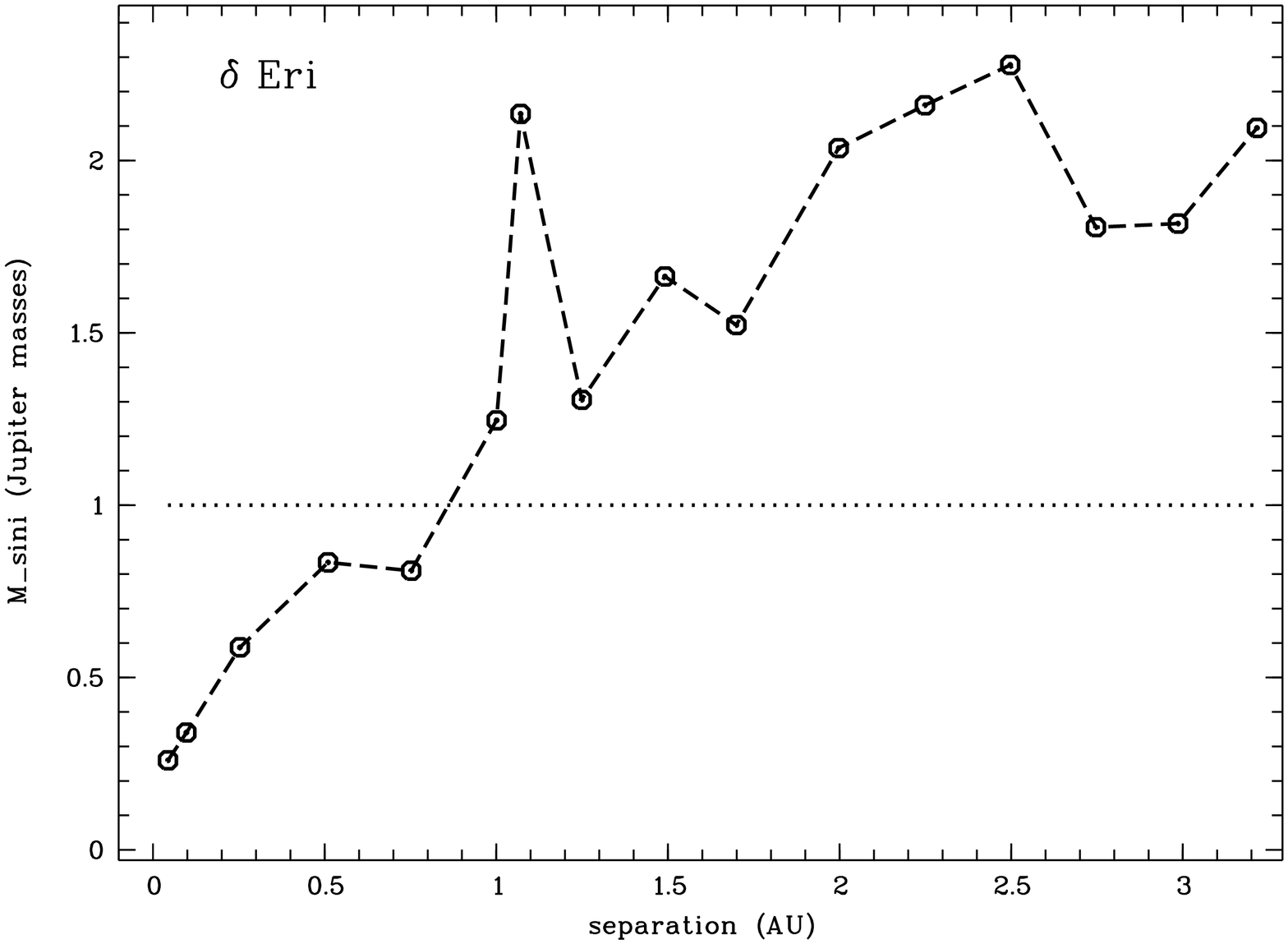,width=9.0cm,height=5.5cm,angle=270}}
   \par
        }
	\caption[]{
	Planetary companion limits for $\alpha$ For, $\zeta^{2}$ Ret, $\epsilon$ Eri and 
	$\delta$ Eri.
	}
  \label{limfig3}
\end{figure}

\begin{figure}
 \centering{
        \vbox{\psfig{figure=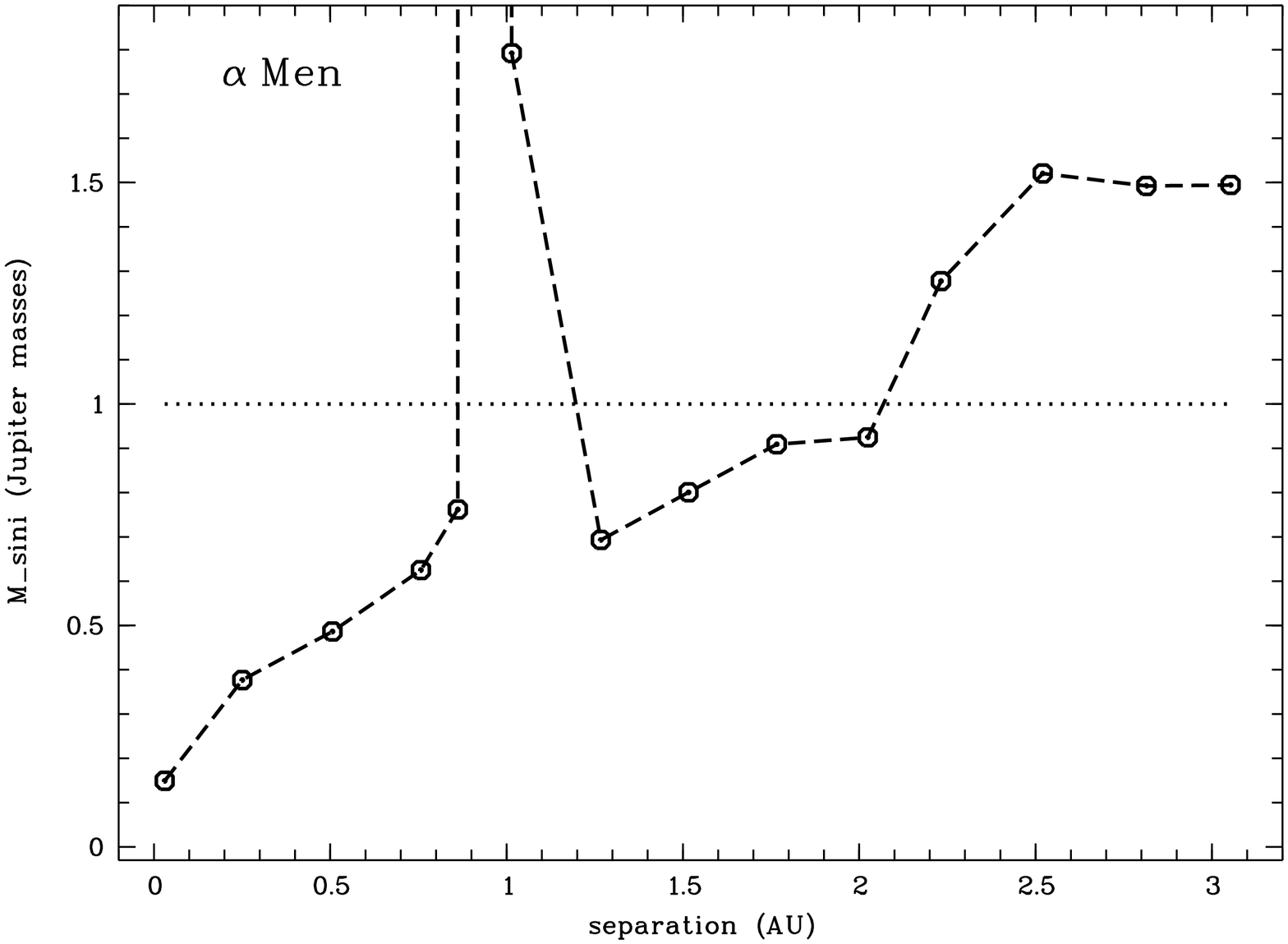,width=9.0cm,height=5.5cm,angle=270}}
        \vbox{\psfig{figure=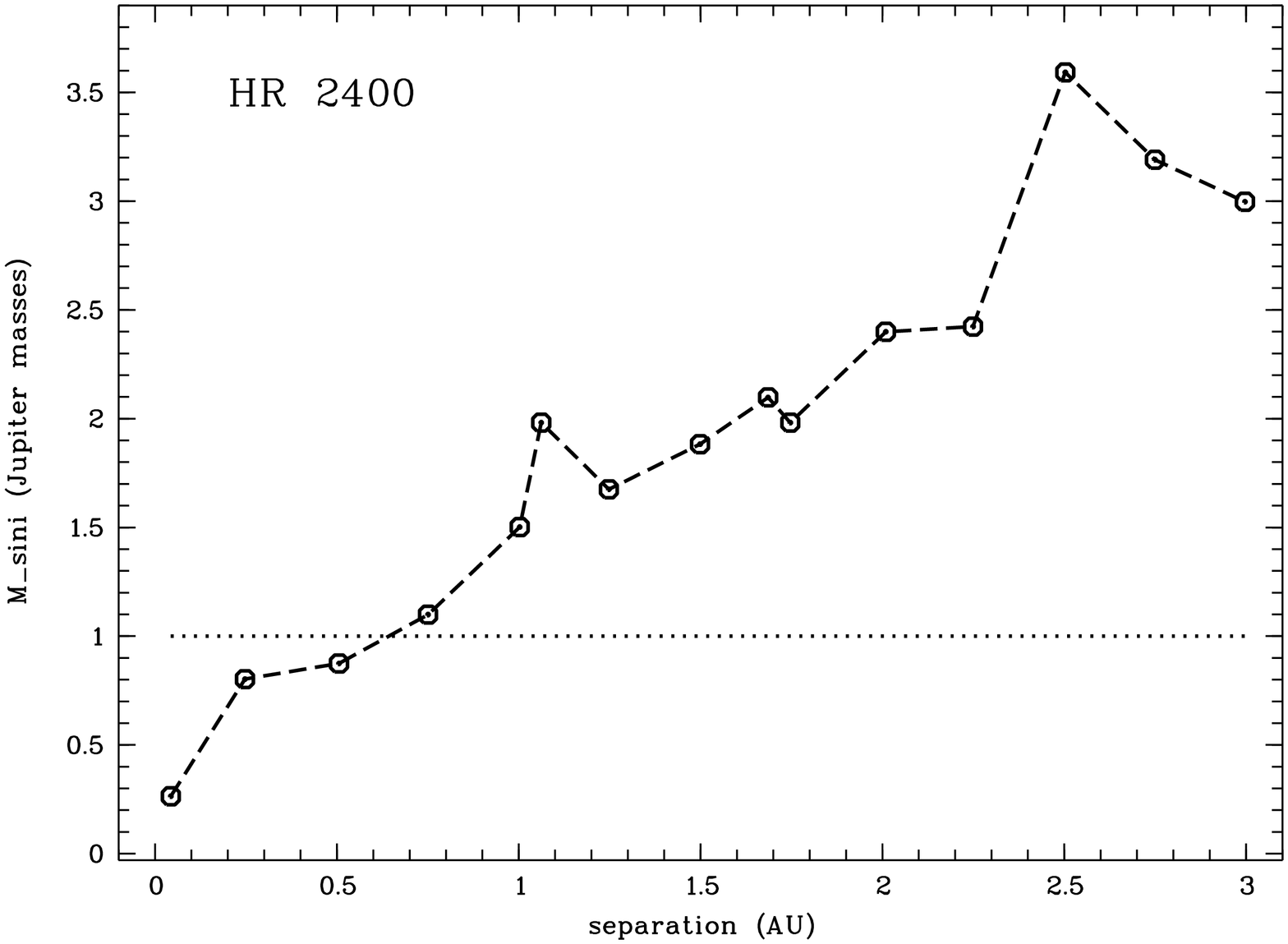,width=9.0cm,height=5.5cm,angle=270}}
        \vbox{\psfig{figure=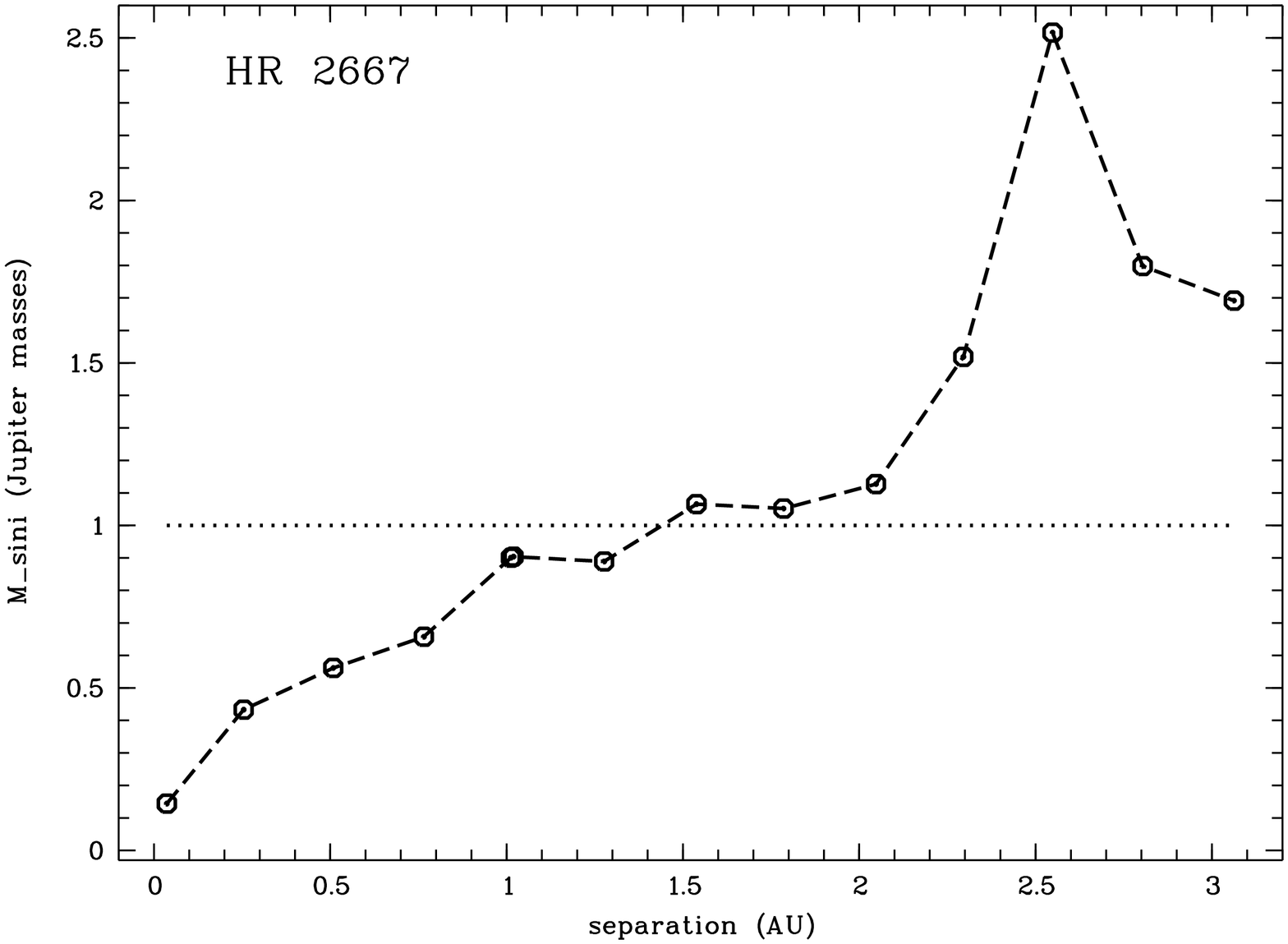,width=9.0cm,height=5.5cm,angle=270}}
        \vbox{\psfig{figure=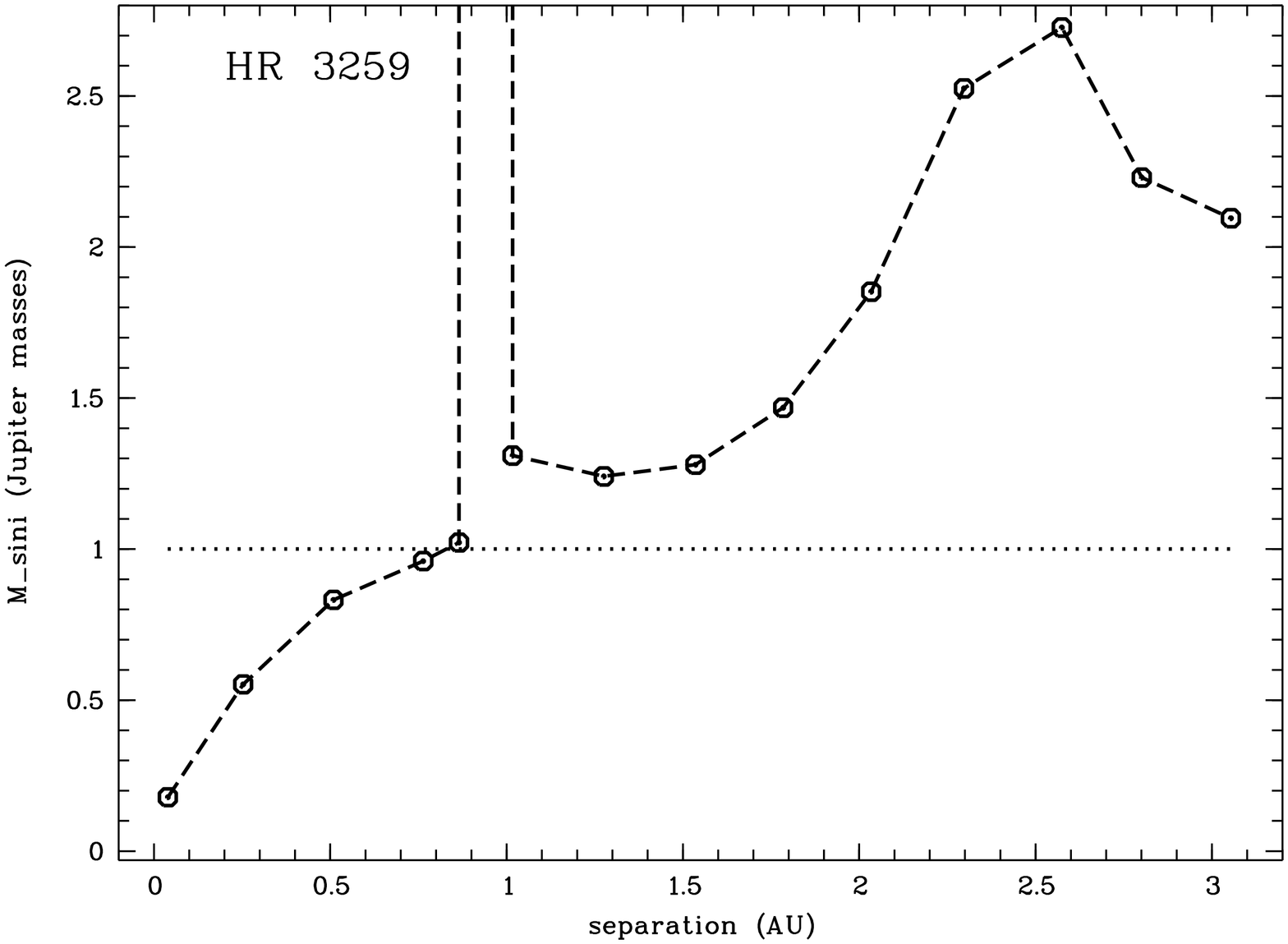,width=9.0cm,height=5.5cm,angle=270}}
   \par
        }
   \caption[]{
	Planetary companion limits for $\alpha$ Men, HR 2400, HR 2667 and HR 3259.
	}
  \label{limfig4}
\end{figure}

\begin{figure}
 \centering{
        \vbox{\psfig{figure=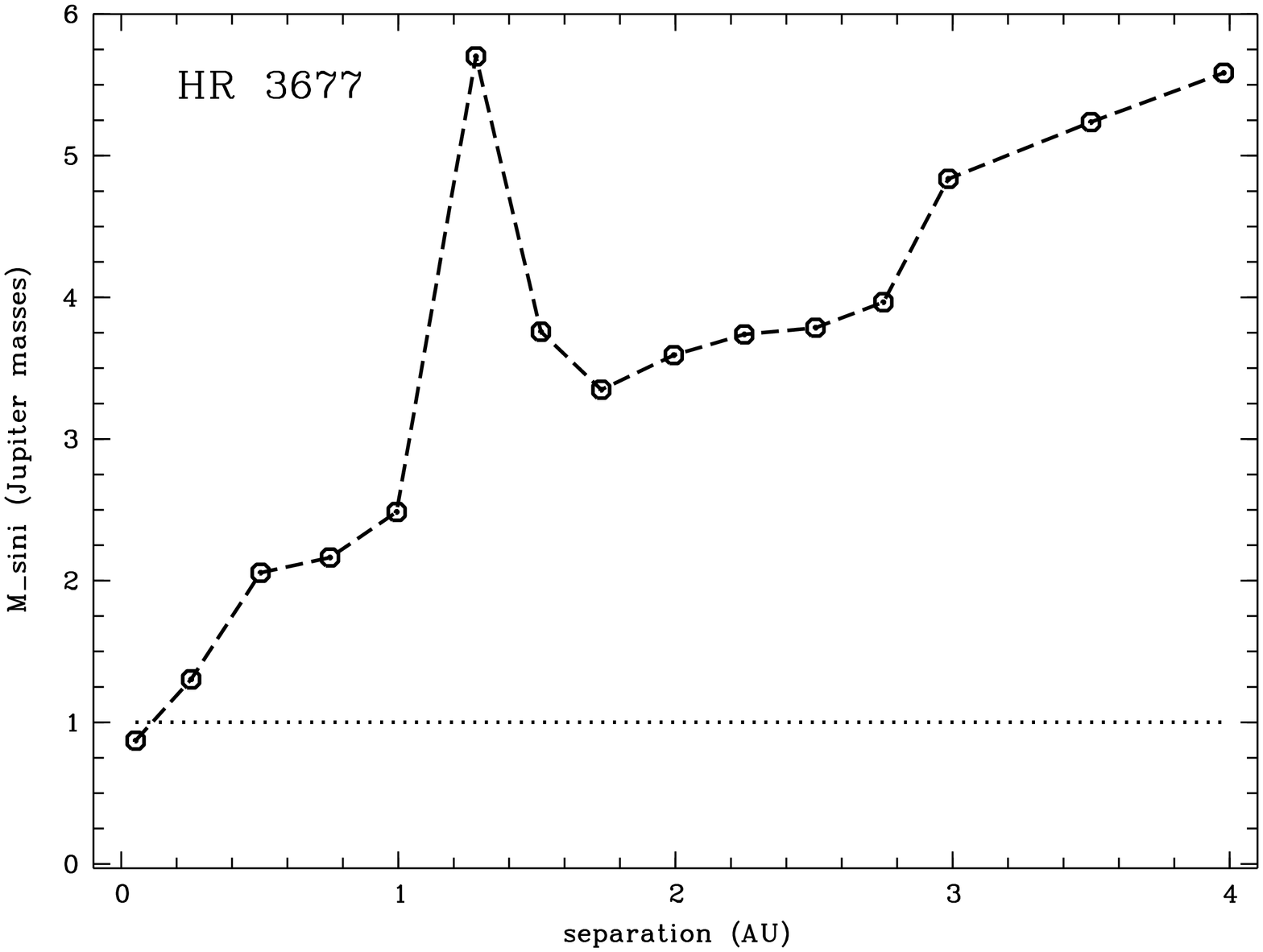,width=9.0cm,height=5.5cm,angle=270}} 
        \vbox{\psfig{figure=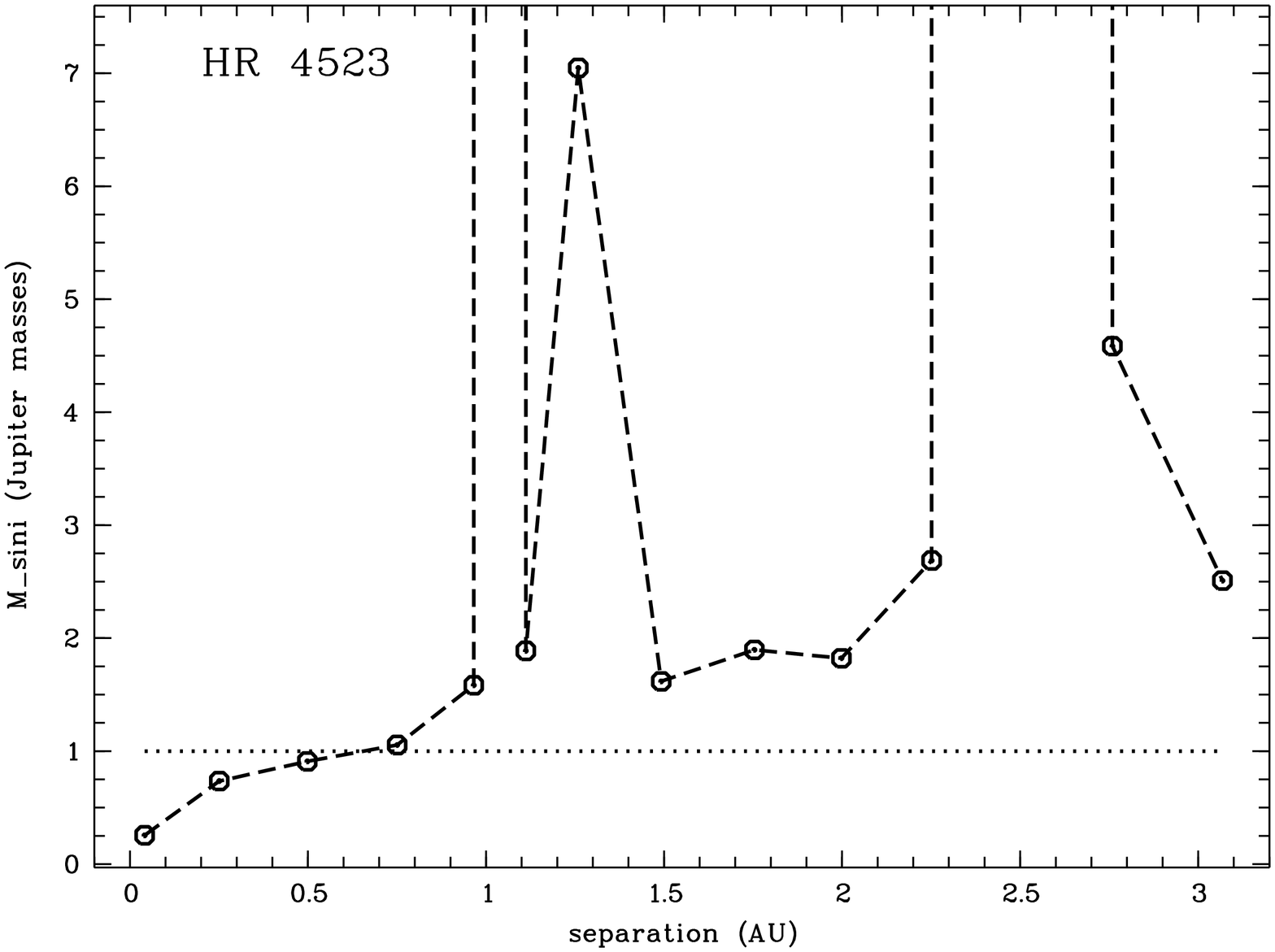,width=9.0cm,height=5.5cm,angle=270}}
        \vbox{\psfig{figure=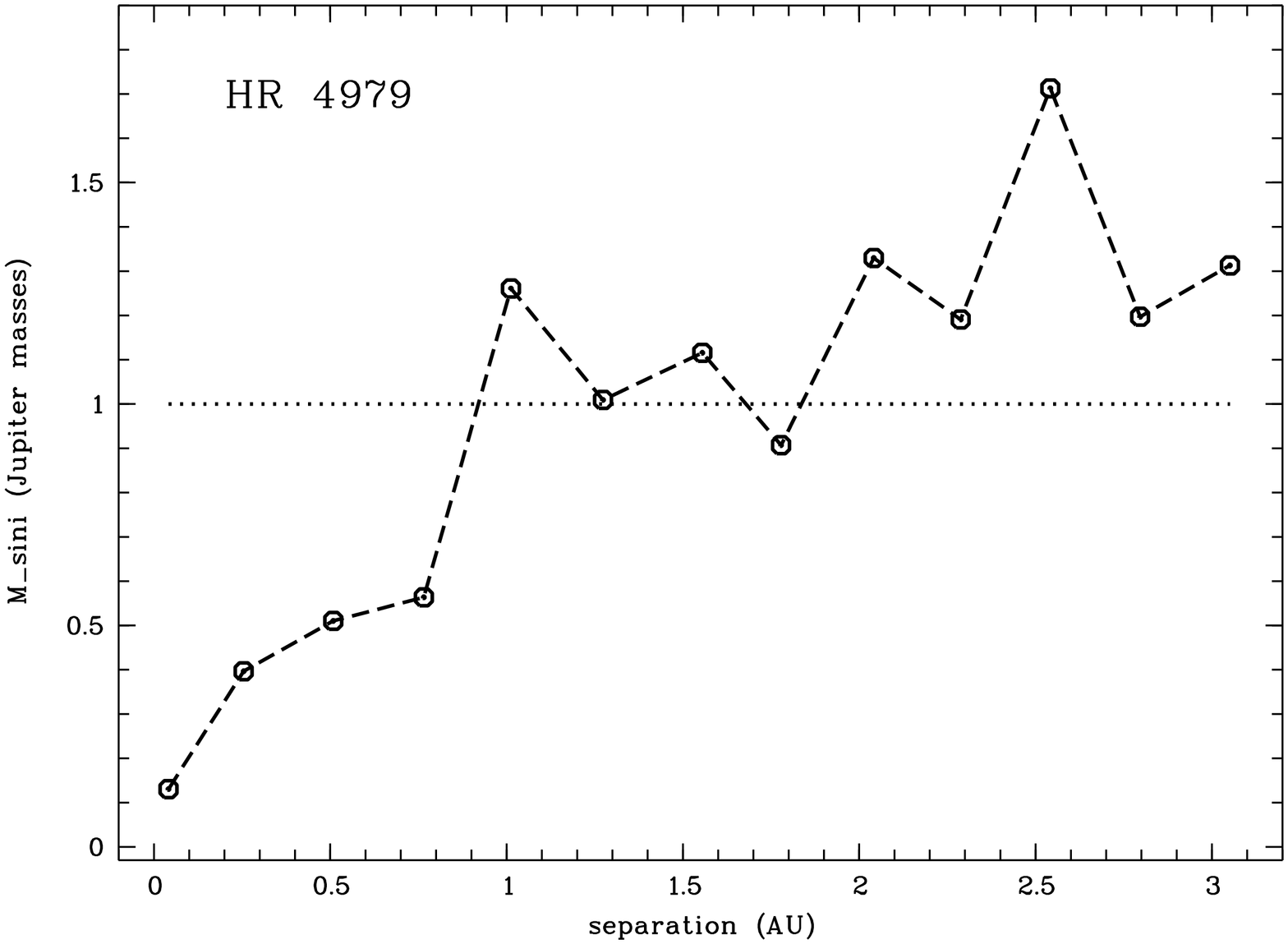,width=9.0cm,height=5.5cm,angle=270}}
        \vbox{\psfig{figure=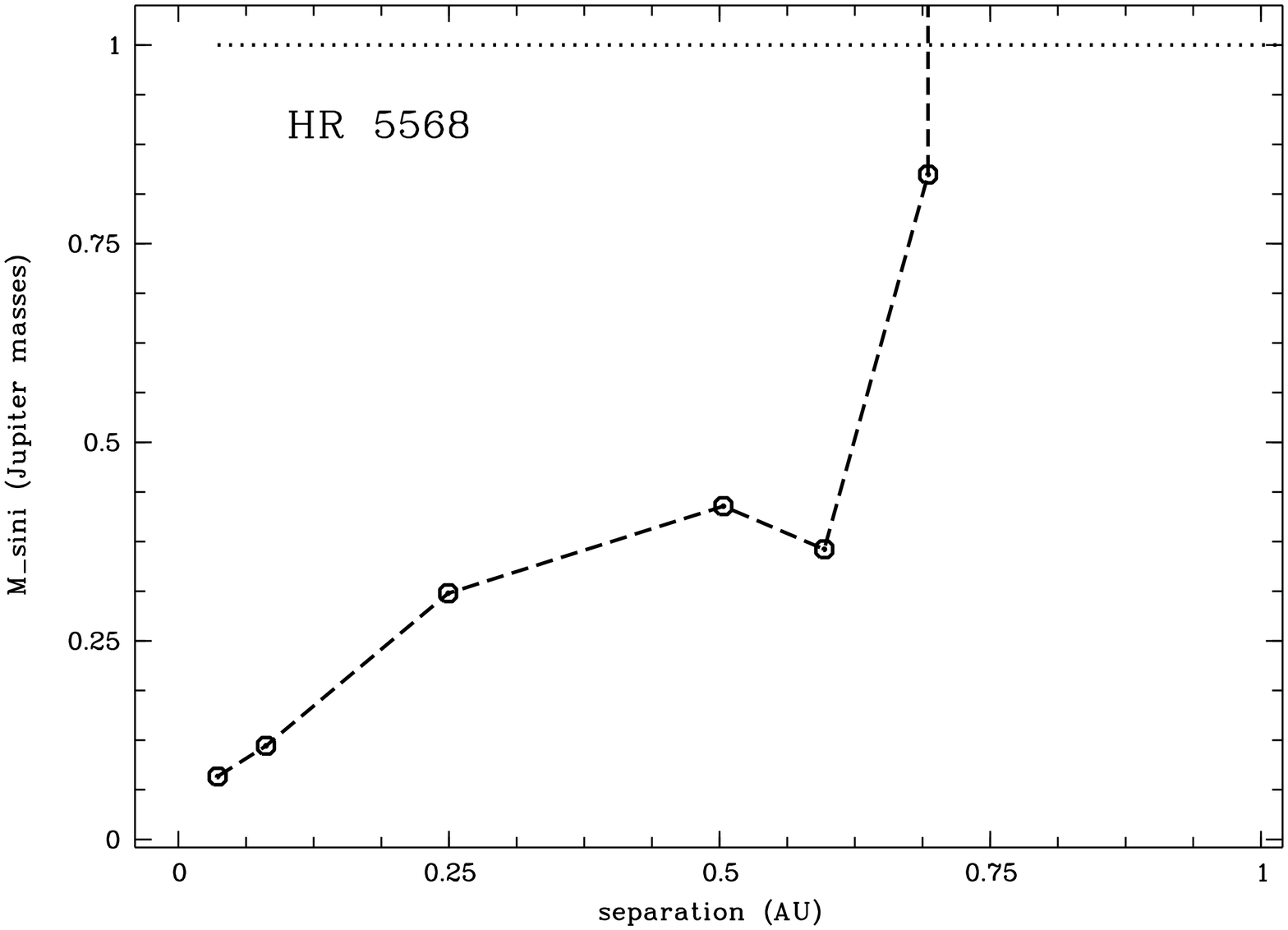,width=9.0cm,height=5.5cm,angle=270}}
   \par
        }
   \caption[]{
	Planetary companion limits for HR 3677, HR 4523, HR 4979 and HR 5568.
       	}
  \label{limfig5}
\end{figure}

\begin{figure}
 \centering{
        \vbox{\psfig{figure=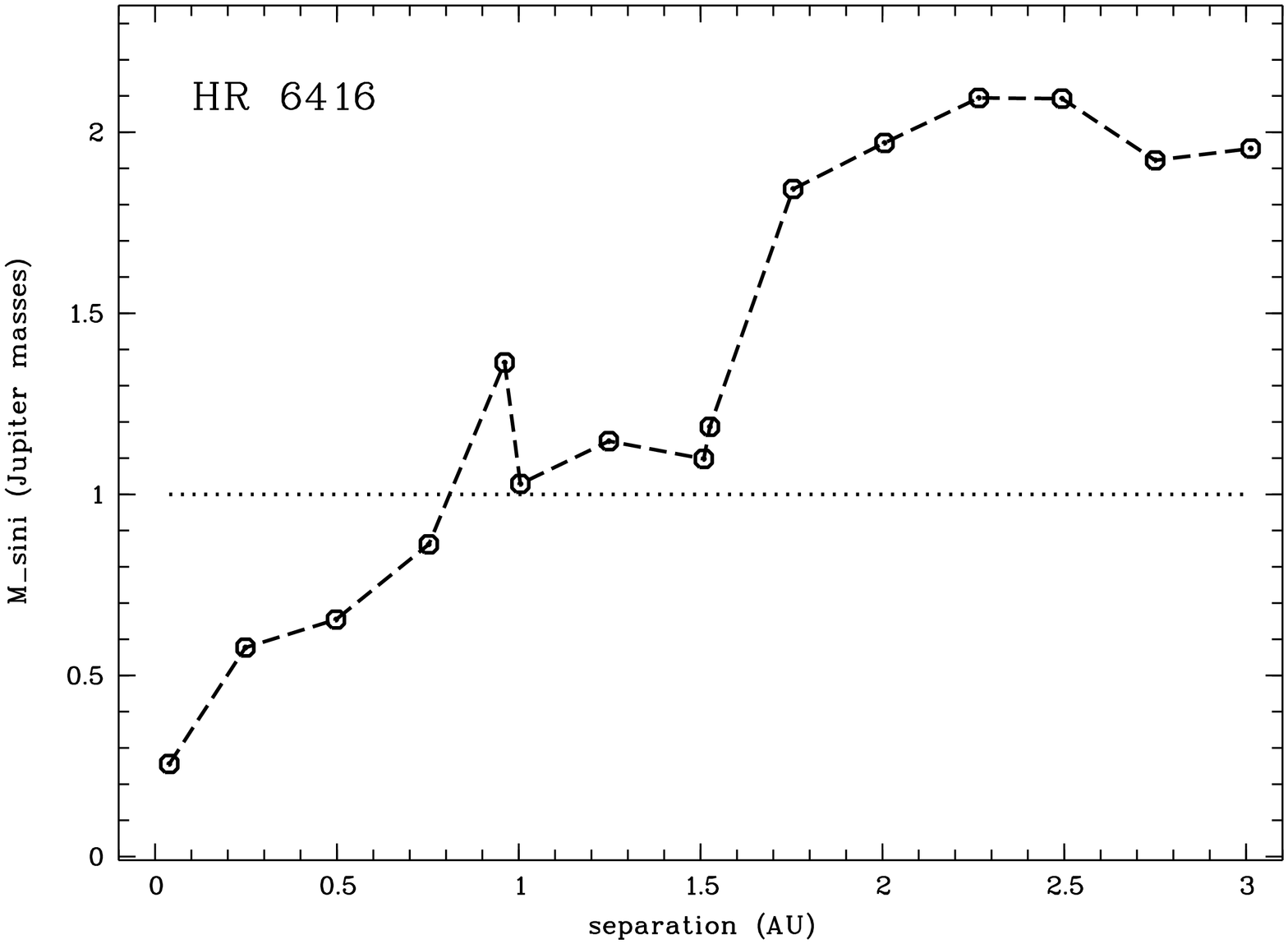,width=9.0cm,height=5.5cm,angle=270}}
        \vbox{\psfig{figure=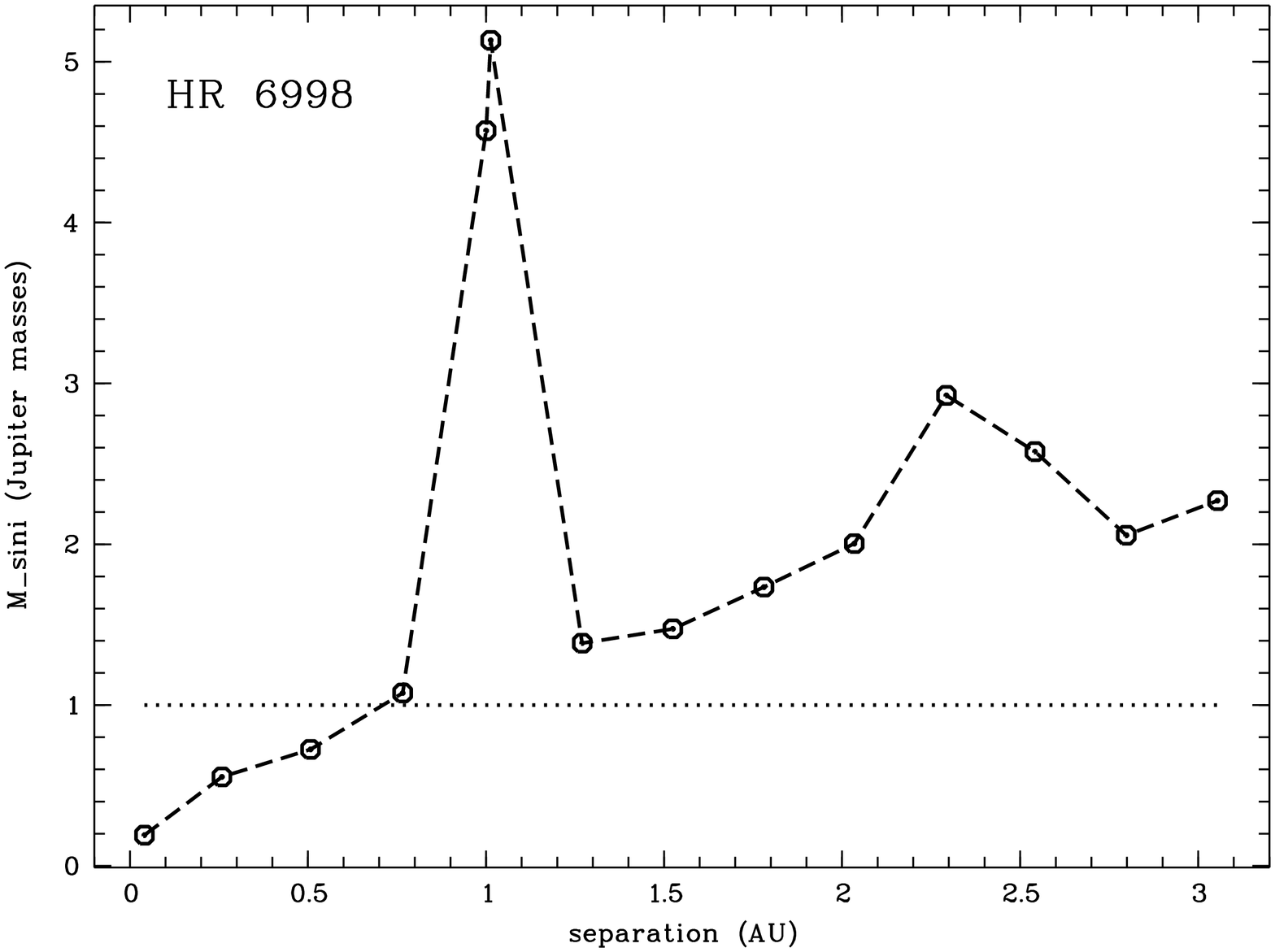,width=9.0cm,height=5.5cm,angle=270}}
        \vbox{\psfig{figure=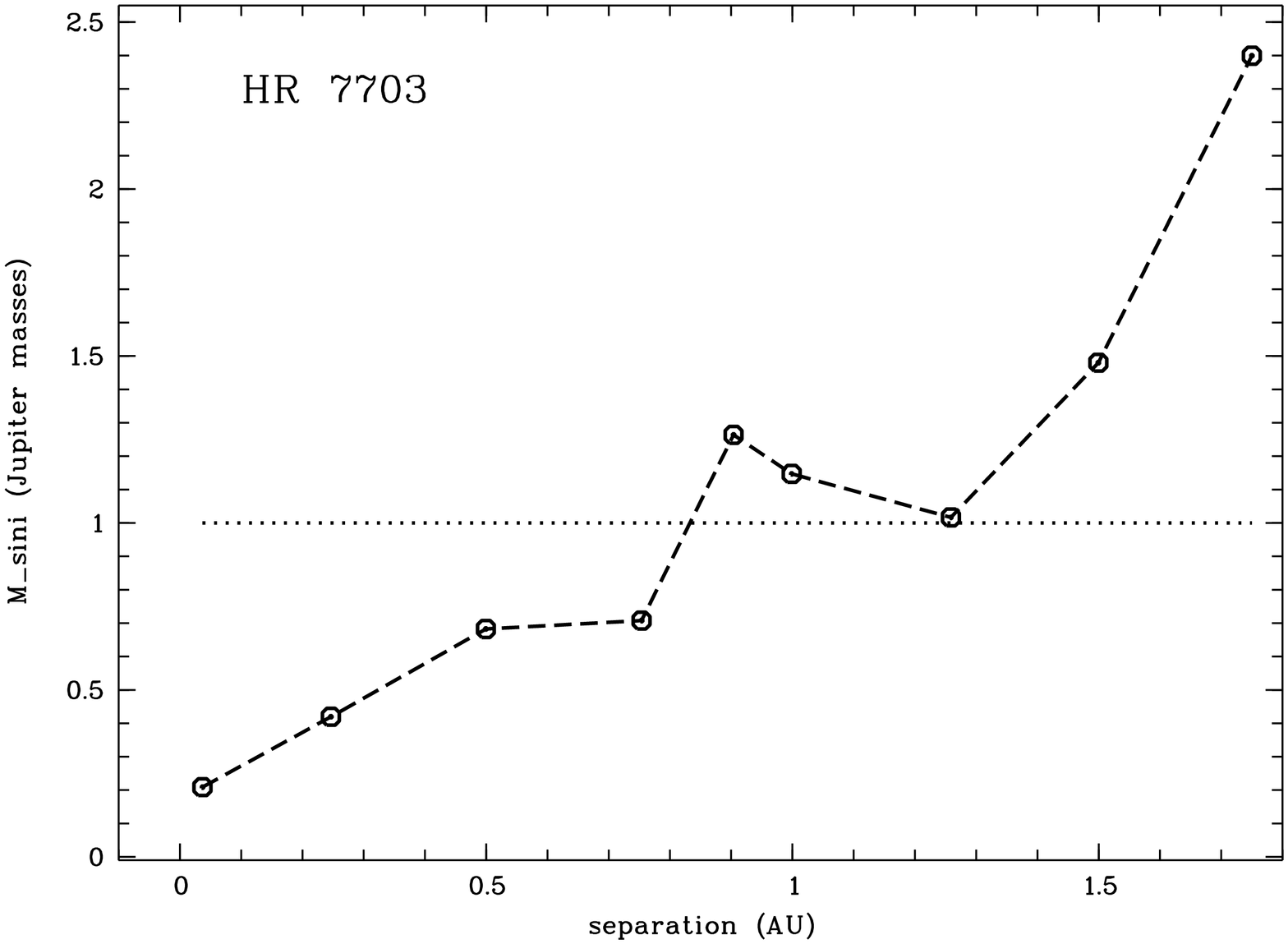,width=9.0cm,height=5.5cm,angle=270}}
        \vbox{\psfig{figure=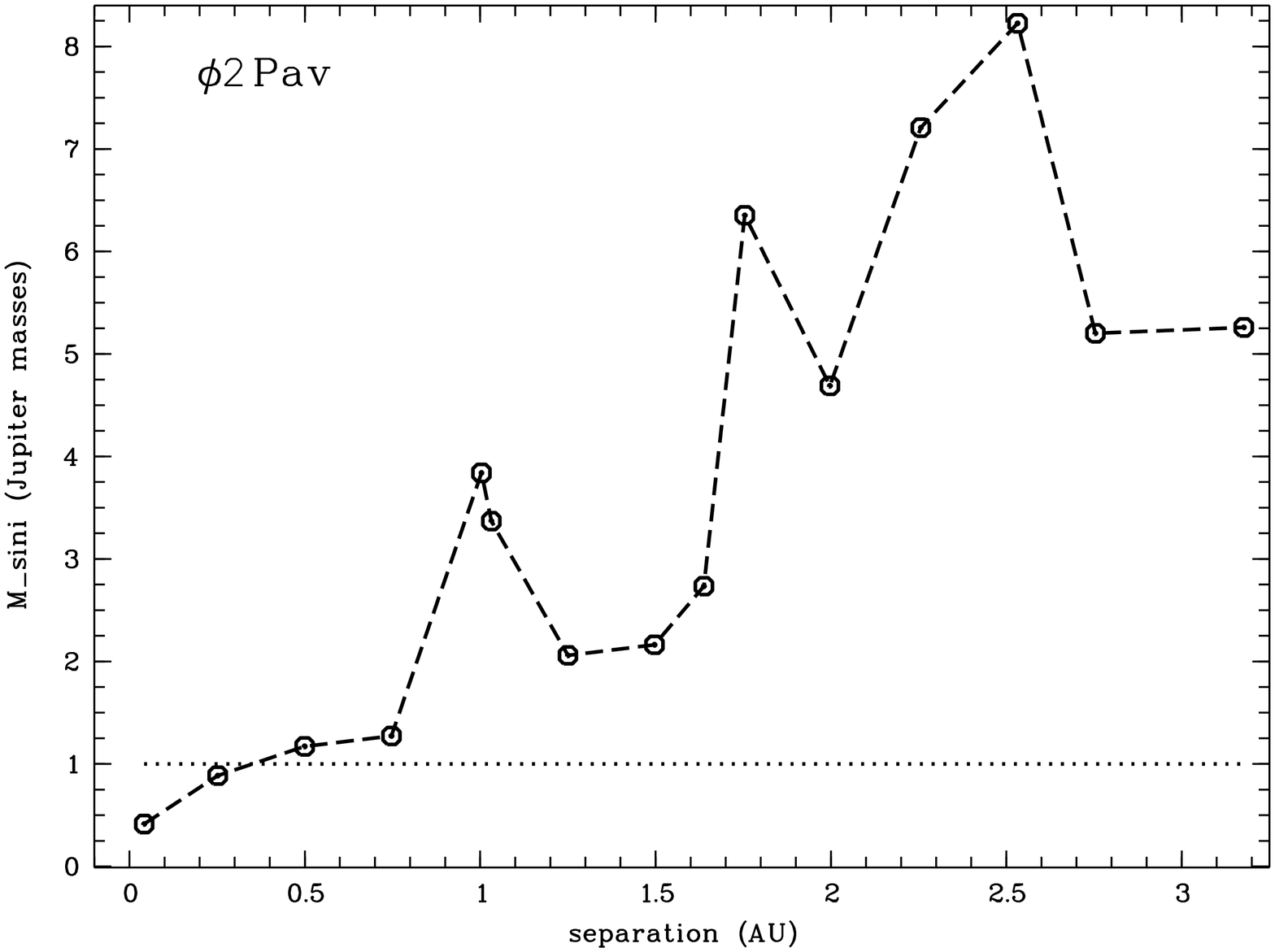,width=9.0cm,height=5.5cm,angle=270}}
   \par
        }
   \caption[]{
	Planetary companion limits for HR 6416, HR 6998, HR 7703 and $\phi^{2}$ Pav.
        }
  \label{limfig6}
\end{figure}

\begin{figure}
 \centering{
        \vbox{\psfig{figure=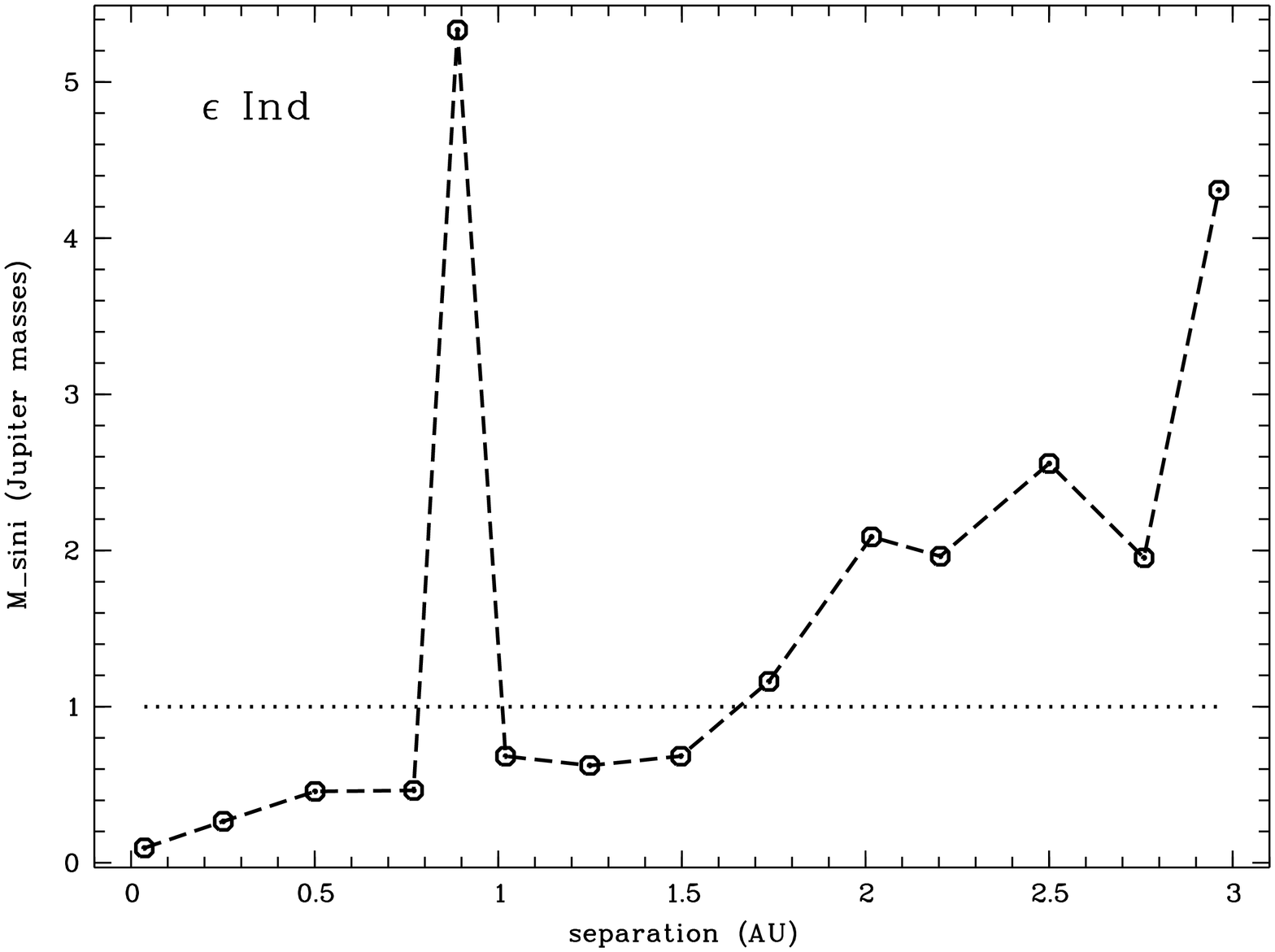,width=9.0cm,height=5.5cm,angle=270}}
        \vbox{\psfig{figure=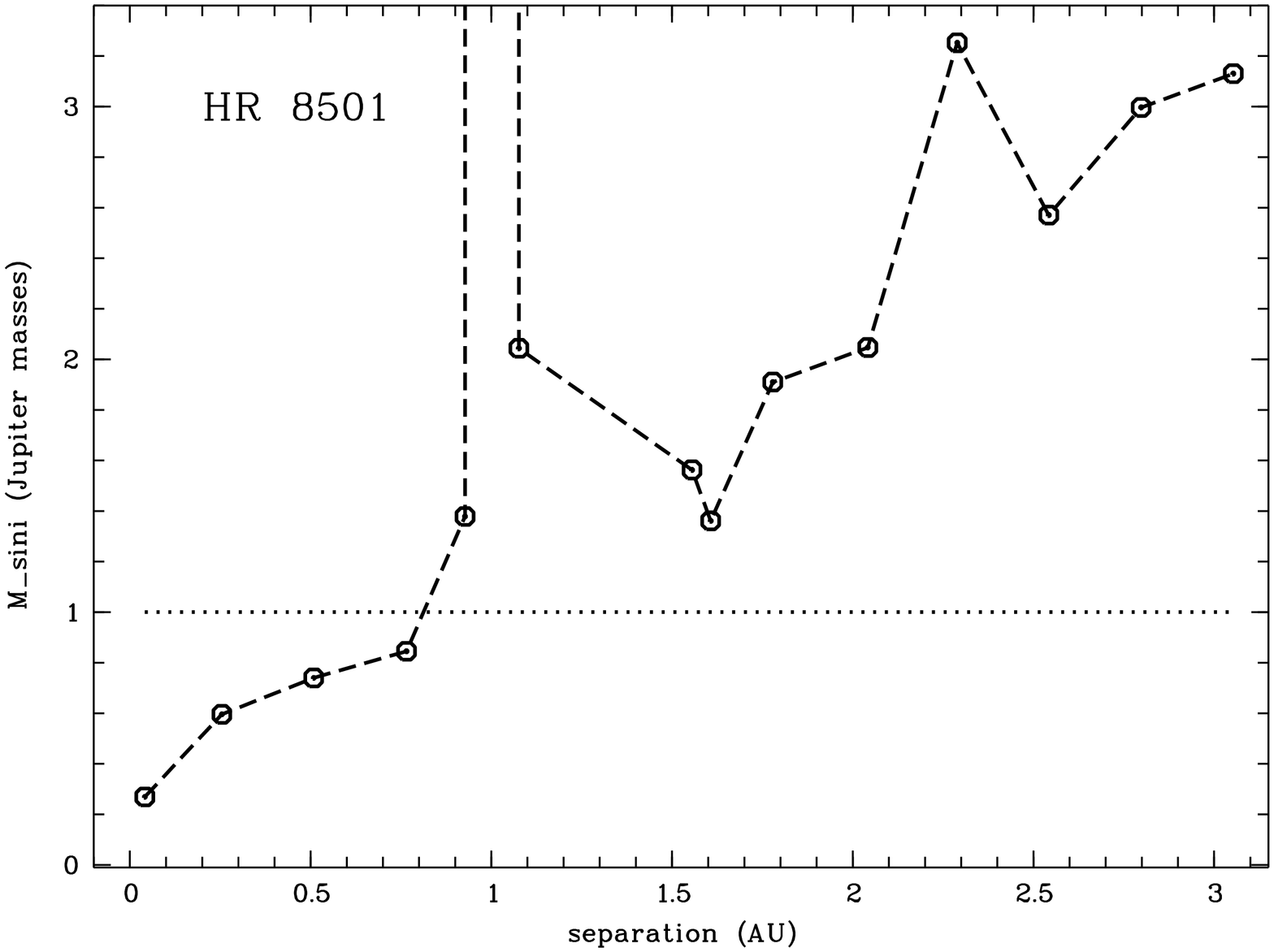,width=9.0cm,height=5.5cm,angle=270}}
        \vbox{\psfig{figure=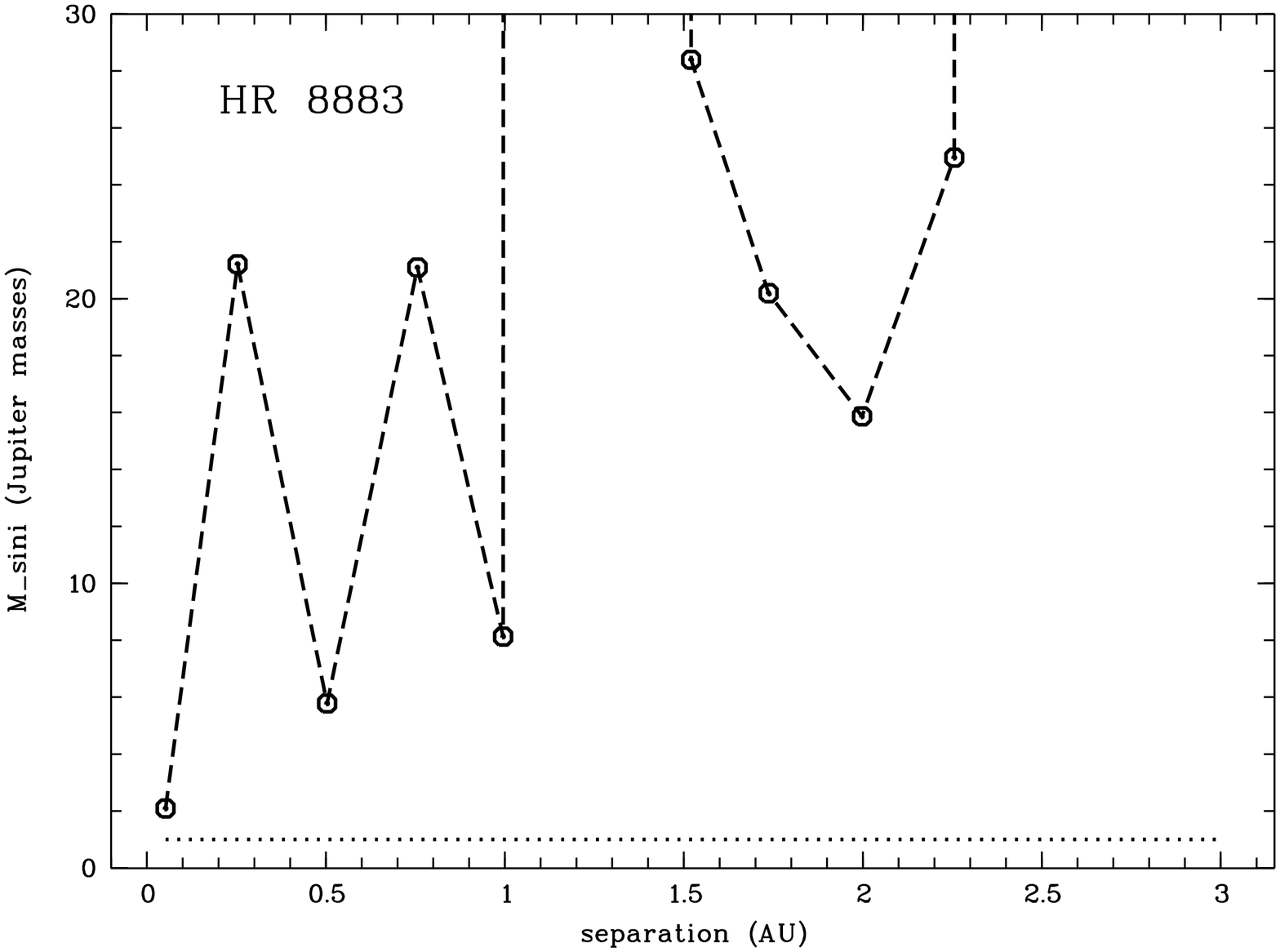,width=9.0cm,height=5.5cm,angle=270}}
	\vbox{\psfig{figure=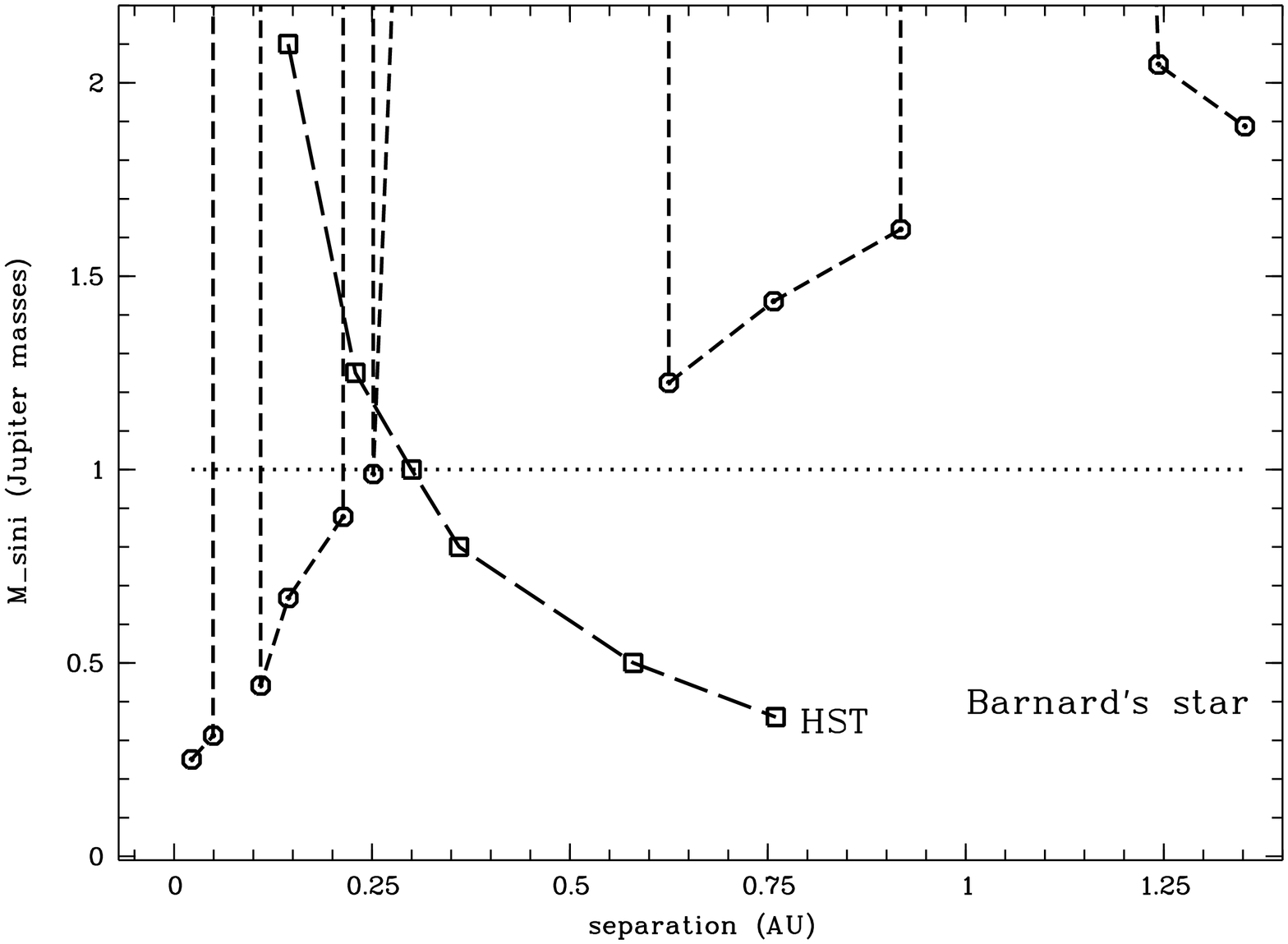,width=9.0cm,height=5.5cm,angle=270}}
   \par
        }
   \caption[]{   
	Planetary companion limits for $\epsilon$~Ind, HR 8501, HR 8883, and Barnard's star.
        }
  \label{limfig7}
\end{figure}

The upper mass-limits for planets we can place according to the method 
described in Sect. 5 are shown here for the 28 examined CES survey 
stars (the limits for planets orbiting either
$\alpha$ Cen A or B were presented in Endl et al. (\cite{michl01}).  
Fig.~\ref{limfig1} to Fig.~\ref{limfig7} display the $m\sin i$ values 
(in units of Jupiter masses) of detectable planetary signals as circles plotted vs. 
separation in AU. We can
exclude all planets in circular orbits with masses above these limits. 
As described in Sect. 5 some stars have windows of non-detectability 
which are plotted as vertical lines. The $m\sin i = 1~{\rm M}_{\rm Jup}$ border
is shown for a better comparison as horizontal dotted line in each figure.   

In the case of $\epsilon$~Eri the location of the long-period planetary
companion from Hatzes et al. (\cite{artie00}) is also indicated (Fig.~\ref{limfig3})
by an asterisk. 
For Barnard's star (Fig.~\ref{limfig7}) we include the HST astrometric limits 
from Benedict et al. (\cite{benedict}).


\end{document}